\documentclass[aps, prb, reprint, superscriptaddress]{revtex4-1}

\usepackage{amsmath, amssymb, times, mathrsfs, hyperref, array, bbm}
\usepackage{graphicx}
\usepackage[usenames,dvipsnames]{xcolor}
\usepackage{braket}

\hypersetup{colorlinks=false}

\bibpunct{[}{]}{,}{n}{}{}

\newcommand{\e}{\mathrm{e}}

\newcommand{\hlt}{\mathcal{H}}
\newcommand{\intg}{\mathbb{Z}}
\renewcommand{\vec}[1]{{\boldsymbol #1}}

\newcommand{\Jz}{J_\text{d}}

\begin{document}
\title{Partial flux ordering and thermal Majorana metals in (higher-order) spin liquids}

\author{Tim Eschmann}
\email[E-mail: ]{eschmann@thp.uni-koeln.de}
\affiliation{Institute for Theoretical Physics, University of Cologne, 50937 Cologne, Germany}
\author{Vatsal Dwivedi}
\affiliation{Institute for Theoretical Physics, University of Cologne, 50937 Cologne, Germany}
\author{Henry F. Legg}
\altaffiliation[Present address: ]{Department of Physics, University of Basel, Klingelbergstrasse 82, CH-4056 Basel, Switzerland}
\affiliation{Institute for Theoretical Physics, University of Cologne, 50937 Cologne, Germany}
\author{Ciar\'an Hickey}
\affiliation{Institute for Theoretical Physics, University of Cologne, 50937 Cologne, Germany}
\author{Simon Trebst}
\affiliation{Institute for Theoretical Physics, University of Cologne, 50937 Cologne, Germany}


\begin{abstract}

In frustrated quantum magnetism, chiral spin liquids are a particularly intriguing subset of quantum spin liquids
in which the fractionalized parton degrees of freedom form a Chern insulator.
Here we study an exactly solvable spin-3/2 model which harbors not only chiral spin liquids but also
spin liquids with higher-order parton band topology 
-- a trivial band insulator, a  Chern insulator with gapless chiral edge modes,
and a second-order topological insulator with gapless corner modes.
With a focus on the thermodynamic precursors and thermal phase transitions associated with these distinct states,
we employ numerically exact quantum Monte Carlo simulations to reveal a number of unconventional phenomena.
This includes a heightened thermal stability of the ground state phases,
the emergence of a partial flux ordering of the associated $\intg_2$  lattice gauge field, and 
the formation
of a thermal Majorana metal regime extending over a broad temperature range. 

\end{abstract}

\maketitle


\section{Introduction}
\label{sec:intro}

The emergence of topological phases from local constraints induced by competing interactions in frustrated quantum magnets has fascinated researchers for decades \cite{Balents2010spin,Savary2017quantum}.
In a groundbreaking conceptual work, Kalmeyer and Laughlin~\cite{KalmeyerLaughlin1987} in the late 80s
put forward
the formation of a bosonic analogue of the fractional quantum Hall state in what has since been termed a {\sl chiral spin liquid}.
First envisioned as resonating valence bond (RVB) ground states of geometrically frustrated Heisenberg antiferromagnets
and relevant to high-temperature superconductivity~\cite{anderson_science_235_1987,wen_prb_39_1989},
such chiral spin liquids have remained elusive for many years. In the past decade, however, tremendous progress has been
achieved in firmly establishing chiral spin liquids as ground states of microscopic models
\cite{Kitaev2006anyons,yao-kivelson,schroeter_prl_99_2007,messo_prl_108_2012,bauer_ncom_5_2014,gong_srep_4_2014}
and their eventual experimental observation in quantized thermal Hall measurements \cite{kasahara_majorana_2018,yokoi2020halfinteger}.
On the theoretical side, conceptual insight has been gained from the unambiguous numerical detection of chiral spin liquids via the calculation
of modular matrices \cite{Rowell2009,Bruillard2016} from the ground state entanglement structure \cite{Zhang2012}  of a number of 
kagome models \cite{bauer_ncom_5_2014,gong_srep_4_2014}.
Analytically, Kitaev showed that a time-reversal symmetry breaking magnetic field gives rise to a non-Abelian chiral spin liquid in his
eponymous model \cite{Kitaev2006anyons}.
The advent of Kitaev materials \cite{Trebst2017} has since produced a direct experimental observation of this state in 
measurements of a half-quantized thermal Hall effect in the spin-orbit entangled Mott insulator $\alpha$-RuCl$_3$ \cite{kasahara_majorana_2018}.

\begin{figure*}[t!]
   \centering
    \includegraphics[width=\linewidth]{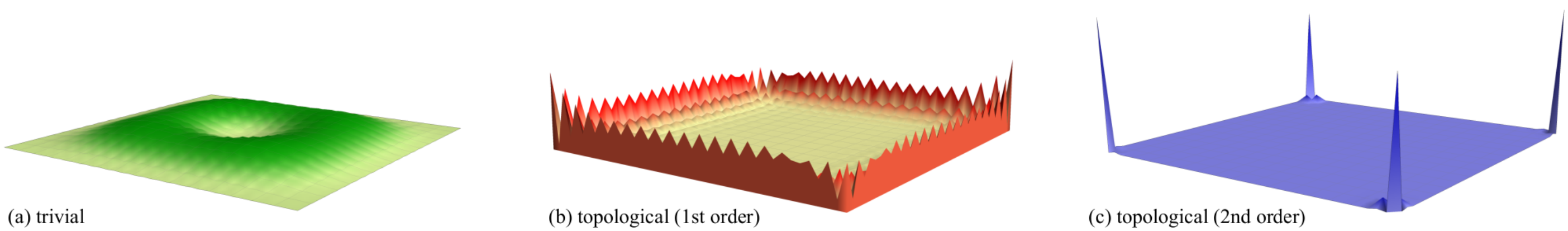}
  \caption{{\bf Variable-order spin liquids} can be conceptualized via the formation of (topological) band structures of emergent partons.
  		Shown here is the reflection of these band structures in the ground-state parton wavefunction $|{\psi_{ij}}|^2$ for
		different settings with (a) trivial topology, (b) conventional (first-order) topology with gapless (chiral) edge modes, and (c) second-order topology with localized corner modes.
		The actual calculations have been performed in the Majorana representation of model \eqref{eq:GammaModel}
		on a $16\times 16$ lattice.
   }
    \label{fig:Wavefunctions}
\end{figure*}

The physical mechanism underlying the formation of a chiral spin liquid can be elegantly formulated using Wen's parton construction \cite{Wen1992}: At low temperatures the original spin degrees of freedom fractionalize into a parton coupled to an emergent lattice gauge field \cite{Savary2017quantum}. For the Kitaev model, this parton decomposition is typically done \cite{Kitaev2006anyons} in terms of Majorana fermions coupled to a $\intg_2$ gauge field
\footnote{An alternative parton decomposition employs complex fermions coupled to a $U(1)$ gauge field \cite{burnell_su2_2011}, which then leads to a description of the Kitaev spin liquid as a nodal superconductor. This picture has been particularly insightful in explaining the in-field behavior of the antiferromagnetic Kitaev honeycomb model which exhibits a Higgs transition to an intermediate gapless $U(1)$ spin liquid \cite{hickey_emergence_2019}.}
-- thereby casting the original interacting spin model to a free (Majorana) fermion problem with a static $\intg_2$ gauge order (at zero temperature).
One can then resort to the tools of topological band theory to classify the band structure of the emergent Majorana fermions, and thereby classify the fundamental topological properties of the spin liquid state.
For the case of the Kitaev honeycomb model, this perspective results in an understanding of the field-induced topological spin liquid as
the formation of a  Chern insulator (in the Majorana band structure) with all bands carrying a non-trivial Chern number $|\nu| = 1$.
Such a state -- a Chern insulator of the emergent Majorana fermions -- is gapped in the bulk, but exhibits chiral gapless modes at its boundary that are topologically protected, as depicted in Fig.~\ref{fig:Wavefunctions}(b). That such a Majorana Chern insulator indeed forms in the Kitaev material $\alpha$-RuCl$_3$ is substantiated by the observation of a {\sl half}-quantized thermal Hall conductance, which is direct evidence for Majorana fermions (and not conventional electrons) carrying the thermal edge currents. Subsequent measurements \cite{yokoi2020halfinteger} of the anomalous character of the quantized thermal Hall effect, which arises even without any {\sl perpendicular} magnetic field component, have brought  verification that the observed state is indeed arising from the formation of topological Chern bands (and not the formation of Landau levels).

In parallel the field of topological band theory has been developed \cite{BernevigHughes}. There one distinguishes {\sl strong} topological insulators (TIs), whose topological features are protected by time reversal and/or charge conjugation symmetries, from {\sl crystalline} TIs whose topology is endowed by certain lattice symmetries. 
A particular example of such crystalline TIs are second-order topological insulators, an instance of {\sl higher-order topology} \cite{neupert_tci_rev}.
An $n^\text{th}$ order TI in $d$ spatial dimensions exhibits ($d-n$)-dimensional topologically protected gapless modes that are localized at the intersection of $n$ boundary planes, while the boundaries of codimension less than $n$ remain fully gapped. A second-order TI in two spatial dimensions thus exhibits topologically protected corner modes, i.e. zero-dimensional gapless modes at the intersection of two boundaries giving rise to a corner. 
Such higher-order topology can also play out in the context of a quantum spin liquid -- with the itinerant fractionalized parton degrees of freedom forming a non-trivial topological band structure. 
The example of a {\sl second-order spin liquid} with topologically protected Majorana corner modes has been
discussed in the context of an exactly solvable spin-3/2 generalization of the Kitaev model \cite{DwivediCornerModes2018}. 
Within this same framework, the chiral spin liquid can be thought of as a first-order spin liquid, since a Chern insulator can be considered an example of a first-order topological insulator \footnote{
Note that, though all of the spin liquid ground states spontaneously break time-reversal symmetry, we reserve the term `chiral spin liquid' for those that exhibit chiral edge states (or, more technically, those that possess a non-zero chiral central charge). }.

In this manuscript, we study the thermodynamic precursors and symmetry-breaking thermal phase transitions leading to the formation of a family of spin liquid ground states which exhibit the full range of parton band topology, second-order, conventional (first-order), and trivial topology, in a generalized Kitaev model. We employ sign-problem free quantum Monte Carlo simulations in the parton basis \cite{nasu2014vaporization}. These numerically exact calculations allow us to track the fractionalization of the original spin degrees of freedom, the formation of gauge order, and the spontaneous breaking of time-reversal symmetry upon entering the different flavors of spin liquid ground states. Our main results include (i) the observation that the {\sl thermal stability} of the chiral spin liquid in our model is enhanced by almost an order of magnitude in comparison with other chiral spin liquid models, with the highest transition temperatures reaching about $1/10$
of the bare coupling strength; (ii) the emergence of {\sl partial flux order} in an intermediate temperature range, accompanied by a characteristic 3-peak signature in the specific heat; and (iii) the formation of a gapless phase at finite temperatures that is best described as a {\sl thermal Majorana metal}. 

Our discussion of these results in the remainder of the manuscript is structured as follows.
In Section \ref{sec:model} we briefly introduce a generalized spin-3/2 Kitaev model, its $\Gamma$-matrix representation, and the underlying five-coordinated Shastry-Sutherland lattice. In discussing its analytical solution at zero temperature, we also introduce the parton basis relevant to our sign-free QMC simulations to explore the thermodynamics at finite temperatures.
The formation of a conventional (first-order) chiral spin liquid is discussed in Section \ref{sec:ShastryChiral}. Our main results on thermal stability, partial flux ordering, and thermal metal formation are all discussed in detail here.
In Section \ref{sec:ShastrySOSL} we then  turn to the formation of a second-order spin liquid, whose zero-temperature properties we previously discussed in Ref.~\onlinecite{DwivediCornerModes2018}. Our focus here is on its thermodynamic properties.
We conclude with an outlook in Section \ref{sec:summary}.


\section{The Shastry-Sutherland Kitaev model}
\label{sec:model}

We start our discussion with a brief review of the generalization of the Kitaev model to the Shastry-Sutherland lattice \cite{WuHungGammaMatrix2009, DwivediCornerModes2018},
its fundamental (lattice) symmetries, the formation of spin liquid ground states of various levels of topology, and its numerical representation in
sign-free quantum Monte Carlo simulations.


\subsection{The model: spin-3/2 and Gamma matrices}

The Kitaev honeycomb model is the paradigmatic example of an exactly solvable quantum spin liquid model. The model consists of spin-$1/2$ degrees of freedom on the sites of a honeycomb lattice interacting via bond-dependent Ising interactions. By representing the spin operators in terms of Majorana fermions, the model can be reduced to a nearest-neighbor hopping model of non-interacting fermions coupled to a \emph{static} $\intg_2$ gauge field. The Kitaev model and its exact solution can be straightforwardly generalized to other lattices with an odd coordination number, $z=2n-1$, wherein the local ``spins'' are decomposed into $2n$ Majorana fermions.

\begin{figure}[t]
   \centering
    \includegraphics[width=0.65\columnwidth]{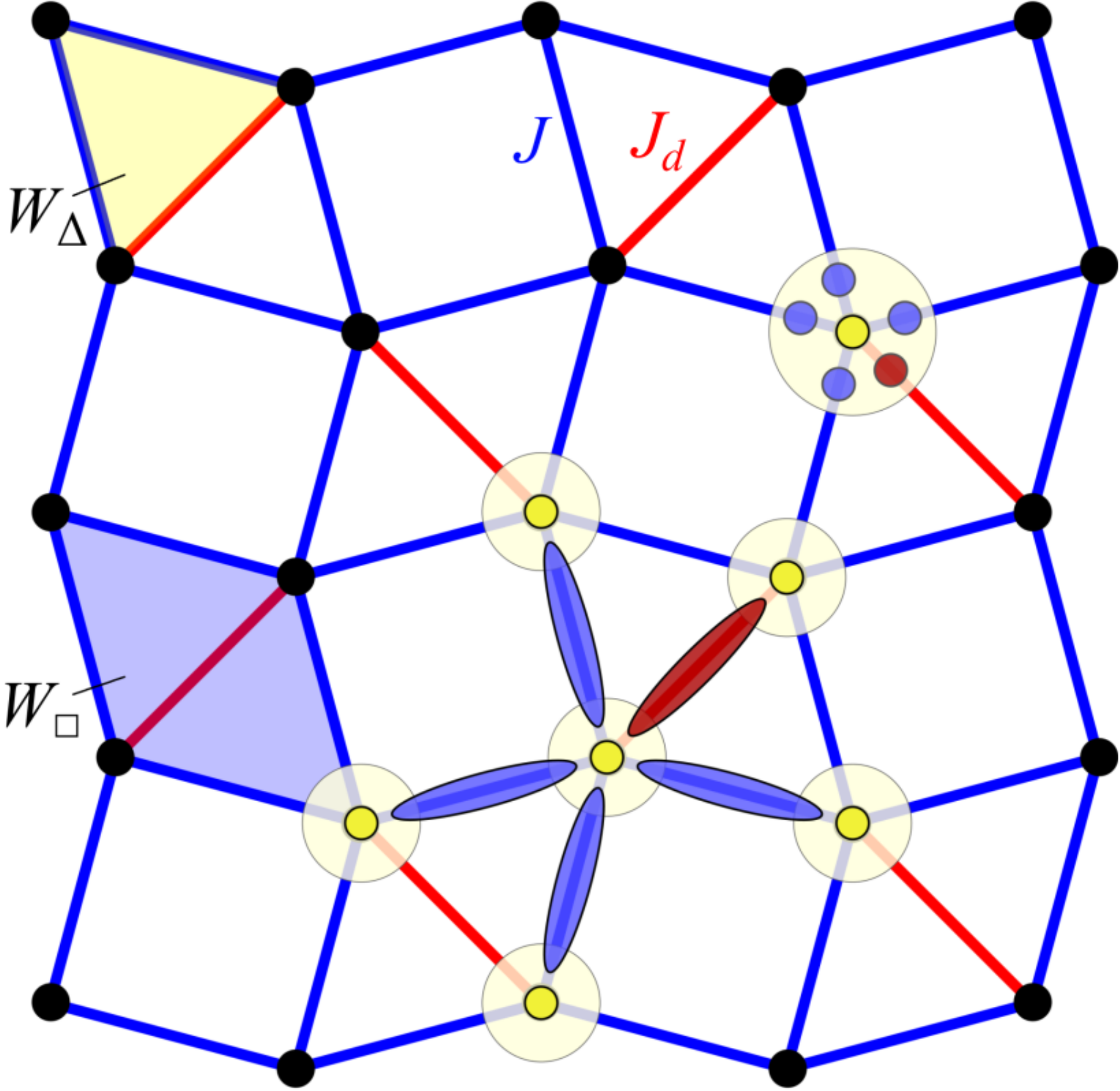}
  \caption{{\bf Shastry-Sutherland lattice.} On this five-coordinated lattice, a generalized version of the Kitaev model can be defined, where anticommuting $ 4 \times 4$ $\Gamma$-matrices on the lattice sites represent spin-$\frac{3}{2}$ degrees of freedom. The horizontal and vertical (blue) bonds carry a coupling $J$ (which we set to unity), and the diagonal (red) bonds a coupling $\Jz$. For a suitable choice of these parameters, the resulting higher-order spin liquid system is shown to possess a topologically non-trivial ground state.
   }
    \label{fig:LatticeDefinition}
\end{figure}

Here, we study such a generalization of Kitaev's honeycomb model to the pentacoordinated Shastry-Sutherland lattice, previously introduced in Refs. \cite{WuHungGammaMatrix2009, DwivediCornerModes2018} (see Fig. \ref{fig:LatticeDefinition}). The lattice is most well-known for the orthogonal dimer model, which was solved by Shastry and Sutherland \cite{shastry-sutherland} and serves as an effective low-temperature model for the transition metal oxide SrCu$_2$(BO$_3$)$_2$ \cite{Kageyama1999}. The generalized Kitaev model is described by the Hamiltonian
\begin{equation}
  \mathcal{H} = - \sum_{\langle j,k \rangle_\gamma} J^\gamma \Gamma_j^\gamma \Gamma_k^\gamma,
  \label{eq:GammaModel}
\end{equation}
where $\gamma = 1, \dots 5$ labels the bond direction and we have a set of five $4\times4$ anticommuting matrices $\Gamma_j^\gamma$ for each site. Physically, the $\Gamma$-matrices can be interpreted as acting on either $j = \frac{3}{2}$ spins or two coupled spin-$\frac{1}{2}$ degrees of freedom, such as spin and orbital degrees of freedom, on each lattice site \cite{DwivediCornerModes2018}.

To solve this model exactly, we represent the $\Gamma$-matrices on each site in terms of six Majorana operators $c_j$ and $\{b^\gamma_j\}$ by setting \footnote{
  Note that this representation doubles the dimension of the local Hilbert space. To remedy this situation, i.e., to project down to the {\sl physical} subspace of this extended Hilbert space, one defines $\Lambda_j = i c_j b_j^1 b_j^2 \dots b_j^5$ and demands that the physical states satisfy $\Lambda_j \ket{\psi} = - \ket{\psi} \, \forall j$. 
}
$\Gamma_j^\gamma = i b_j^\gamma c_j$. The ``bond Majoranas'' $b_j^\gamma$ are recombined into bond operators $\hat{u}_{jk} = i b_j^\gamma b_k^\gamma$.. Since all $\hat{u}_{jk}$ commute with the Hamiltonian, they can be replaced by their eigenvalues $u_{jk} = \pm 1$. We are thus left with a hopping model of Majorana fermions $\{c_i\}$ coupled to a static $\intg_2$-gauge field $u_{jk}$, described by the Hamiltonian 
\begin{align}
  \hlt = \frac{i}{2} \sum_{j,k} J^\gamma u_{jk}^\gamma c_j c_k. 
  \label{eq:MajoranaHamiltonian}
\end{align}
The spectrum of $\hlt$ must be invariant under $\intg_2$ gauge transformations, and can thus depend only on the $\intg_2$ fluxes $W_p$ associated with plaquettes $p$, defined as
\begin{equation}
  W_p = \prod_{\langle j,k \rangle \in p} \left( -i u_{jk} \right),
\end{equation}
where, as a convention, the product here is taken with a clockwise orientation. The Shastry-Sutherland lattice has {\sl two} types of elementary plaquettes, \emph{viz}, square plaquettes with flux $W_\square = \pm 1$, and triangular plaquettes with flux $W_\triangle = \pm i$.

\begin{figure}[t]
   \centering
    \includegraphics[width=\columnwidth]{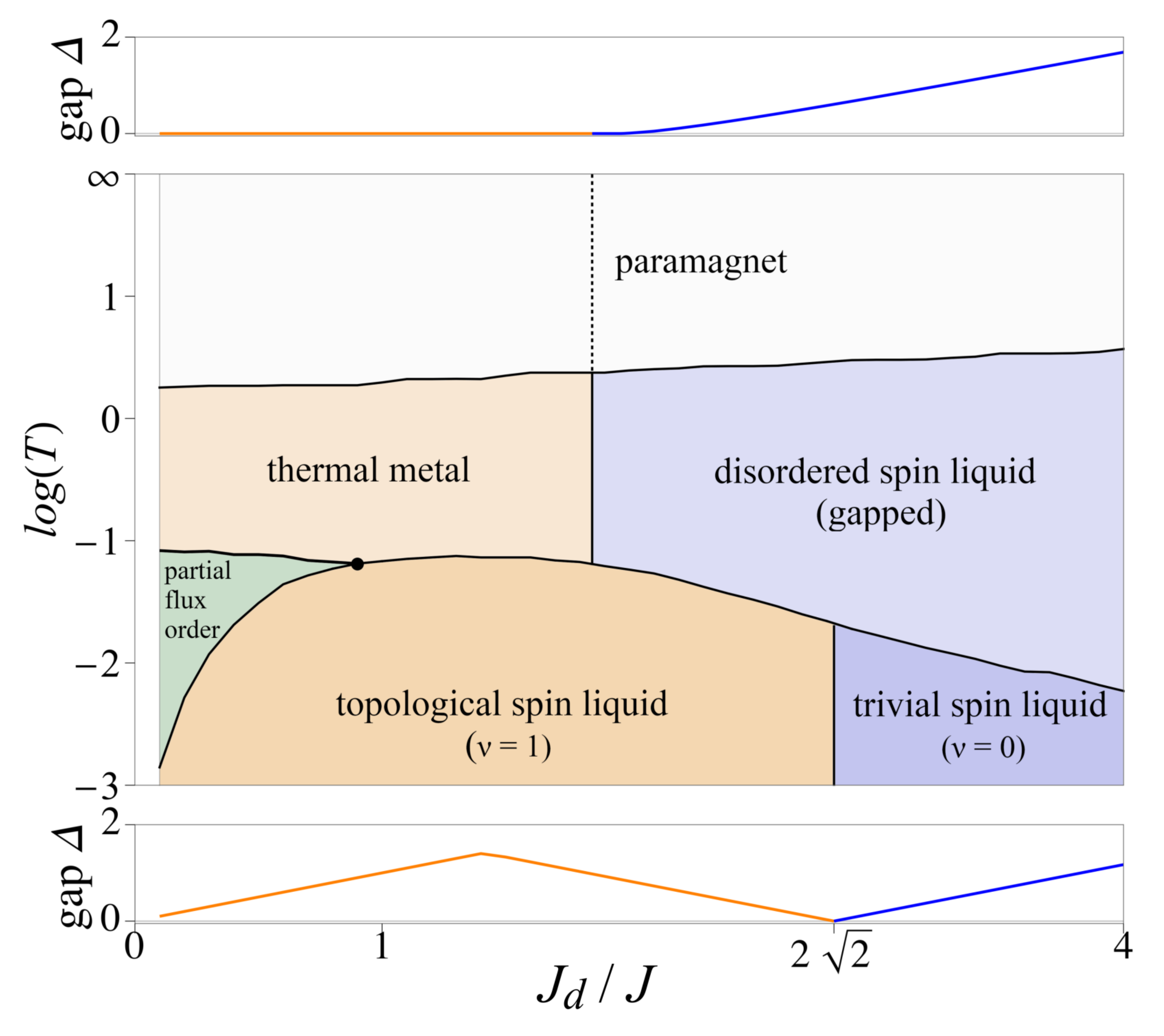}
  \caption{{\bf Schematic phase diagram} of the Shastry-Sutherland Kitaev model. In this model, all possible states of symmetry class D -- a chiral spin liquid, a trivial insulator and a thermal metal -- are realized. For $\Jz < 1$, we find that the ordering of triangle and square plaquettes is decoupled, resulting in the occurrence of a {\it partial flux order} regime. The intermediate temperature regime is separated into the gapless {\it thermal metal} and a gapped phase. In the Majorana band structure, this distinction survives up to infinite temperature (dotted line in the paramagnetic regime). }
    \label{fig:SchematicPhaseDiagram}
\end{figure}


\subsection{Sign-problem free sampling of flux configurations}
\label{sec:qmc}

To determine the ground state, one primary task is to find the flux configuration that minimizes the total energy. This is generally a non-trivial matter and one that can almost never be resolved in an analytically exact fashion -- with the honeycomb Kitaev model being the most notable exception, for which a theorem by Lieb \cite{Lieb1994} can be invoked. For the general case one can, however, rely on a numerically exact treatment by performing quantum Monte Carlo (QMC) sampling of the gauge field, which in the  Majorana decomposition introduced above is possible without encountering a sign problem \cite{nasu2014vaporization,DwivediCornerModes2018} (and yet keeping track of all the Majorana physics).

In such a sign-problem-free QMC approach, the simulation is performed in the Majorana basis, where the $\intg_2$ gauge degrees of freedom $u_{jk}$ can be sampled as classical Ising variables, with the Majorana quantum physics entering in the weights of the Markov chain sampling. For a fixed gauge configuration, the noninteracting Majorana Hamiltonian (Eq.~\eqref{eq:MajoranaHamiltonian}) can be numerically diagonalized, a computation which scales as $N^3$, where $N$ is the number of lattice sites \cite{nasu2014vaporization, mishchenko_prb_96_2017}.  
All thermodynamic observables can then be obtained without needing to introduce an additional imaginary-time dimension or other non-trivial mapping schemes for the quantum problem. The QMC method is thus guaranteed to be sign-problem-free on any lattice geometry. A comprehensive discussion of the technicalities associated with this QMC approach to the Shastry-Sutherland Kitaev model can be found in Ref.~\cite{DwivediCornerModes2018}.

Performing such a quantum Monte Carlo (QMC) simulation readily demonstrates that the ground state flux order corresponds to $W_\square = -1$ for all square plaquettes. However, though $W_\square = W_\triangle^2$, the value of $W_\triangle$ itself is not fixed. The two possibilities $W_\triangle = \pm i$ are related by time-reversal symmetry (TRS) and degenerate in energy. At zero temperature, the ground state must thus spontaneously break TRS by selecting either $W_\triangle = i$ or $W_\triangle = - i$ for all triangular plaquettes. This spontaneous breaking of time-reversal symmetry is generally true for any Kitaev-type model on a lattice containing plaquettes with an {\sl odd} number of bonds, as first noted by Kitaev \cite{Kitaev2006anyons}, and later elucidated within a concrete spin-1/2 model by Yao and Kivelson \cite{yao-kivelson}. This is precisely what is seen in our thermodynamic QMC data for finite temperatures, as we will discuss in detail below.

\subsection{Symmetries and the ground-state phase diagram}
On a conceptual level, the possible ground-state phases of the (Majorana) Hamiltonian can be inferred from the tenfold classification of topological insulators and superconductors \cite{Schnyder2008classification,Kitaev2009periodic}. The Majorana Hamiltonian, by design, obeys a particle-hole symmetry (PHS) which squares to $+1$, while time-reversal symmetry is broken spontaneously, as discussed earlier. Thus, this Majorana Hamiltonian belongs to symmetry class $D$. 
In two spatial dimensions, this symmetry class $D$ allows for a $\intg$ invariant, \emph{viz}, a Chern number, signaling the possibility of the formation of topological order. Such a topological state is precisely the chiral spin liquid (with a gapless chiral edge mode) discussed in the introduction.

If we further restrict to the case where all the ``square lattice'' bonds are of equal strength (which we set to $J=1$), the only remaining parameter is  $\Jz$, the strength of the diagonal ``dimer" bonds of the lattice. As a function of varying $\Jz$ the Majorana band structure exhibits two gapped phases: a topological phase with Chern number $C=\pm 1$ for $\Jz < 2\sqrt{2}$ and a trivial one for $\Jz > 2\sqrt{2}$. The two gapped phases are separated by a gap closing at $(\pi,\pi)$, as illustrated in the lower panel  of the schematic phase diagram in Fig.~\ref{fig:SchematicPhaseDiagram}. The $C=\pm 1$ phase is a \emph{chiral spin liquid} with a chiral Majorana edge mode and bulk Ising anyon topological order. On the other hand, the $C=0$ phase, while still spontaneously breaking TRS, is fully gapped and possesses the same Abelian topological order as the toric code. We refer to this phase as a ``trivial'' spin liquid.


\section{Trivial and chiral spin liquids}
\label{sec:ShastryChiral}

Coming to the actual results of our finite-temperature analysis of the Shastry-Sutherland Kitaev model, we first concentrate on the
thermodynamic behavior above the phase transition from the topological (first-order) spin liquid (with Chern number $\nu = \pm 1$)
to the trivial spin liquid (with Chern number $\nu = 0$). 


\subsection{Thermodynamics}
\label{eq:ShastryThermalPhaseTransition}

\begin{figure}[t]
   \centering
    \includegraphics[width=0.9\columnwidth]{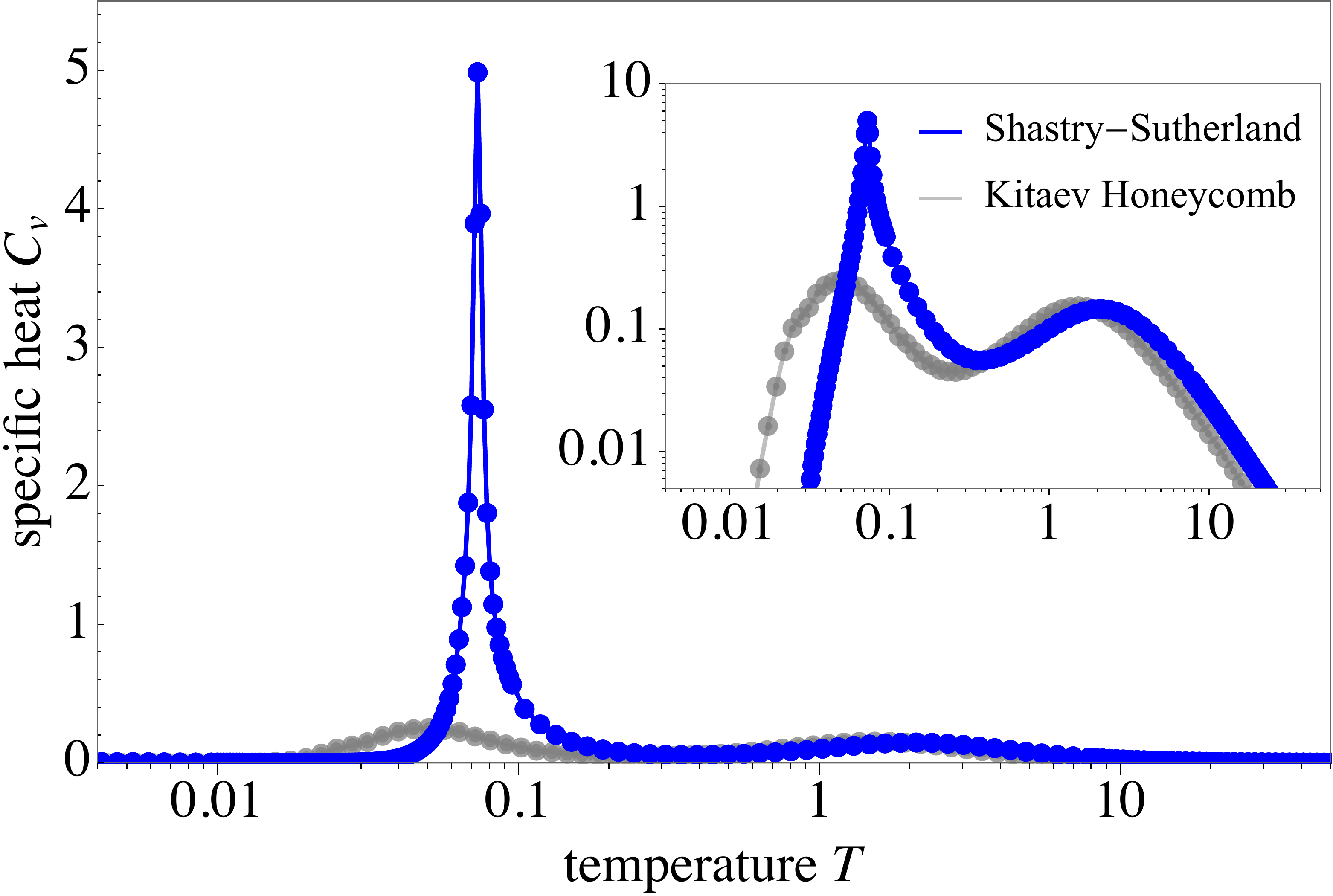}
   \caption{{\bf Double-peak signature in the specific heat} $C_v(T)$ of the Shastry-Sutherland Kitaev model.
    		The higher temperature peak at $T' \sim 2$ is the signature of a thermal crossover indicating spin fractionalization,
		while the low-temperature peak (here at $T_c \sim 0.1$) is the signature of a thermal phase transition, associated with spontaneous breaking of time-reversal symmetry.
		This phase transition happens at a higher temperature scale than the thermal crossover of the Kitaev honeycomb model, which is seen from a comparison of the
		two specific heat curves. In the Kitaev honeycomb model, the ground state does not show any spontaneous symmetry breaking. Data shown is for the coupling
		parameter $\Jz = 1.2$ and linear system size $L = 10$ for the Shastry-Sutherland Kitaev model and isotropic coupling $J_x = J_y = J_z = 1$ and $L =
		16$ for the Kitaev honeycomb model.}
    \label{fig:SpHDoublePeak}
\end{figure}

A central quantity distinguishing the various finite-temperature phases of our model is the specific heat $C_v(T)$ and its characteristic {\sl multi-peak} structure, as illustrated in Fig.~\ref{fig:SpHDoublePeak}. For the closely related Kitaev spin liquids, it is well established \cite{nasu2014vaporization} that one finds two well-separated peaks in the specific heat -- a smooth high-temperature peak, indicating a thermal crossover, corresponding to the (local) fractionalization of spins and a second low-temperature peak associated with the freezing of the $\intg_2$ gauge field. We observe a similar two-peak structure of the specific heat in our model for a broad range of parameters $\Jz \gtrsim 1$ as well. Plotted in Fig.~\ref{fig:SpHDoublePeak} is a characteristic $C_v(T)$ trace in comparison with data for the Kitaev honeycomb model for similar system sizes. Both systems show a shallow high-temperature crossover around the value of the elementary coupling strength, in our case $T \sim 2$, whose  shape and height are essentially independent of the system size -- indicating a purely local crossover. In fact, this is where the spin fractionalization happens and the system releases precisely half of its entropy \cite{nasu2014vaporization,DwivediCornerModes2018}. A more pointed distinction is found in the low-temperature peak -- this is a sharp peak for the model at hand (that sharpens with increasing system size), while it is a more shallow feature in the Kitaev honeycomb model. This is an immediate reflection of the fact that in the model at hand there is a true phase transition occurring at this lower temperature -- the spontaneous breaking of time-reversal symmetry upon entering the chiral spin liquid regime, while in contrast the Kitaev honeycomb model exhibits a finite-temperature crossover at this lower temperature scale at which, for a given system size, the $\intg_2$ gauge field freezes into its ground-state configuration. While the latter is a true phase transition in three spatial dimensions \cite{2020EschmannThermalClassification}, it remains a thermodynamic crossover in two spatial dimensions \cite{Read1991,Senthil2000}.

\begin{figure}[t]
\includegraphics[width=\columnwidth]{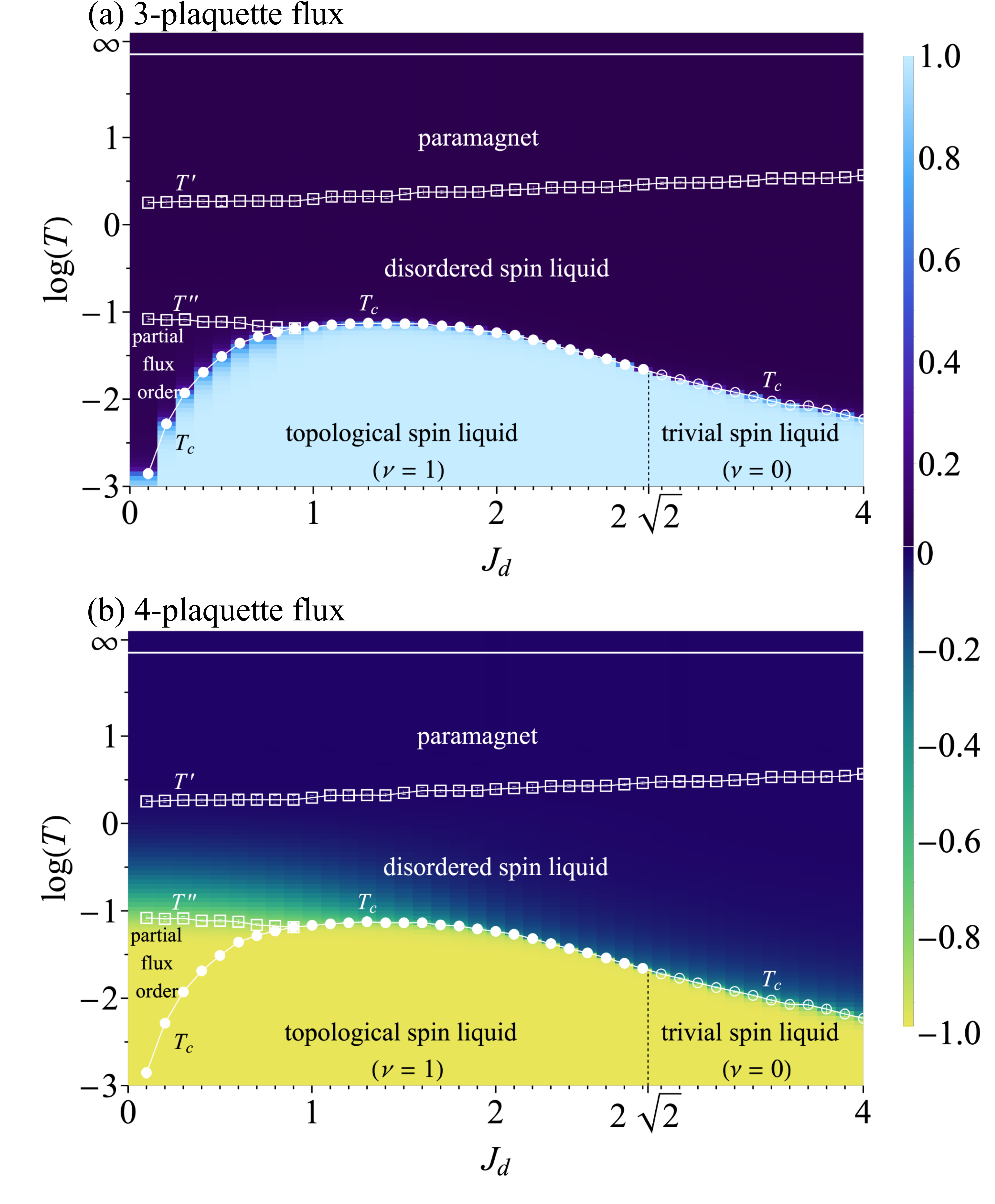}
\caption{{\bf Thermal phase diagram and flux ordering} as a function of $\Jz$.
	     The color coding represents the average 3-plaquette flux $|\overline{W}_\triangle|$ (a)
	     and the average 4-plaquette flux $\overline{W}_\square$ (b).
	     The (white) data points denote the transition/crossover temperatures determined from the multi-peak structure of the specific heat,
	     with the filled (open) circles indicating the thermal phase transition for the chiral (trivial) spin liquid phase.
	     For $\Jz \lesssim 1$, we encounter an additional ``partial flux order" phase, where the square plaquettes are already ordered,
	     while the triangle plaquettes remain disordered.
	     A thermal crossover separates this phase from the regime with full flux disorder (disordered $\intg_2$ spin liquid).
}
\label{fig:PhaseDiagramShastry}
\end{figure}

The key distinction between these model systems is that the Shastry-Sutherland lattice is non-bipartite and exhibits elementary (triangular) plaquettes with an {\em odd} number of bonds. For the emergent Majorana fermions this results in an ambiguous situation in which they can pick up a phase $e^{\pm i \pi/2}$ upon hopping around such a triangular plaquette, endowing the latter with a flux $W_\triangle = \pm i$ as discussed in the previous section. By spontaneously breaking time-reversal symmetry, one of the two possible signs is chosen, with the system simultaneously undergoing an Ising-type phase transition.
Such a time-reversal symmetry breaking thermal phase transition was first observed in the Yao-Kivelson model (on a decorated honeycomb lattice) \cite{Nasu2015} at a temperature scale $T_c \sim 10^{-2} J$. This is also the typical temperature scale for thermal phase transition in 3d Kitaev models. 
In comparison, the critical temperature scale of the model at hand is elevated, see Fig.~\ref{fig:SpHDoublePeak}, with its maximum close to one-tenth of the Kitaev coupling (at $\Jz \sim 1.2$).
One might speculate that this enhanced transition temperature is a reflection of the higher coordination number of the Shastry-Sutherland lattice $(z=5)$  in comparison to conventional Kitaev models on tricoordinated lattice geometries. This idea, however, does not hold up when further generalizing our model to a 7-coordinated lattice (by placing additional diagonal bonds on the lattice), which has a transition temperature of the same order of magnitude as the Shastry-Sutherland case. 

\subsection{Partial flux ordering}
\label{ssec:ShastryPartialFluxOrdering}

\begin{figure}
\includegraphics[width=0.45\columnwidth]{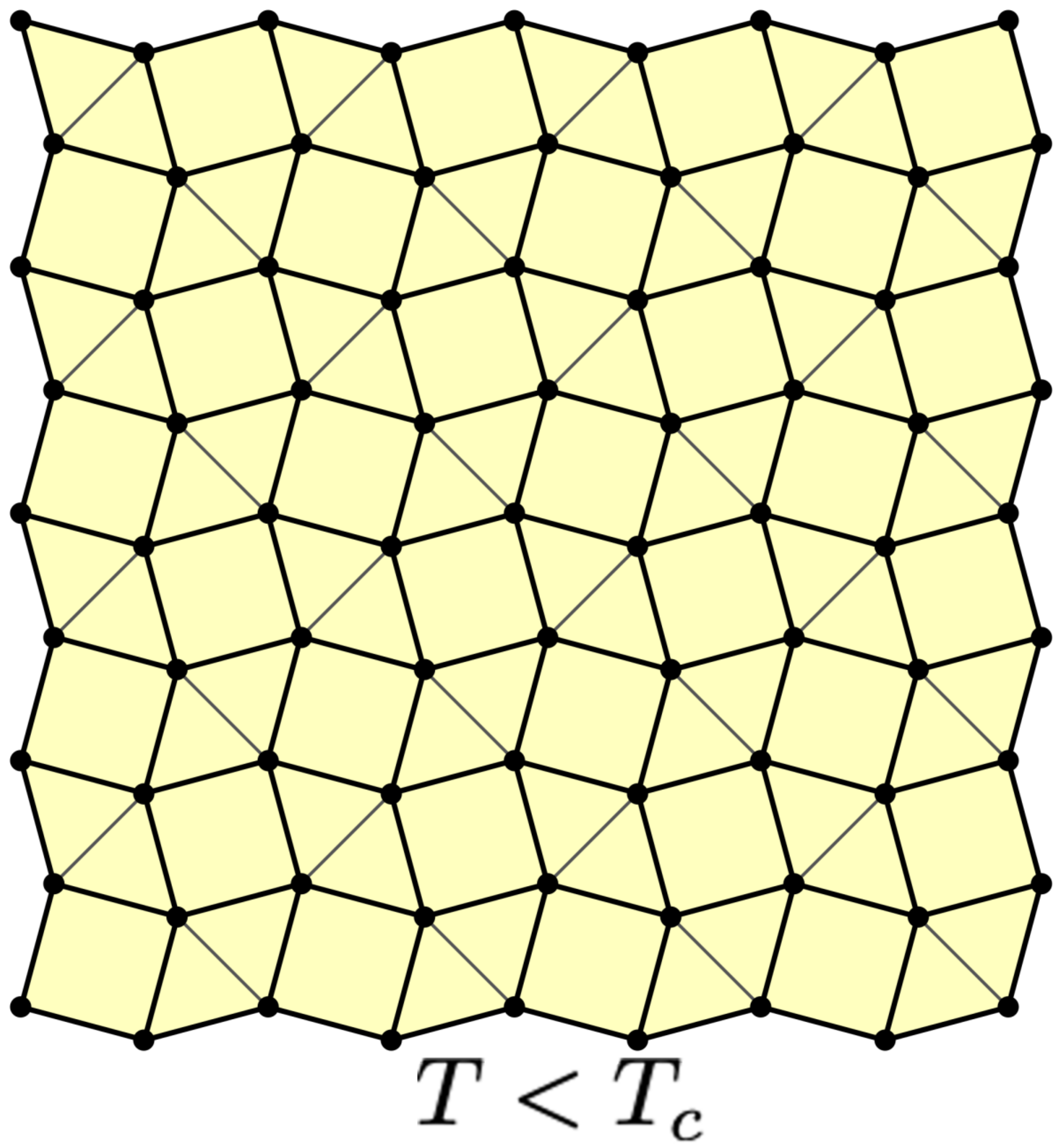}
\hspace{0.5cm}
\includegraphics[width=0.45\columnwidth]{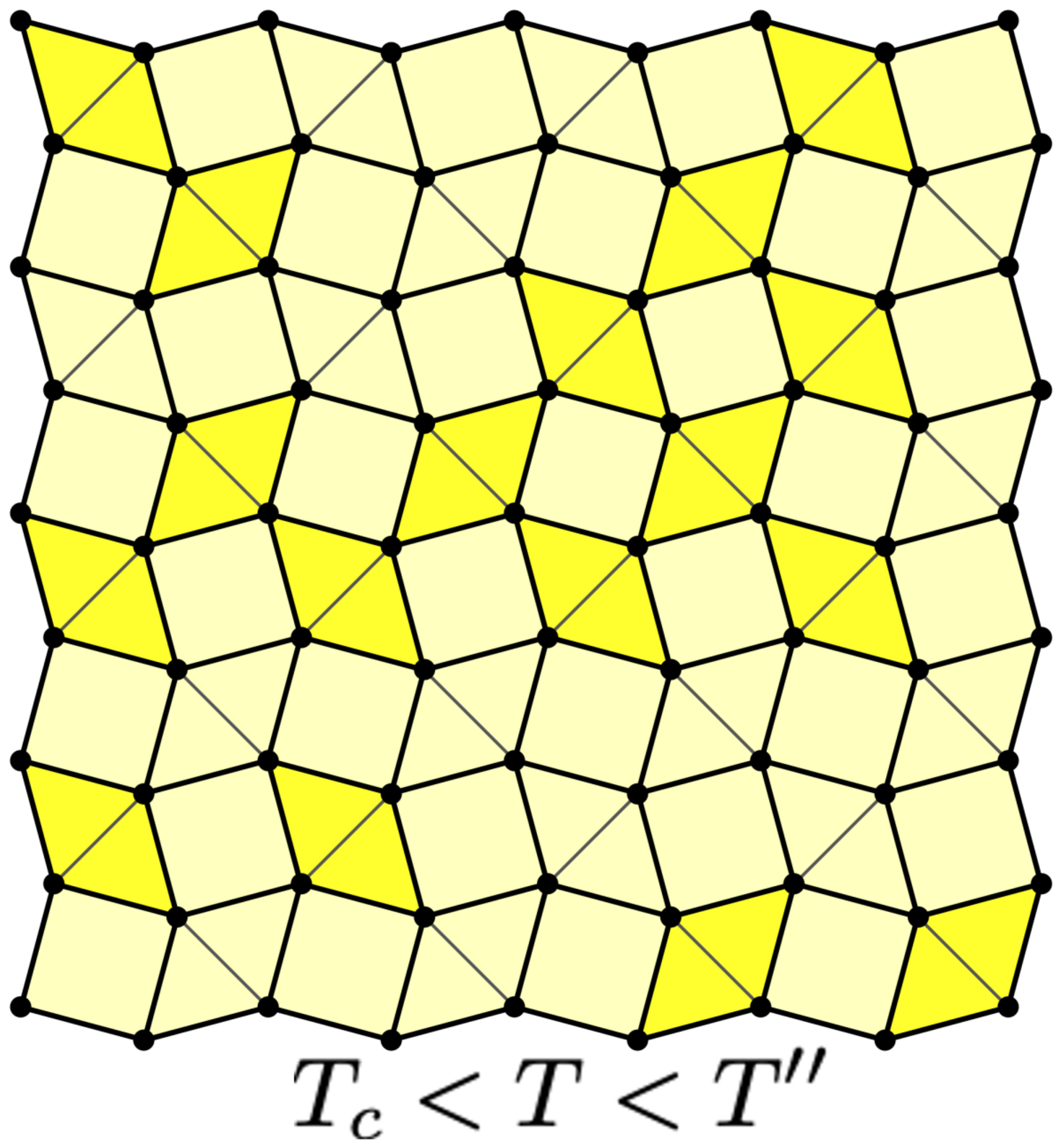}
\caption{{\bf Partial flux ordering.} For $\Jz \lesssim 1$, the square plaquettes assume an ordered $\pi$-flux configuration at $T''$. In this regime, the triangle plaquettes remain disordered, assuming pairwise fluxes $W_\triangle = \pm i$ (darker / lighter yellow). At $T_c \sim 0.1 \Jz$, also the triangle plaquettes order into one of the two homogeneous flux configurations.
}
\label{fig:PartialFluxOrderingVisualization1}
\end{figure}

Upon closer inspection, the formation of flux order at low temperatures turns out to be slightly more intricate.
As noted above, the spontaneous breaking of time-reversal symmetry is intimately connected with the flux ordering of the triangular plaquettes.
With two such triangular plaquettes constituting a single square plaquette, this also implies ordering for the latter.
The corresponding flux satisfies $W_\square = W_{\triangle_1} \cdot W_{\triangle_2}$ = +1, independent of the actual assignment of the triangular plaquettes
\footnote{Such a $\pi$-flux ground state for square plaquettes is also generally in line with the expectation from Lieb's theorem on ground-state flux assignments in bipartite lattices with certain mirror symmetries \cite{Lieb1994}, tough it does not strictly apply to the lattice geometry at hand.}.
Thus, an ordering of the 3-plaquettes implies ordering of at least half the 4-plaquettes. The converse is, however, not true; we can have ordering of the 4-plaquettes while the 3-plaquettes remain disordered, resulting in a partial flux ordering. This is exactly what we observe for a limited parameter range $0 < \Jz \lesssim 1$, where the 4-plaquette fluxes order at a higher temperature $T''$ than the critical temperature $T_c$ for the 3-plaquettes fluxes. This is illustrated in Fig.~\ref{fig:PhaseDiagramShastry} where we plot the phase diagram of our model by color-coding the flux of the 3-plaquettes (top panel) and 4-plaquettes (bottom panel) as function of $\Jz$. A schematic rendering of the indermediate partial flux ordering is provided in Fig.~\ref{fig:PartialFluxOrderingVisualization1}.

\begin{figure*}[t]
	\includegraphics[width=0.66\columnwidth]{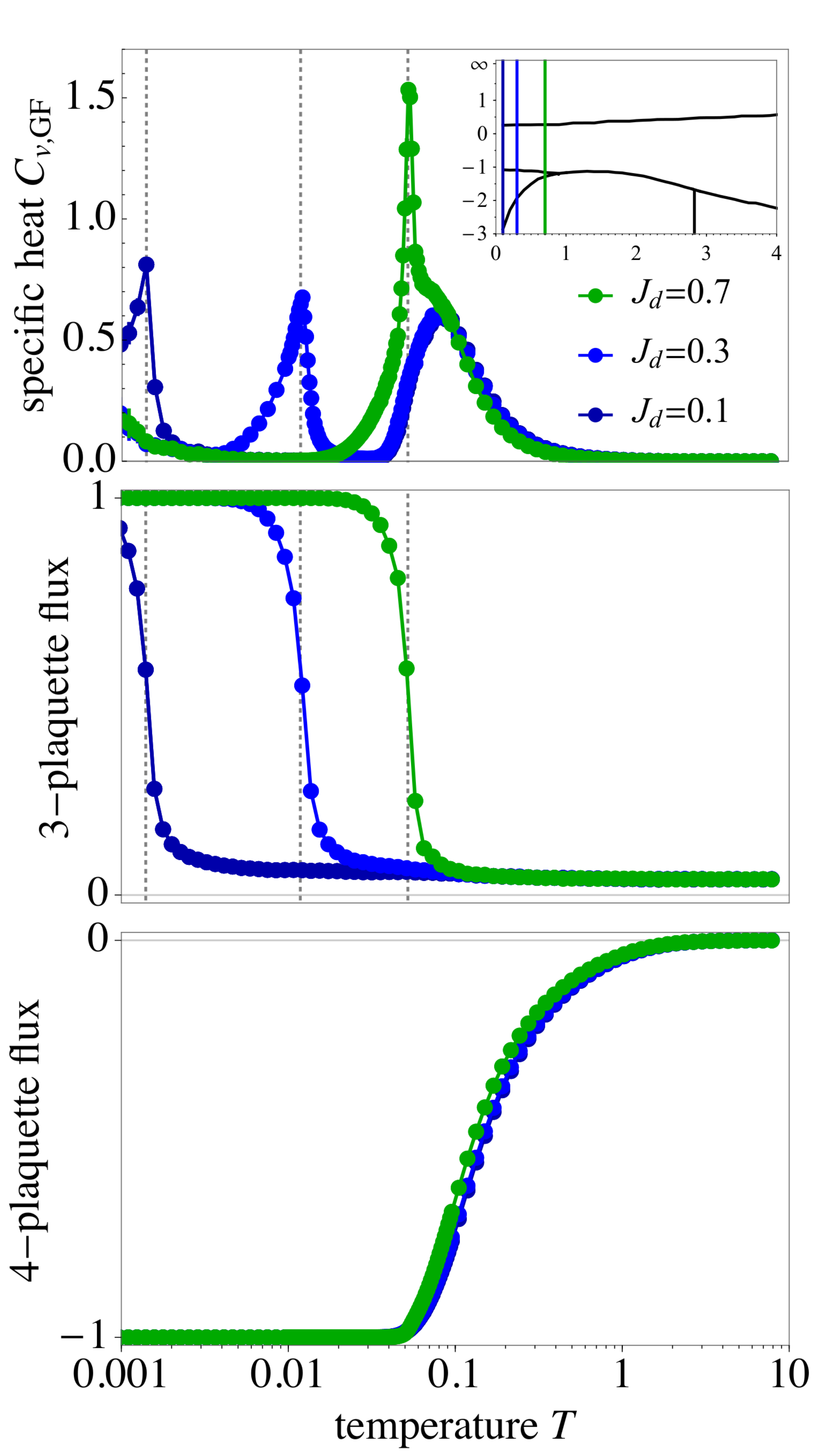}
	\includegraphics[width=0.66\columnwidth]{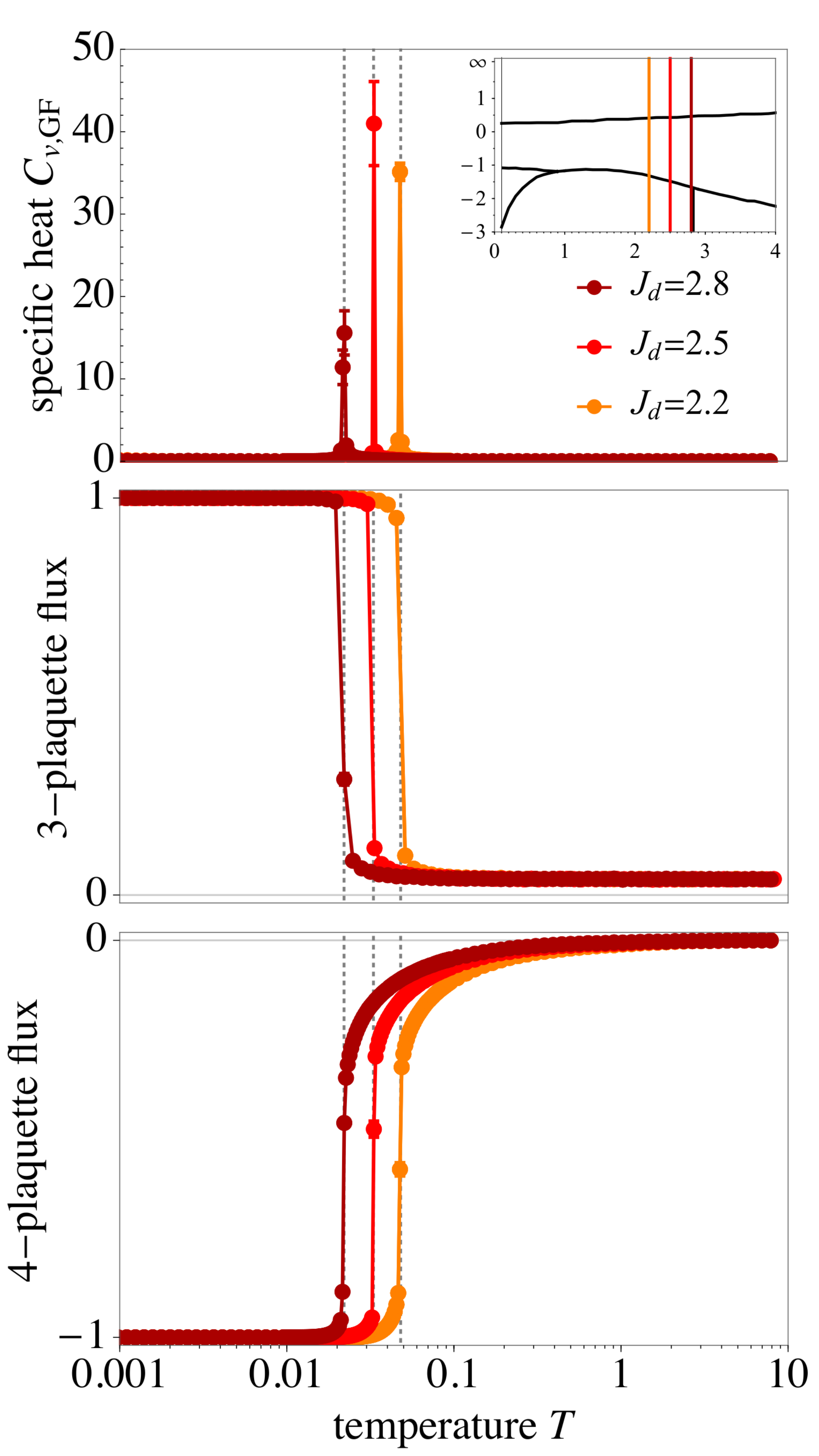}
	\includegraphics[width=0.66\columnwidth]{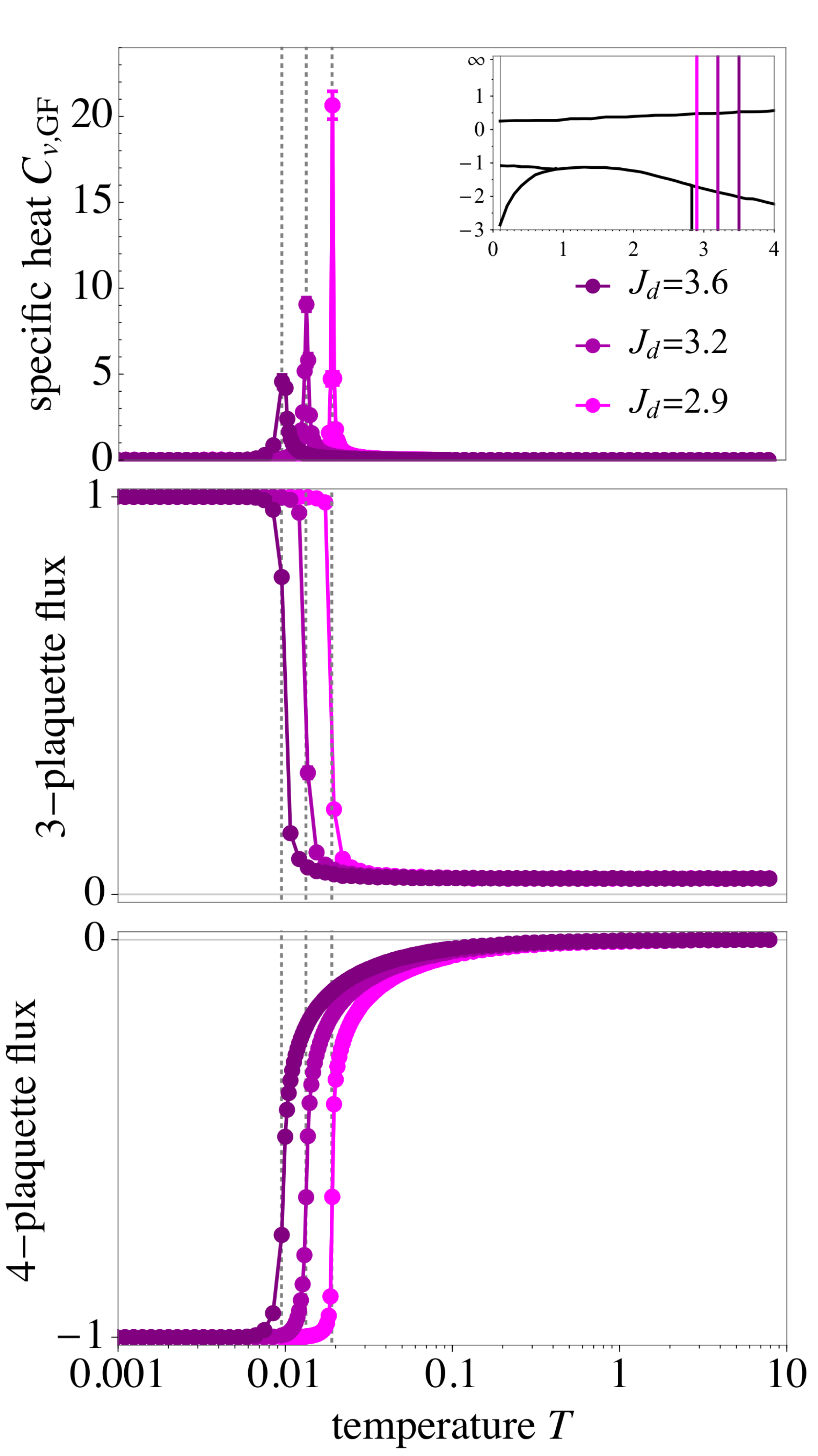}
\caption{{\bf Thermodynamics.} (a) The specific heat $C_{v, {\rm GF}}(T)$ ($\intg_2$ gauge field contribution),  (b) the average 3-plaquette flux $|\overline{W}_\triangle|$, and (c) and the average 4-plaquette flux $\overline{W_\square}$ for various values of $\Jz$. The low-temperature peak in $C_{v, {\rm GF}}(T)$ indicates a thermal phase transition (dashed lines), which is associated with the ordering of the fluxes $W_\triangle = \pm i$ on the triangular plaquettes of the Shastry-Sutherland lattice. The difference between the transition temperature for the 3-plaquettes and the 4-plaquettes in the left column indicates an intermediate partial flux order.
}
\label{fig:Thermodynamics}
\end{figure*}

The evolution of this partial flux ordering with varying coupling strengths can be seen in the  sequence of data sets for vertical cuts through the phase diagram provided in Fig.~\ref{fig:Thermodynamics}. The top row shows the gauge contribution to the specific heat (i.e.~omitting the Majorana contribution resulting in the higher temperature crossover). In the partial flux ordering regime $0 < \Jz \lesssim 1$, one finds that the low-temperature specific heat peak actually splits into two parts -- with the lower temperature peak indicating the true thermal phase transition associated with time-reversal symmetry breaking and 3-plaquette flux ordering, see also the medium row of panels. While this transition quickly moves to zero temperature as $\Jz \to 0$, there remains a signature in the specific heat around $\mathcal{O}(1/10)$, which upon closer inspection is the thermal crossover associated with the ordering of the remaining half of the 4-plaquettes, see the lower row of panels.

Taking a step back, we conclude that the emergence of a partial flux ordering is, for a limited range of parameters, a precursor phenomenon to the formation of a low-temperature topological chiral spin liquid.


\subsection{Thermal Majorana metal}
\label{sec:ThermalMetal}

The thermodynamic signatures discussed so far -- the finite-temperature, time-reversal symmetry breaking phase transition as well as the emergence of a partial flux ordering and its associated thermal crossover -- are both closely connected to the underlying lattice gauge theory.
From a more conceptual perspective, these aspects are an interesting variation to the gauge physics that has been intensely studied
in the context of two- and three-dimensional Kitaev models \cite{nasu2014vaporization,2020EschmannThermalClassification}.
We now turn to additional thermodynamic aspects of our model at hand, which are genuinely rooted in the physics of the emergent Majorana fermions.
The most notable feature here is the formation of a {\sl thermal metal} regime above the transition to the topological chiral spin liquid as
illustrated in the schematic phase diagram of Fig.~\ref{fig:SchematicPhaseDiagram}, and as evidenced in the numerical observations of Fig.~\ref{fig:PhaseDiagramShastryChern}.
This thermal metal regime, principally located in the intermediate temperature regime $T_c \lesssim T \lesssim T'$
(i.e.~between spin fractionalization and the time-reversal symmetry breaking transition),  does not extend over the entire parameter space of our model, but marks a relatively sharp transition for a critical value
of $\Jz$ that is considerably shifted in comparison with the zero-temperature transition between the chiral and trivial spin liquids.
We rationalize these numerical observations using analytical arguments based on the analytical (self-consistent) Born and T-matrix approximations along with a numerical computation using transfer matrices. We thus explain how the {\sl entire} gapless thermal metal phase emanates from the quantum critical point between the chiral and trivial spin liquid at zero temperature.


\subsubsection*{Numerical observations}

\begin{figure}[t]
   \centering
    \includegraphics[width=\columnwidth]{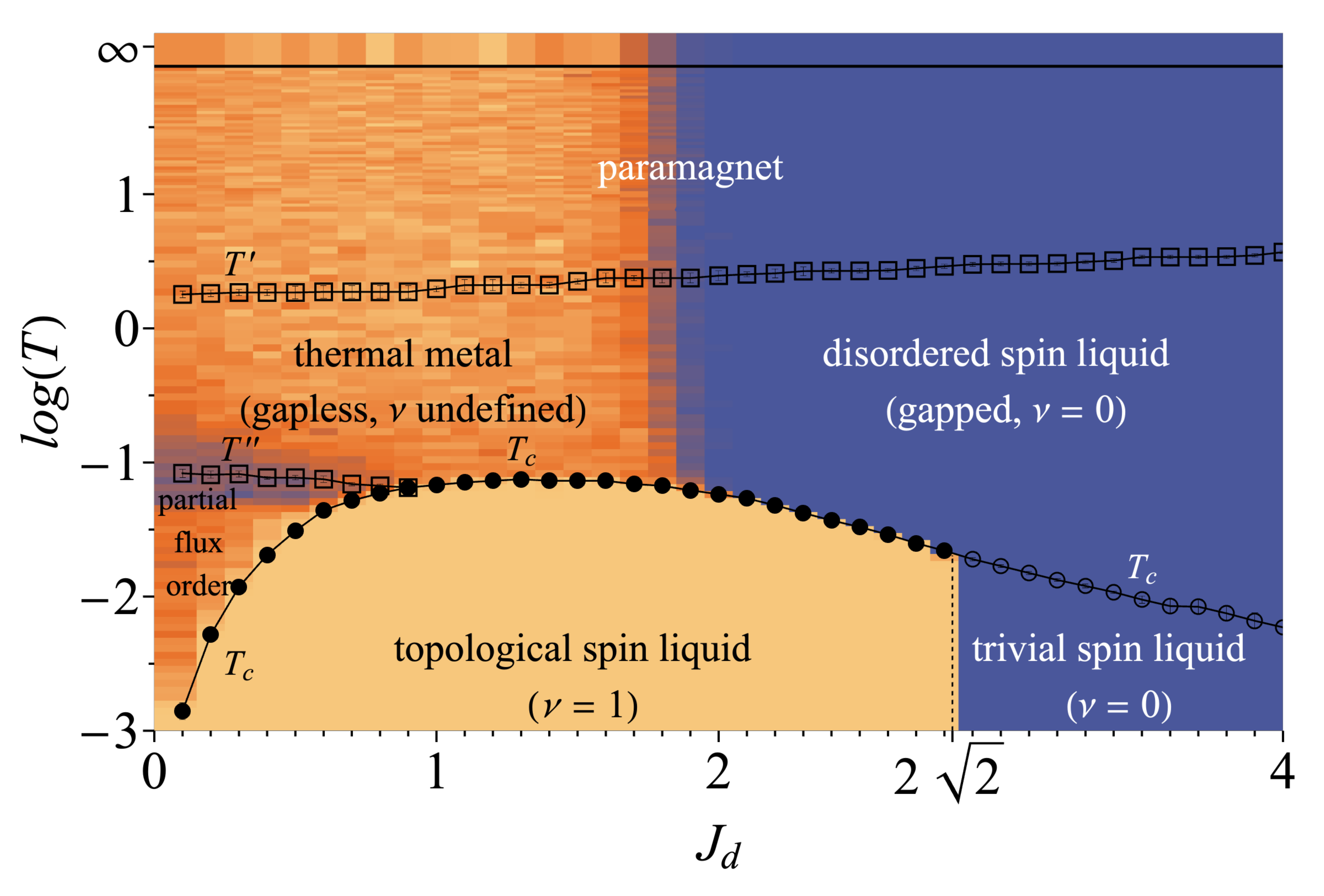}
   \caption{{\bf Thermal metal.} Color-coded is the average Chern number $|\nu|$ for the Majorana band structure overlaid on the finite-temperature phase diagrams. The average Chern number distinguishes the chiral spin liquid ground state ($|\nu| = 1$, orange) from the trivial phase ($|\nu|$ = 0, blue) below the thermal phase transition. In the intermediate temperature regime, the Chern number is only well-defined in the gapped regimes for $\Jz \gtrsim 1.9$ (where it assumes a zero value, indicated in blue). }
  \label{fig:PhaseDiagramShastryChern}
\end{figure}

In our numerics, the existence of two distinct regimes above the low-temperature chiral spin liquid phases is most evident in calculations of
the {\sl average Chern number} $|\nu|$ of the Majorana band structures encountered when sampling (and averaging over) the different gauge configurations for a given temperature.
The behavior of this average Chern number as a function of $\Jz$ is plotted in Fig.~\ref{fig:PhaseDiagramShastryChern}(a),
which immediately reveals three distinct regimes: For a horizontal cut in the lowest temperature regime, 
the average Chern number jumps from $\braket{|\nu|} = 0$ in the trivial phase to $\braket{|\nu|} = 1$ in the topological phase, as expected.
Unanticipated is probably the behavior for a horizontal cut in the intermediate-temperature regime, where our simulations also reveal a relatively sharp (vertical) boundary at $\Jz \approx 1.9$, 
at which the average Chern number mimics the low-temperature behavior, jumping from $\braket{|\nu|}  = 0$ for large couplings $\Jz \gtrsim 1.9$, to  $\braket{|\nu|}  \approx 0.6$ for $\Jz \lesssim 1.9$ (see also the scans of $\braket{|\nu|}$ shown in Figs.~\ref{fig:ChernNumbers1} and \ref{fig:ChernNumbers2} of the Appendix).
This clearly separates two different thermodynamic regimes, which seemingly persist not only up to the fractionalization crossover scale but beyond into the paramagnetic regime extending all the way up to infinite temperatures.

One important aspect in interpreting this numerical result, in particular the non-quantization of the average Chern number for $\Jz \lesssim 1.9$, is to ask whether the Majorna band structure actually remains gapped -- a prerequisite for the proper calculation of a Chern number -- also at finite temperatures. In fact, this is precisely what distinguishes the two regimes: while the system remains gapped for $\Jz \gtrsim 1.9$, it becomes gapless for smaller couplings.
That the phase for $\Jz \lesssim 1.9$ is in fact gapless can be shown in a straightforward manner by computing the gap in the Majorana spectrum averaged over flux configurations sampled with a uniform distribution, or equivalently, at infinite temperature. As shown in the top panel of the schematic phase diagram of Fig.~\ref{fig:SchematicPhaseDiagram}, the gap in the Majorana spectrum indeed vanishes for small $\Jz$ and slowly opens only for $\Jz \gtrsim 1.9$, as compared to the scenario of a gap closing and re-opening for $\Jz = 2\sqrt2 \approx 2.8$ in the zero-temperature case.
The Chern number results are thus only well-defined in the gapped regimes of the phase diagram, $\Jz \gtrsim 1.9$ and $T<T_c$. For detailed scans of the average Chern number we refer to Appendix~\ref{app:chern}.

\begin{figure}[t]
   \centering
    \includegraphics[width=\columnwidth]{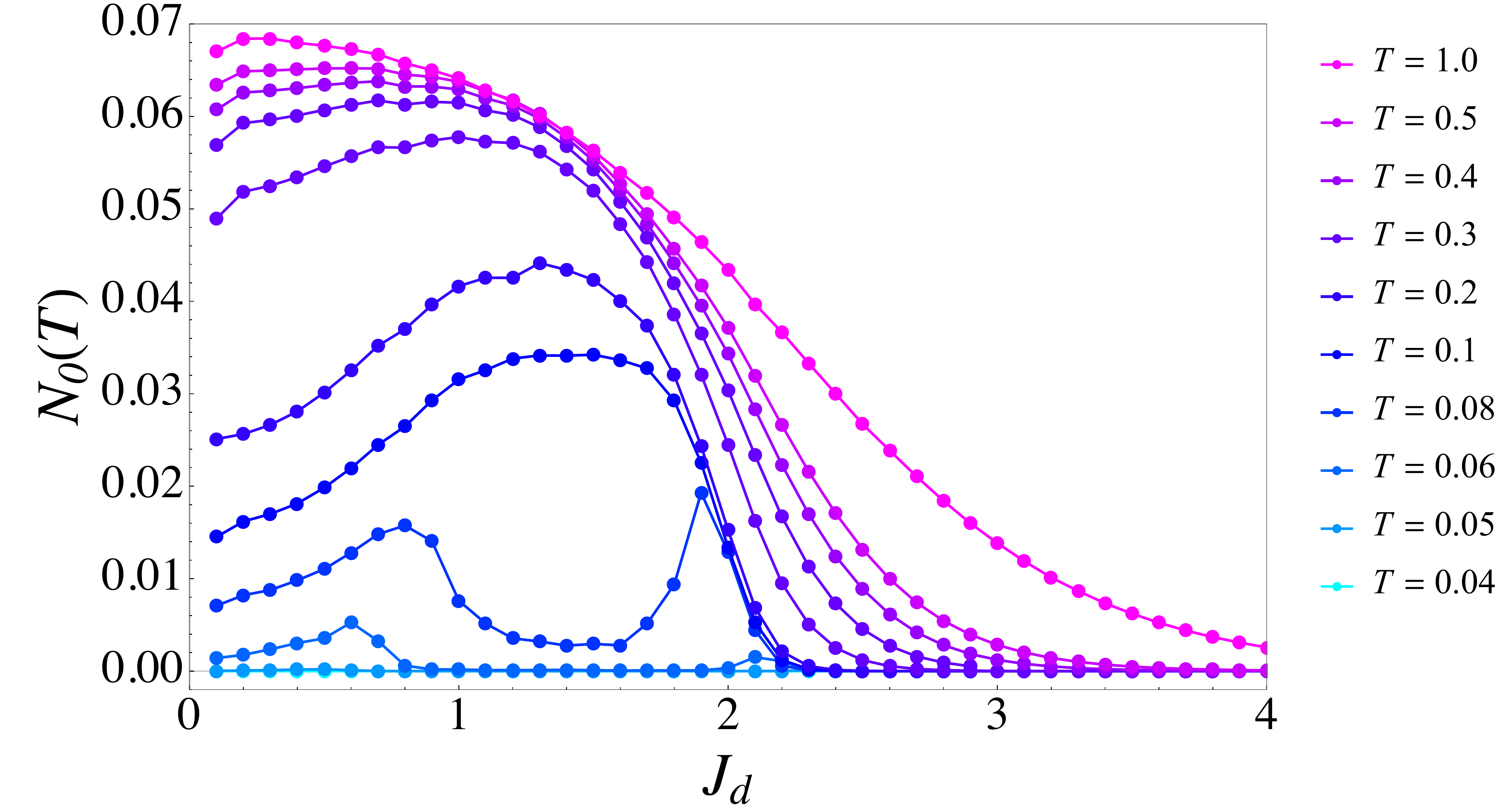}
   \caption{The {\bf effective density of states} at zero energy for finite temperatures, $N_0(T)$ as defined in Eq.~\eqref{eq:NumberOfStates},
   		as a function of $\Jz$ .
		The nearly vanishing number of states for $\Jz \gtrsim 1.9$ indicates a gapped states (at all temperatures),
		while the finite number of states for $\Jz \lesssim 1.9$ and elevated temperatures signals a gapless state.
		Note the special double-peak structure in the parameter regime of $T = 0.04 - 0.06$, which is related to the partial flux ordered phase
		and indicates a finite band gap also for this phase.
   }
    \label{fig:NumberOfStates}
\end{figure}

Having established the principal gapless character of the finite-temperature spectrum for $\Jz \lesssim 1.9$, a more meaningful quantity to calculate is the {\sl effective number of states} accessible to the system at a given temperature
\begin{equation}
	N_0(T) = -\int_{0}^\infty d E  \rho(E) \, \partial_E n_F(E, T) ,
	\label{eq:NumberOfStates}
\end{equation}
where $\rho(E)$ is the Majorana density of states and $n_F(E,T)$ is the Fermi function, whose derivative has a peak at $E = 0$ of width $\sim T$. Plotting $N_0(T)$ as a function of $\Jz$ for various temperatures in Fig.~\ref{fig:NumberOfStates},  we indeed see that the number of available state becomes finite for small $\Jz$, indicating a gapless state, and vanishes only near the transition point $\Jz \approx 1.9$, indicating a truly gapped state for larger $\Jz$.

We note in passing that the partial flux ordered phase also appears to be gapped, as $N_0(T) \to 0$ in the corresponding parameter regime, i.e., for $\Jz \lesssim 0.3$ and $T \lesssim 0.04$. This gap opens at the crossover to the thermal metal phase, as indicated by the increase in $N_0(T)$. This is also indicated in the Chern number plot of Fig.~\ref{fig:PhaseDiagramShastryChern}(a), where a narrow blue stripe above the partially flux-ordered phase indicates the thermal crossover points associated with the ordering of square plaquettes.
Thus, for $\Jz < 1$, the Kitaev Shastry-Sutherland system starts in the gapped chiral spin liquid ground state at $T = 0$, undergoes a thermal phase transition into a gapped partial flux-ordered phase, and the gap finally collapses at the crossover to the intermediate temperature phase.

\begin{figure}[t]
   \centering
    \includegraphics[width=\columnwidth]{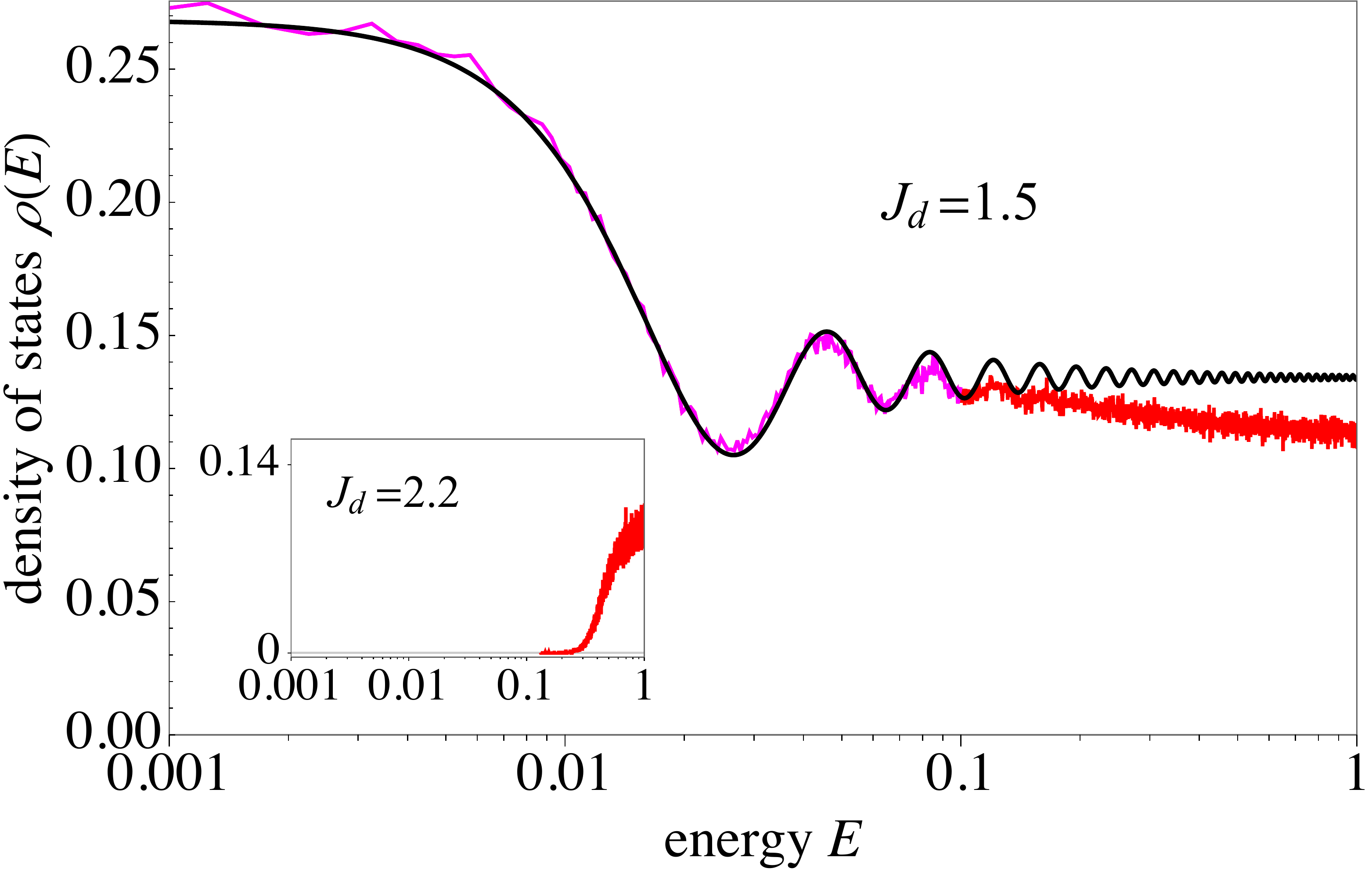}
   \caption{Characteristic oscillations in the {\bf low-energy density of states} $\rho(E)$ for the thermal metal regime
    		(computed for flux configurations sampled at $T = 1.86$ and coupling $\Jz = 1.5$).
		The ringing of $\rho(E)$ follow the sinusoidal function of Eq.~\eqref{eq:DOSRinging}, predicted by random matrix theory for the
		 metallic state in symmetry class D \cite{Altland1997classification, Zirnbauer1996, ThermalMetal3}.
		 This unambiguously shows that the gapless disordered $\intg_2$ spin liquid phase, observed above the thermal phase transition,
		 can be effectively described as a thermal metal, although time-reversal symmetry is only broken below the phase transition. Inset: $\rho(E)$ for the gapped intermediate-temperature regime
    		(computed for flux configurations sampled at $T = 1.86$ and coupling $\Jz = 2.2$)
   }
  \label{fig:ShastryDOS}
\end{figure}


\subsubsection*{Symmetry-class considerations}
To interpret these numerical observations, it is useful to remind oneself of the symmetry class classification of free fermion systems \cite{Altland1997classification,Zirnbauer1996} and its application to topological states of matter \cite{Schnyder2008classification,Kitaev2009periodic}. When considering the Majorana fermion representation of our model, the system naturally exhibits particle-hole symmetry (due to the self-conjugation of Majorana fermions) and broken time-reversal symmetry, which is implied by the non-bipartite geometry of the lattice as discussed earlier. This puts our model into symmetry class D. In the 10-fold way classification of topological states of matter \cite{Schnyder2008classification,Kitaev2009periodic}, this symmetry class allows, in two spatial dimensions, for the formation of both a trivial and a topological insulator (with vanishing and integer Chern number, respectively). In addition, it is well established that in the presence of disorder this symmetry class allows for the formation of a gapless  ``thermal metal'' phase \cite{ThermalMetal1, ThermalMetal1b, ThermalMetal1c, ThermalMetal2, ThermalMetal2b, ThermalMetal3}.

In the Kitaev Shastry-Sutherland model, all three of these phases are found to be realized. In the low-temperature regime, where time-reversal symmetry is spontaneously broken and the gauge field is ordered, we realize the two ``clean" phases of symmetry class D -- the trivial insulator (with Chern number $\nu = 0$) for $\Jz > 2\sqrt{2}$ and the topological insulator (with Chern number $\nu = \pm 1$) for  $\Jz < 2\sqrt{2}$.
Going to the intermediate-temperature regime, i.e. above the ordering transition at $T_c$, a more subtle situation arises: Here the $\intg_2$ gauge field does not exhibit any order, effectively providing a static disorder potential for the Majorana system. And while time-reversal symmetry (TRS) is not broken on a global level, which strictly speaking does not put the system into symmetry class D anymore, TRS is broken for any given static disorder configuration: Each particular flux configuration must have a \emph{fixed} value of flux ($\pm i$) across the 3-plaquettes, and hence breaks time-reversal symmetry. The {\sl ensemble} of Majorana Hamiltonians thus always corresponds to the symmetry class D at any temperature, even if the true state of the Kitaev model is itself time-reversal invariant. This allows the system to principally form the thermal metal phase of symmetry class D.

This thermal metal phase is indeed realized in the intermediate temperature regime for $\Jz \lesssim 1.9$, i.e. in the parameter regime that our numerical experiments indicated as gapless. The very nature of the thermal metal phase can be probed explicitly via a calculation of the low-energy density of states $\rho(E)$ of the Majorana fermions, which is supposed to show a characteristic ``ringing" \cite{ThermalMetal3}. For $E\to 0$, the latter takes the universal form
\begin{equation}
	\rho(E) = \alpha + \frac{\sin(2 \pi E L^2) }{ 2 \pi E L^2 }
	\label{eq:DOSRinging}
\end{equation}
with a single fitting parameter $\alpha$.
This is indeed what we find upon numerical inspection, as shown in Fig.~\ref{fig:ShastryDOS} for a parameter set deep in the thermal metal regime
($\Jz = 1.5$ and $T = 1.86$). Similar behavior has first been reported for the honeycomb Kitaev model in a magnetic field in a broad temperature regime above the thermal crossover into its chiral spin liquid ground state \cite{Self2019}.


\subsubsection*{Analytical perspective}

The essential physics of the formation of the thermal metal phase can be understood analytically by analyzing our Majorana hopping model with \emph{uncorrelated} $\intg_2$ flux disorder
\footnote{
  For the full Kitaev model at finite temperature, the disorder is far from uncorrelated, since the probability of a disorder realization is the Boltzmann weight corresponding to all the vison excitations required to arrive at it from the ground state flux configuration.
}.
Explicitly, this implies that arbitrary disorder configurations can be obtained by starting from the ground state flux configuration and flipping each plaquette flux with a probability $p$, independent of all other fluxes. This \emph{disorder density} $p$ then roughly corresponds to the temperature in the full Kitaev model. In particular, $p=0$ is the ground state ($T=0$), while $p=1/2$ is the state with completely random fluxes ($T=\infty$).

\begin{figure}[t]
  \includegraphics[width=\columnwidth]{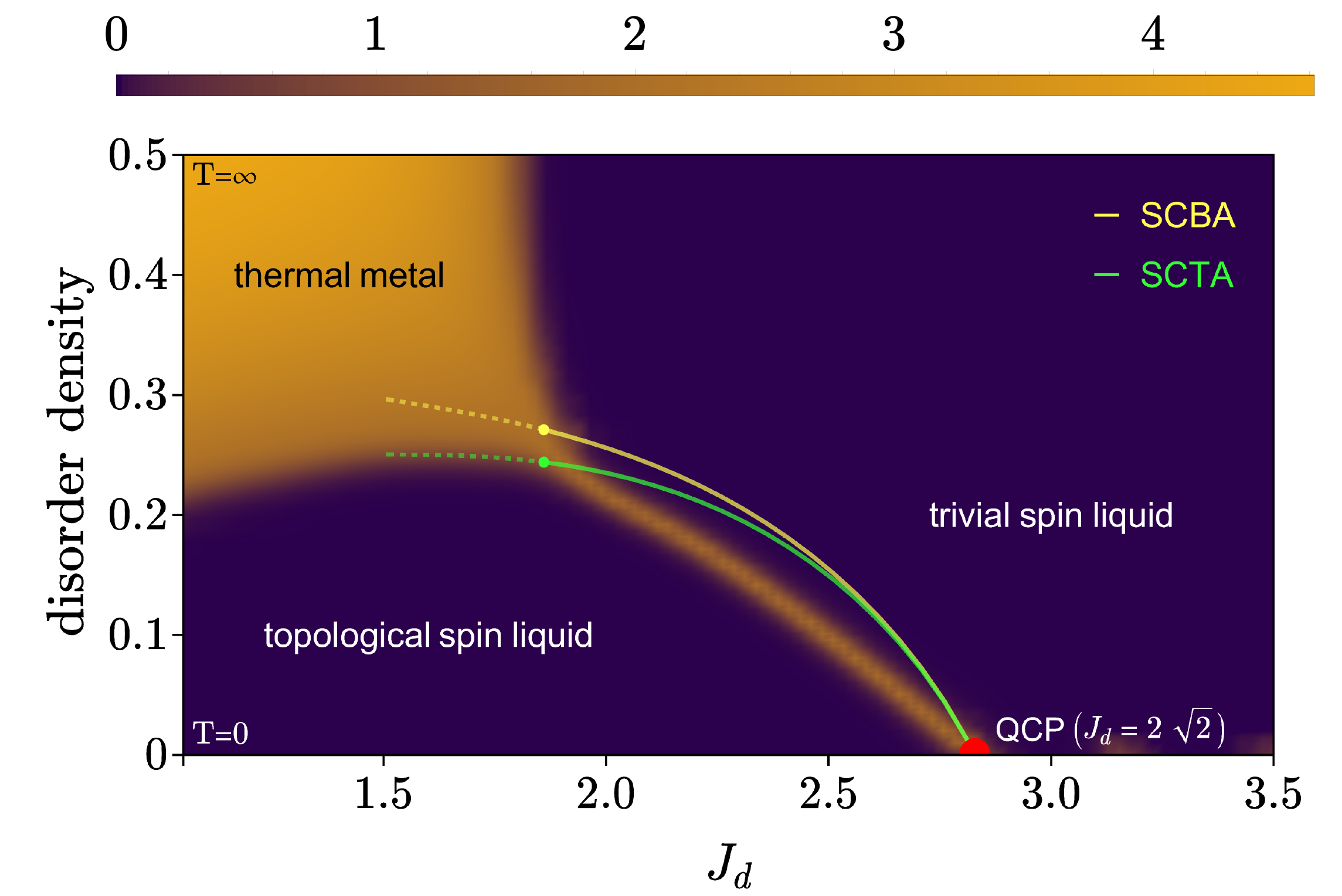}
  \caption{{\bf The thermal metal phase as an extension of the $T=0$ quantum critical point.}
  Shown is the effective number of gapless channels for the Majorana hopping model on a cylinder, overlaid with the phase boundaries obtained from self-consistent Born (SCBA) and T-matrix (SCTA) approximations. The $y$-axis denotes the fraction of plaquettes with zero flux, so that $p=0$ and $p=0.5$ corresponds to zero and infinite temperatures, respectively. Due to the Ising transition, this entire range is compressed in the phase diagram w.r.t, temperature (Fig.~\ref{fig:PhaseDiagramShastryChern}), where we see the phase boundary jump abruptly from $\Jz \approx 2.8$ to $\Jz \approx 1.9$.}
  \label{fig:Tmat}
\end{figure}

We analytically study the effect of this flux disorder by computing the self-energy as a power series in the disorder density -- which corresponds to a series in the number of scattering events -- using the self-consistent Born approximation (SCBA) as well as the self-consistent T-matrix approximation (SCTA). We find that the net effect of the disorder is a renormalization of the bare coupling strength $J$ (which we have set to 1 so far), so that the new phase boundary occurs at
\begin{equation}
  \Jz = 2\sqrt2 J_\text{eff}(p) \approx 2\sqrt2 \big(1 - \Sigma_p(0,\vec{k}_0) \big),
  \label{eq:critical_pt}
\end{equation}
where $\vec{k_0 = (\pi,\pi)}$ is the gap closing point at $\Jz = 2\sqrt2$ in the clean limit (see Appendix~\ref{app:Tmatrix} for a detailed derivation). In Fig.~\ref{fig:Tmat}, we plot this phase boundary as a function of the disorder density. We find that for $p \lesssim 0.25$, the disorder-induced renormalization of $J$ leads to a continuous shift of the phase boundary between the trivial and topological gapped phases towards smaller $\Jz$, i.e., to a net \emph{suppression} of the chiral spin-liquid phase\footnote{
  We contrast this to a similar effect in topological insulators with Anderson disorder, where disorder leads to an \emph{enhancement} of the topological phase. This difference is due to the purely imaginary disorder matrix in the present case, in contrast to a purely real one for topological insulators.
}.

The most intriguing result of this computation is that the {\sl entire} thermal metal phase can be thought of as emerging from the renormalization of the quantum critical point at $\Jz = 2\sqrt2$. This can be understood by a simple scaling argument: The leading term in the perturbation series for $\vec\Sigma_p(0,\vec{k}_0)$ corresponding to scattering events from $n$ impurities scales as $\left(4 p J^2 \right)^n$. Thus, for small $p$, we neglect the $n > 1$ terms to write $J_\text{eff}$ as a linear function of $p$. Within this ``non-crossing'' approximation, Eq.~\eqref{eq:critical_pt} has a {\sl single} solution for a given $p$, leading to a single gapless \emph{point} for each disorder density which we can approximate using the self-consistent Born (SCBA) and T-matrix approximations (SCTA). On the other hand, for $p \gtrsim 1/ 4$, we can no longer ignore the higher-order diagrams, so that Eq.~\eqref{eq:critical_pt} has, in fact, an {\sl infinite} number of solutions, accounting for an entire gapless \emph{region} in the phase space, which is precisely the thermal metal phase.

Since the perturbative approach is no longer useful in the high disorder density regime, we complement it with a numerical computation of the transmission coefficient. In particular, we compute the disorder-averaged generalized transfer matrix \cite{TransferMatrix} for the Majorana model on a cylinder geometry with $L_x \gg L_y$. The eigenvalues of this transfer matrix are related to the inverse localization lengths $\xi_i$, which can be used to compute the transmission coefficient as $g = \sum_{i=1}^{2 L_y} \text{sech}^2 (L_x/\xi_i)$. This can be thought of as the effective number of gapless channels, and is proportional to conductance for complex fermions \cite{mackinnon-kramer}.

This approach effectively captures the contributions from {\sl all} $n$-impurity scattering events, and thus yields a reliable estimate of the phase boundary also for large disorder densities. In Fig.~\ref{fig:Tmat} we plot $g$ as a function of $\Jz$ and $p$, overlaid with the phase boundaries obtained from SCBA/SCTA. We see a qualitative agreement between the numerical and analytical computations in the low density regime $p \ll 1/4$, verifying the effective renormalization of $J$. For $p \gtrsim 1/4$, we clearly see the spreading of the phase boundary into a gapless region extending up to $\Jz\approx1.8$, which is also in good agreement with the results obtained using QMC for the full Kitaev model (Fig.~\ref{fig:PhaseDiagramShastryChern}).


\subsubsection*{Physical interpretation}

In closing our discussion of the thermal metal, we need to address the question whether this thermal metal is a legitimate physical phase of our spin model or a mere artefact of our calculations relying on a Majorana decomposition. The relevance of this question acutely presents itself in the observation that the putative thermal metal regime seems to possibly extend beyond the intermediate-temperature range all the way up to infinite temperatures, as possibly suggested by the calculation of the average Chern number plotted in Fig.~\ref{fig:PhaseDiagramShastryChern}.

The answer to this question is two-fold. It sensitively depends on whether the Majorana fermions correspond to actual physical degrees of freedom of the system (after spin fractionalization) or whether they remain mere mathematical objects that can always be invoked in a spin decomposition. This fine distinction is well known from the solution of the honeycomb Kitaev model \cite{Kitaev2006anyons}, whose intricacy lies precisely in the fact that the spin decomposition ``becomes real" in the low-temperature phase and describes the emergent fractionalized degrees of freedom.
It is also exactly this distinction which distinguishes the intermediate-temperature regime from the high-temperature paramagnet in our model.
The intermediate-temperature regime is defined as the temperature regime sandwiched between the fractionalization crossover at $T'$ and the low-temperature gauge ordering transition at $T_c$, 
in which the emergent fractionalized degrees of freedom of itinerant Majorana fermions moving in the background of a static, but still disordered $\intg_2$ gauge field, form. In this regime,
our line of arguments outlined above fully applies and we conclude that in this temperature regime the thermal metal is a true physical phase.

The high-temperature paramagnet, on the other hand, is different. Here the Majorana decomposition of the spin operators is a mathematical possibility, but there is no deeper physical meaning associated with it (or any other spin decomposition using, e.g., complex fermions or other parton degrees of freedom). As such our numerical observation of a finite average Chern number and other class D physics in the Majorana representation is a direct reflection of the choice of decomposition, but not physically relevant.

We return to the question of how the thermal metal regime in the intermediate-temperature regime can be probed in the original spin system in our discussion section at the end of the manuscript.


\section{Second-order spin liquid}
\label{sec:ShastrySOSL}

\begin{figure}[t]
   \centering
    \includegraphics[width=0.66\columnwidth]{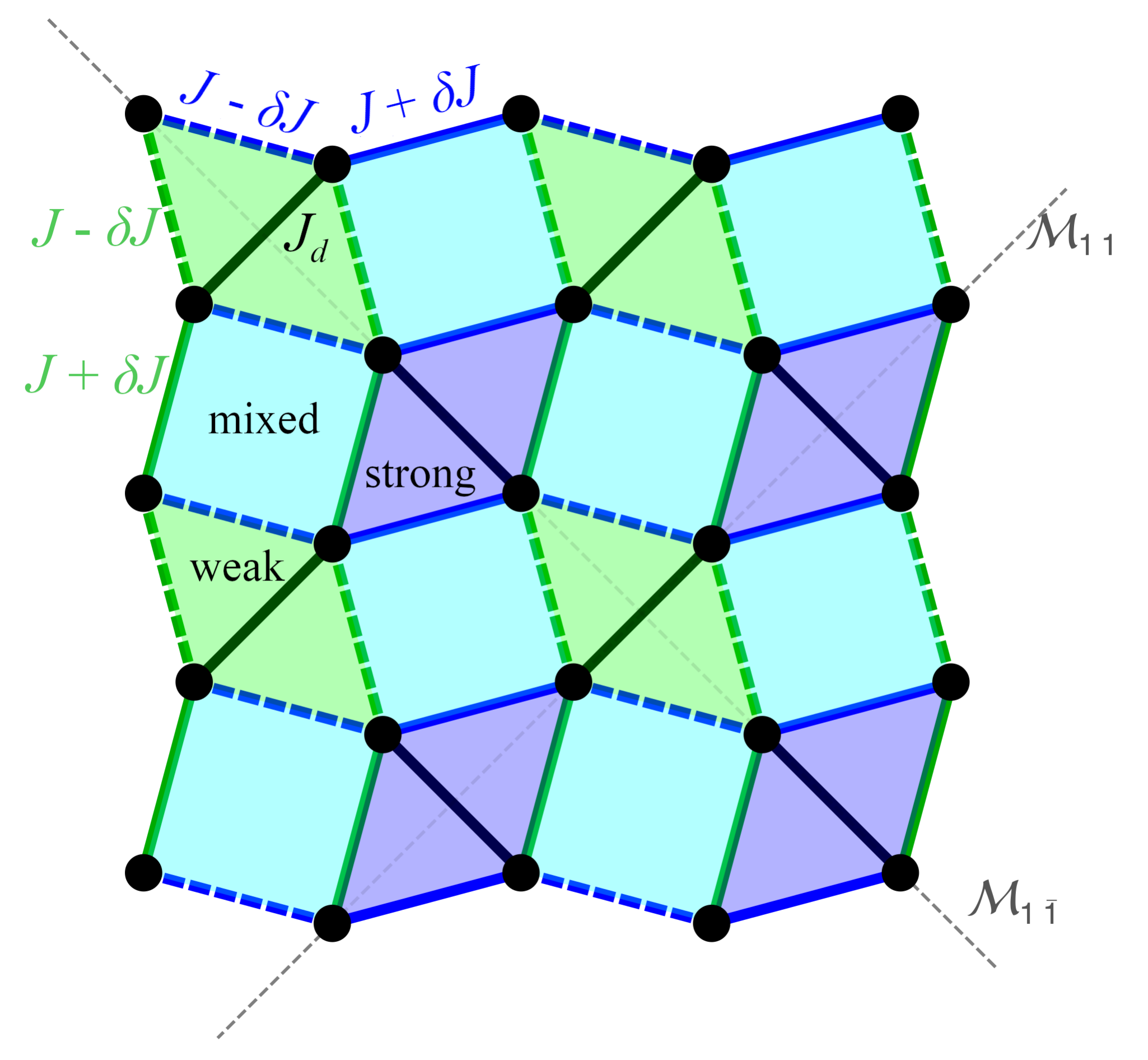}
  \caption{Kitaev Shastry-Sutherland mode with {\bf staggered plaquette couplings}. 
  		The dashed/solid lines correspond to staggered couplings $J \pm \delta J$, respectively, 
		which generates a hierarchy among the square plaquettes: 
		A quarter of the squares possess either only weak (green) or strong (blue) couplings on the edge bonds, 
		while the remaining half of plaquettes has two weak and two strong bonds (cyan).
		The dotted gray lines denote the two mirror axes.  
   }
    \label{fig:StaggeredModel}
\end{figure}

The physics of the Kitaev Shastry-Sutherland model becomes even richer when considering a {\sl staggering} of the plaquette couplings,
which has been shown \cite{DwivediCornerModes2018} to induce another variant of spin liquid ground state.
This ``second-order'' spin liquid (SOSL) exhibits a Majorana band structure in the ground state that, akin to a second-order topological insulator \cite{neupert_tci_rev}, is gapped in the bulk but exhibits gapless {\sl corner modes} (i.e., $d-2$ dimensional zero modes, as opposed to the usual $d-1$ dimensional zero energy modes). These corner modes are a manifestation of the formation of a {\sl symmetry-enriched} topological order that is protected by two mirror symmetries of the lattice (indicated in Fig.~\ref{fig:StaggeredModel}).

In the following, we will turn to the thermodynamics accompanying the formation of such a second-order spin liquid. In doing so, we will concentrate on a representative choice of coupling parameters deep in the SOSL regime. Specifically, we introduce the plaquette staggering
$\delta J = 0.7$ (resulting in weakly and strongly coupled plaquettes with coupling strength $J-\delta J$ and $J+\delta J$, respectively) and vary the relative coupling of the diagonal versus plaquette bonds $\Jz / J$.
At zero temperature, the phase diagram again consists of two phases \cite{DwivediCornerModes2018}: a chiral spin liquid for $\Jz/J > 2\sqrt2$ and a second-order spin liquid phase for $\Jz/J < 2\sqrt{2}$.

\subsection{Thermodynamics}
\label{ssec:SOSLThermalTransitions}

\begin{figure}[t]
   \centering
    \includegraphics[width=\columnwidth]{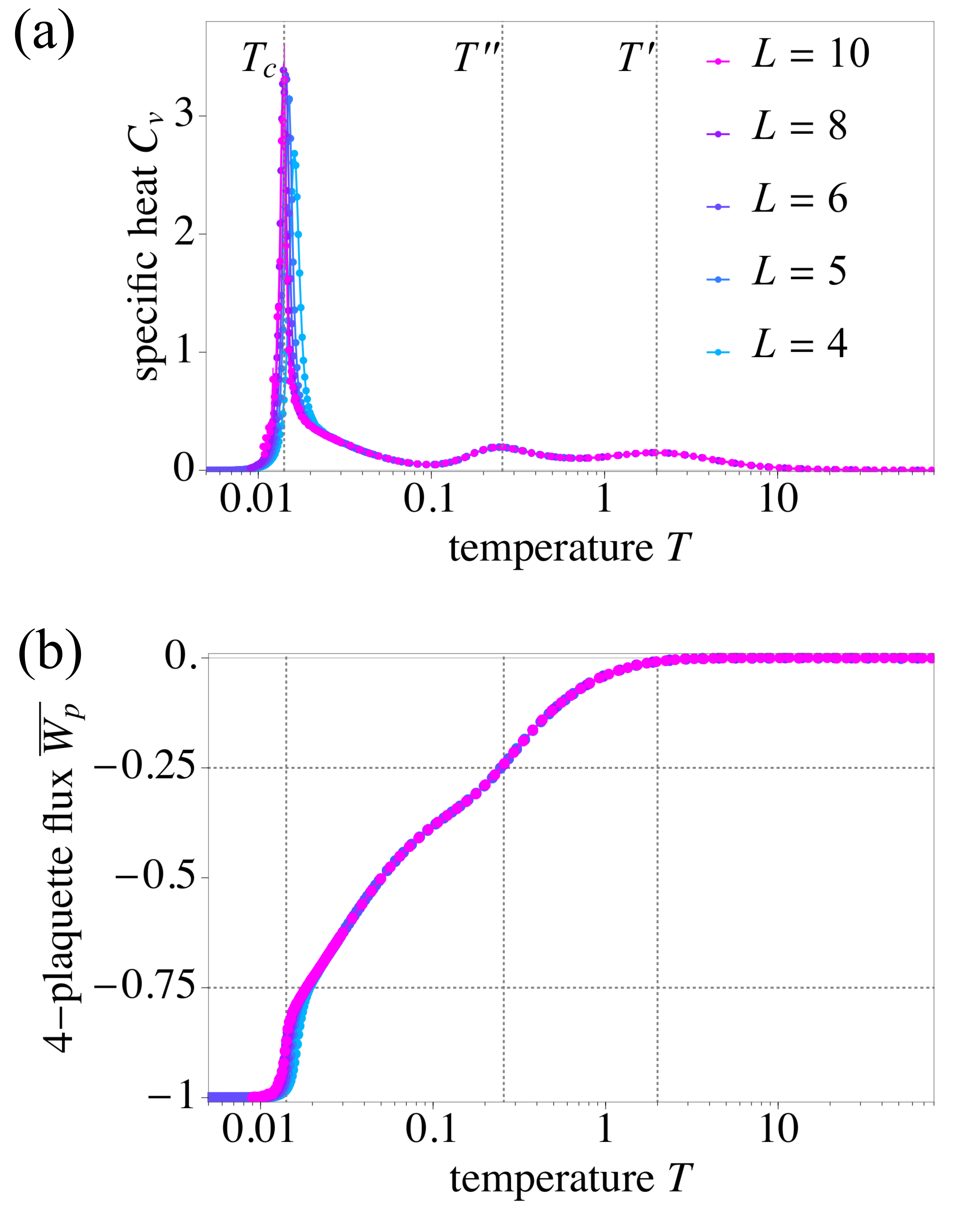}
  \caption{{\bf Signatures of a three-step thermodynamic transition.} Partial flux ordering in the SOSL phase (here, $J = 0.9$ and $\delta J = 0.4$). The specific heat $C_v$ shows a three-peak structure (a). While the high-temperature crossover is associated with spin fractionalization and the low-$T$ phase transition with spontaneous breaking of time-reversal symmetry, the intermediate crossover indicates a partial flux-ordering of square plaquettes (b), which is a consequence of staggered bond couplings $J \pm \delta J$. This choice of coupling generates a hierarchy between the square plaquettes of the lattice, which results in different ordering temperature scales.
  }
     \label{fig:SOSLCv}
\end{figure}

\begin{figure}[t]
\includegraphics[width=\columnwidth]{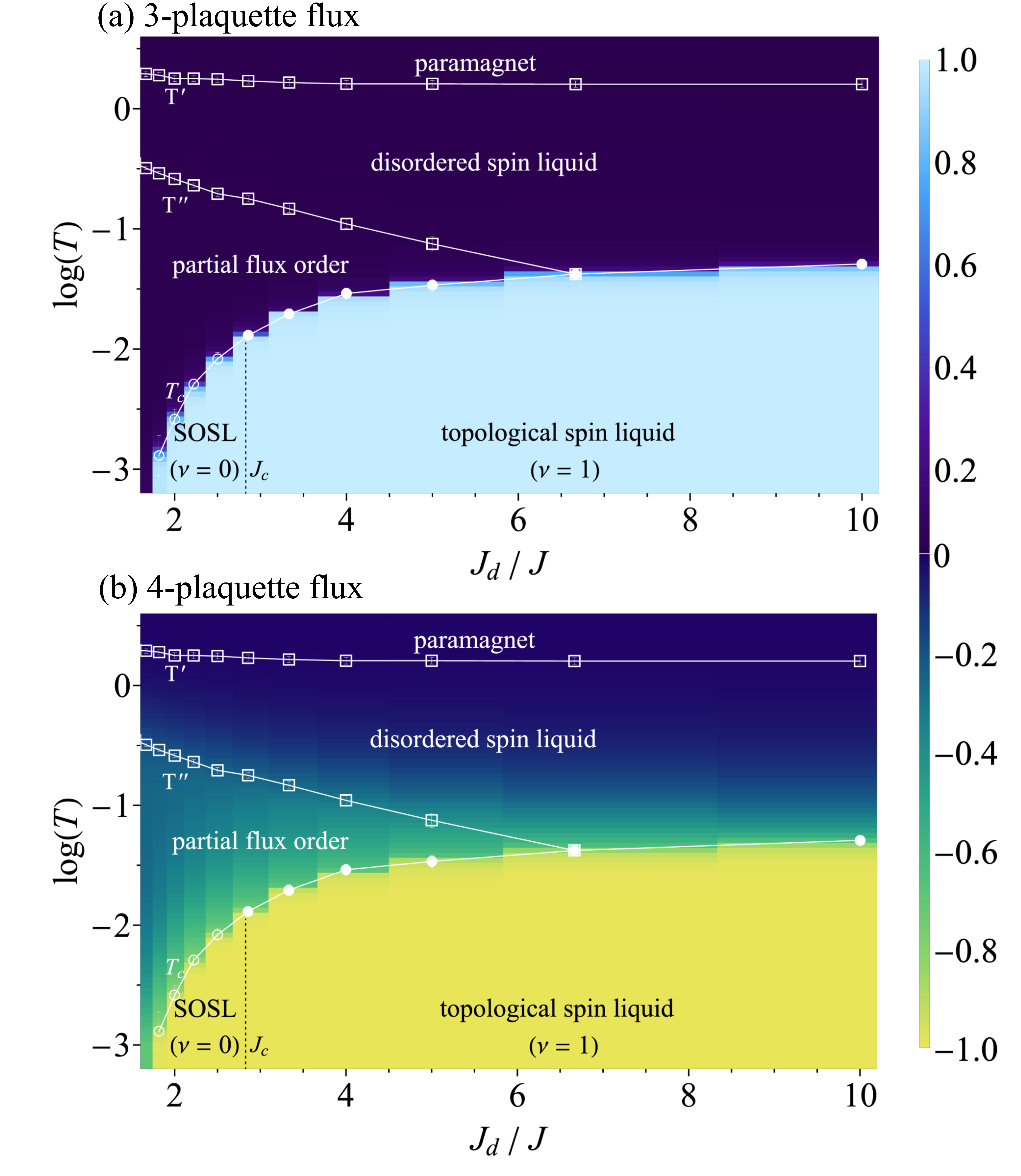}
\caption{{\bf Thermal phase diagram and flux ordering} (1) for fixed staggering parameter $\delta J = 0.7$ and $\Jz / J \in [1.53, 10]$. The transition temperatures of the system are given as a function of the coupling ratio $\Jz/J$ (white data points). The density plots show the 3-plaquette flux $|\overline{W_\triangle}|$ (a), and the 4-plaquette flux  $\overline{W_\square}$ (b). Filled (open) circles indicate phase transition temperatures $T_c$ in the topological spin liquid (SOSL) regime. Open squares indicate thermal crossovers associated with partial flux ordering ($T''$) and spin fractionalization ($T'$).}
 \label{fig:PhaseDiagramShastrySOSL}
\end{figure}

The principle thermodynamic signatures above the formation of this SOSL are similar to what we have seen for the more conventional spin liquids discussed in the previous section: The specific heat $C_v(T)$ exhibits a characteristic multi-peak structure, with a high-temperature local crossover at $T^\prime \sim 2 \Jz$  in this case, and a sharp low-temperature peak, at $T_c$, marking a true thermal phase transition associated with spontaneous breaking of time-reversal symmetry. For $\Jz / J \lesssim 6.7$ the lower peak splits into two peaks, the lower of which marks the symmetry-breaking transition and the upper one, at $T^{\prime\prime}$, marking a crossover into another partial flux ordered state. This is summarized in the finite-temperature cuts of Fig.~\ref{fig:SOSLCv}, which clearly indicate the multi-peak structure of the specific heat, and the thermal phase diagrams of Fig.~\ref{fig:PhaseDiagramShastrySOSL}, which shows the extent and boundaries of the different regimes.

The position of the low-$T$ transition in temperature space monotonously decreases if $\Jz / J$ is decreased. In the chiral spin liquid phase, it shows the particularly high value $T_c \sim 0.1 \Jz$ for large $\Jz / J$, a phenomenon which is discussed above. For $\Jz / J \rightarrow J_c = 2 \sqrt{2}$, the transition temperature reaches the order of magnitude $T_c \sim 10^{-2} \Jz$, and, in the SOSL phase, it is further lowered to $T_c \rightarrow 10^{-3} \Jz$. We note that for $\Jz / J < 1.8$, the transition temperature $T_c$ moves below the temperature range of our QMC simulations. In this limit, where the coupling $J - \delta J$ on half of the lattice bonds approaches 0, it is expected that the transition temperature rapidly decreases to lower temperature scales \cite{DwivediCornerModes2018}. 

\begin{figure}[h!]
\includegraphics[width=0.32\columnwidth]{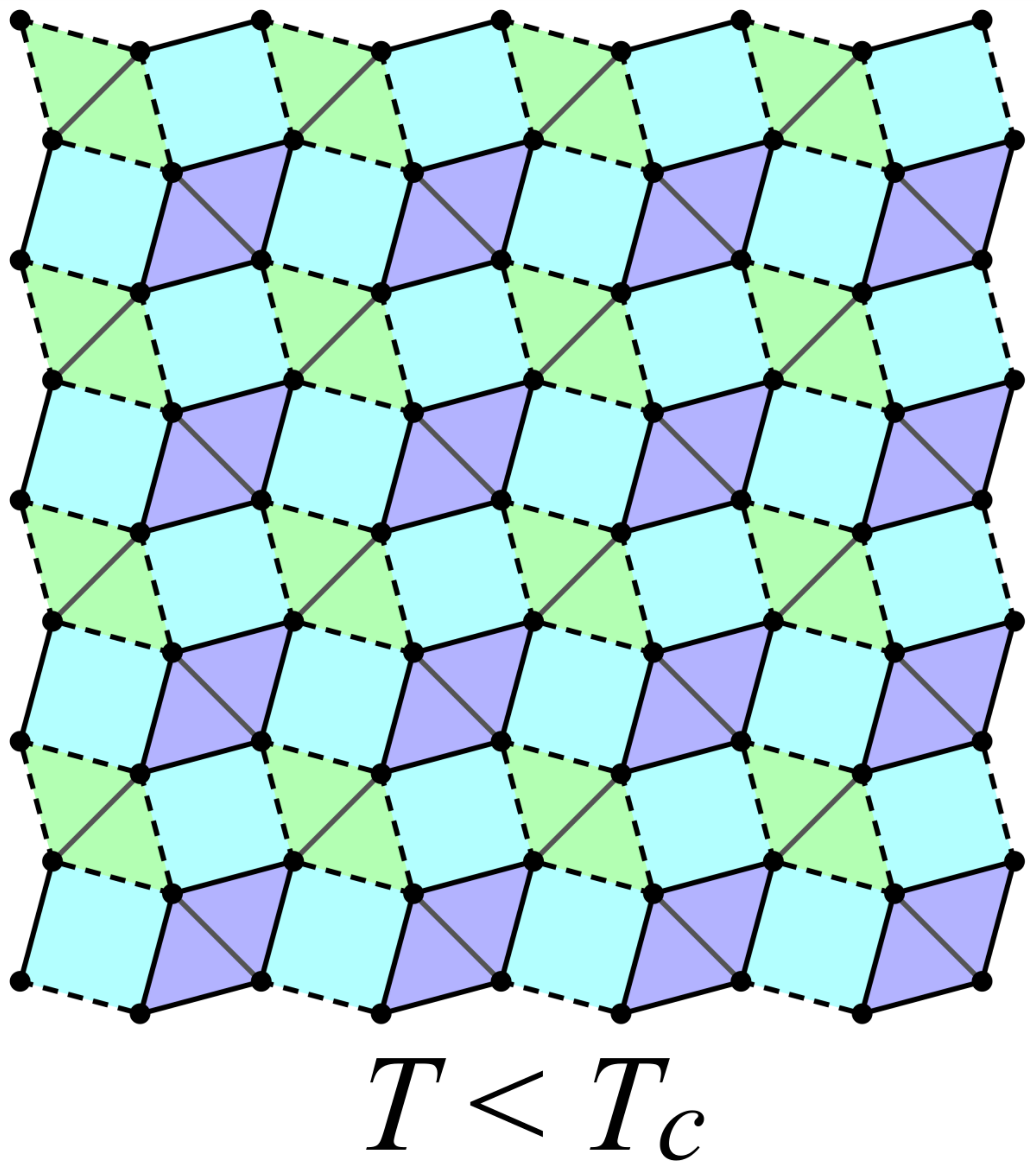}
\includegraphics[width=0.32\columnwidth]{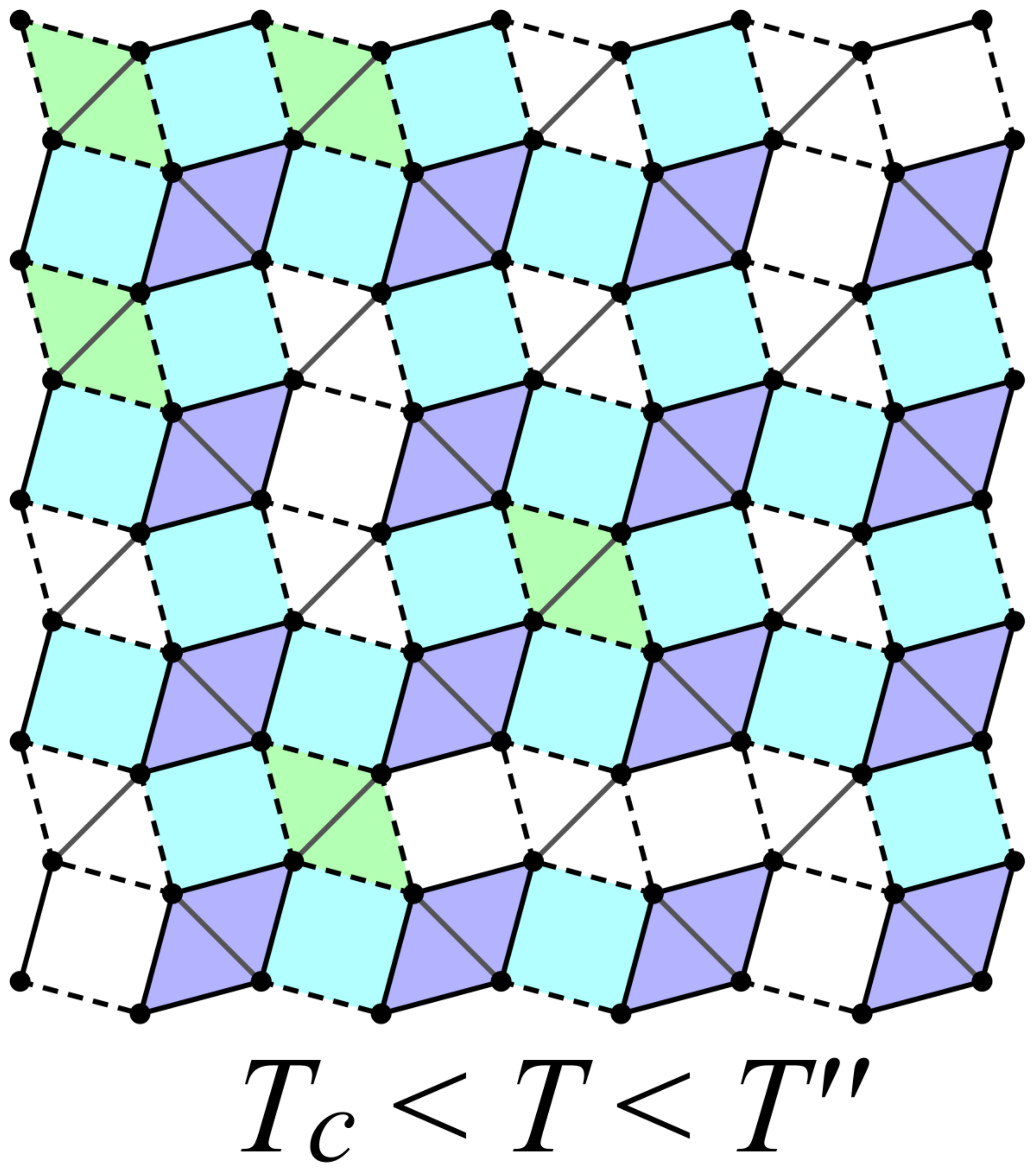}
\includegraphics[width=0.32\columnwidth]{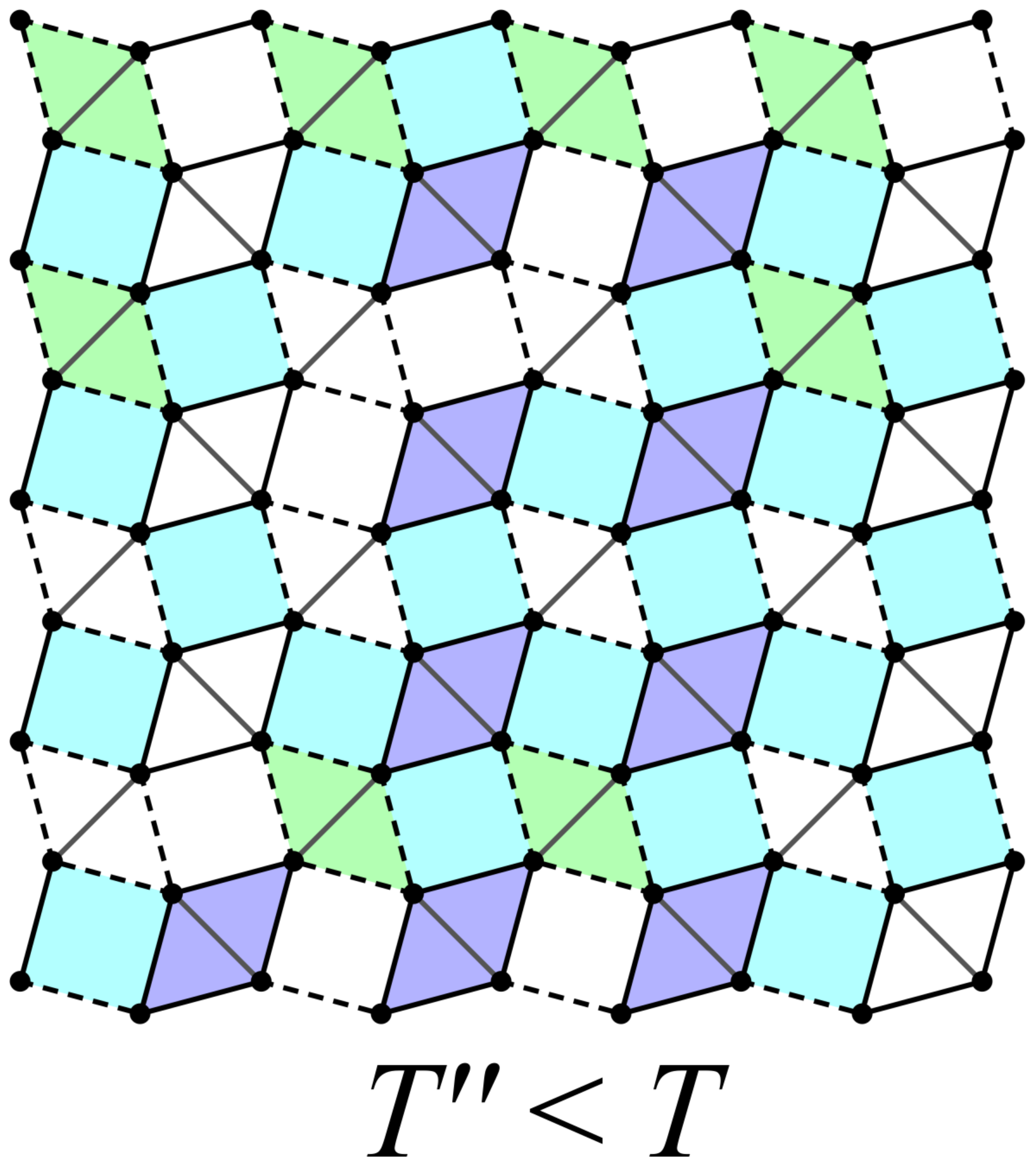}
\caption{{\bf Staggered partial flux ordering.} Here, the partial flux ordering is determined by the staggering of bond couplings (dashed / solid lines).At $T < T_c$, all square plaquettes have a $\pi$-flux (green). At $T_c$, the plaquettes with only weak edge couplings $J - \delta J$ are the first to become disordered, with $\pi$- and $0$-fluxes (green / red). Within this partially flux-ordered phase at $T_c < T < T''$, also the squares with two weak and two strong bond couplings become disordered. Finally, at $T''$, also the square plaquettes with strong couplings disorder. 
}
\label{fig:PartialFluxOrderingVisualization2}
\end{figure}


\subsection{Partial flux ordering}
\label{ssec:SOSLPartialFluxOrder}

The emergence of a partial flux ordering in an intermediate temperature range is again a precursor phenomenon for the formation of spin liquid ground states that, in the presence of staggered plaquette couplings, comes in even more variations.
This is readily illustrated by measurements of the three- and four-plaquette fluxes, overlaid as color-coding in the thermal phase diagrams of Fig.~\ref{fig:PhaseDiagramShastrySOSL}, which reveal a partially flux ordered regime in the parameter range $\Jz / J \lesssim 6.7$ and above the thermal phase transition $T^{\prime\prime}>T>T_c$. However, in this case, the pattern of partial ordering is very different to the one previously discussed in Section \ref{sec:ShastryChiral} for the original model. While there, we observed a region in which the square plaquettes were {\it fully} ordered, and the triangular plaquettes remained disordered, here the partial flux order is characterized by {\it partial} order of the square plaquettes, with the triangular plaquettes again remaining disordered. This can be clearly seen in the thermal phase diagram with Fig.~\ref{fig:PhaseDiagramShastrySOSL}(a) showing the three-plaquette flux $|\overline{W_\triangle}|=0$ and Fig.~\ref{fig:PhaseDiagramShastrySOSL}(b) showing the square-plaquette flux $-0.75 \lesssim \overline{W_\square} \lesssim -0.25$ within the partial flux ordered regime.

What is the reason for this new behavior? When looking at the model with staggered couplings, $J \pm \delta J$, we see that the lattice is now composed of {\sl three} different kinds of square plaquettes: (i) one quarter are ``strong plaquettes'', which contain a diagonal $\Jz$ bond and four ``strong'' bonds with coupling $J + \delta J$, (ii) one quarter are ``weak'' plaquettes, which contain a diagonal $\Jz$ bond and four ``weak'' bonds with coupling $J - \delta J$, (iii) while the remaining half are ``mixed'' plaquettes, which do not contain any diagonal bond and are made up of two ``strong'' bonds and ``two'' weak bonds. This hierarchy of couplings, and thus vison gaps, for the three different kinds of square plaquettes is precisely what underlies the emergence of the partial flux ordering.  

If we consider a lattice of $N_p$ total square plaquettes, the behavior seen within the partially flux-ordered regime can be explained as follows. Starting from the lowest temperatures, for $0<T<T_c$, {\sl all} square plaquettes are in an ordered $\pi$-flux phase, with $W_\square=-1$ for all $N_p$ square plaquettes. At $T_c$ the triangular plaquettes become disordered, triggering the recovery of time-reversal symmetry, and, also at $T_c$, the $N_p/4$ ``weak'' square plaquettes similarly become disordered. This loss of $N_p/4$ plaquettes explains the drop of $\overline{W_\square}$ from $-1$ to $-3/4$ at $T_c$, shown in the lower panel of Fig.~\ref{fig:SOSLCv}. Further increasing the temperature within the regime $T_c<T<T^{\prime\prime}$, the $N_p/2$ ``mixed'' plaquettes gradually disorder for higher temperatures, resulting in a smooth change of $\overline{W_\square}$ from $-3/4$ just above $T_c$ to $-1/4$ just below $T^{\prime\prime}$, see again the lower panel of Fig.~\ref{fig:SOSLCv}. Finally, at $T^{\prime\prime}$, the remaining $N_p/4$ ``strong'' plaquettes disorder, resulting in a fully disordered flux state with $\overline{W_\square}=0$ for all plaquettes for $T>T^{\prime\prime}$.


\subsection{Thermal Majorana metal}

\begin{figure}
   \centering
    \includegraphics[width=\columnwidth]{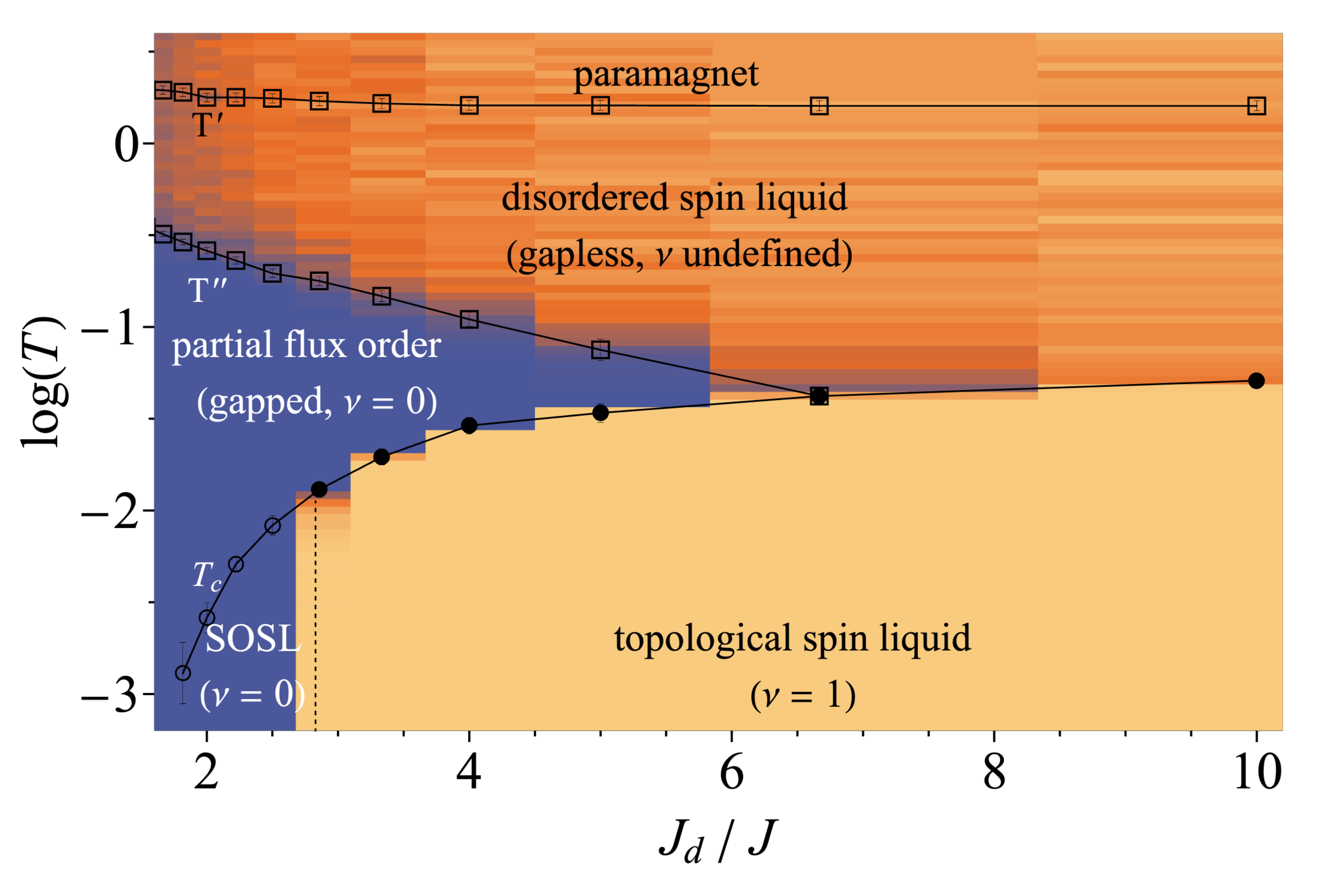}
     \caption{{\bf Thermal metal.} 
     		The color coding shows the average Chern number $|\nu|$ across the thermal phase diagram of the staggered Kitaev Shastry-Sutherland
		model. For the low-temperature ground-state phases the Chern number is $|\nu| = 1$, in the topological spin liquid phase 
		and, as expected, $|\nu| = 0$ in the second-order spin liquid (SOSL).
		In the intermediate temperature regime, we find that the average Chern number vanishes in the partially flux-ordered phase
		in the temperature range $T_c < T < T''$ for the respective range of couplings. 
		Above $T''$ (and $T_c$ for large $\Jz$, respectively) the average Chern number is finite, but not quantized around $|\nu| \approx 0.8$.
		The black data points in indicate the respective transition temperatures of the system as a function of the coupling ratio $\Jz/J$,
		extracted from specific heat traces akin to the ones shown in Fig.~\ref{fig:SOSLCv}. }
    \label{fig:PhaseDiagramShastrySOSLChern}
\end{figure}

Turning to signatures of Majorana physics in the thermodynamic behavior, we find evidence for the formation of a thermal metal regime 
also for the staggered model, similar to what we discussed extensively for the original Kitaev Shastry-Sutherland model in Sec. \ref{sec:ThermalMetal} above. The numerical evidence for the emergence of such a phase again is the observation of the vanishing of the Majorana gap, which is reflected in the finite (but not quantized) average Chern number illustrated in Fig.~\ref{fig:PhaseDiagramShastrySOSLChern}. The broad temperature regime where this average Chern number does not vanish (beyond the topological ground state phases) sharply sets in above the (partially) flux ordered phases, i.e. in the regime where the $\intg_2$ fluxes are effectively disordered and thereby create a disorder potential for the Majorana fermions (which form a nearly gapless band structure). Like in the original model, the average Chern number remains finite also in the high-temperature paramagnet and we refer to our previous discussion on the physical interpretation of this observation at the end of section \ref{sec:ThermalMetal}.

\section{Summary}
\label{sec:summary}

One of the most intriguing phenomena associated with the formation of quantum spin liquids is the emergence of novel, fractionalized
quantum mechanical degrees of freedom -- a quasiparticle, generally referred to as a parton, coupled to the gauge field of a deconfined
lattice gauge theory. For the spin-$3/2$ Kitaev Shastry-Sutherland model studied in this manuscript, these emergent degrees of freedom are
itinerant Majorana fermions and a $\intg_2$ lattice gauge field, with all spin liquids coming in the form of
varying
levels of topology -- trivial, first- and second-order, akin to the classification of higher-order topological insulators.

With a focus on the thermodynamic signatures of this fractionalization, we have observed in our numerically exact (sign-free)
quantum Monte Carlo simulations characteristic fingerprints of both the underlying gauge physics and the emergent Majoranas.
Below the thermodynamic crossover where these fractionalized degrees of freedom (locally) come to live, the gauge physics manifests 
itself through a sequence of (partial) flux ordering transitions, with the transition into the ground-state manifold being accompanied by (global)
time-reversal symmetry breaking. The latter is also mandated (already on a local level) by the Majorana physics for non-bipartite lattice geometries. 

Another striking manifestation of the interplay of Majorana physics and the $\intg_2$ gauge structure is the formation of a thermal
metal regime in an intermediate temperature range where the gauge field is intrinsically disordered. Notably, this gapless phase 
emerges only for a limited parameter regime of the Kitaev Shastry-Sutherland model, with a sharp transition to a gapped regime. 
This principle distinction might also make the thermal metal phase observable in experimental studies, e.g. in heat transport
measurements which have been demonstrated, both theoretically \cite{Nasu2015thermal,Yoshitake2016fractional,Gao2019} and experimentally \cite{Do2017incarnation,Kasahara2018,kasahara_majorana_2018,yokoi2020halfinteger}, to be sensitive probes of the 
Majorana physics. The gapless versus gapped character of the intermediate-temperature phases should also reflect itself in
four-spin correlation functions that probe the algebraic versus exponential decay of the bond-energy bond-energy correlations.


\acknowledgments
We acknowledge partial support from the Deutsche Forschungsgemeinschaft (DFG) -- project grants 277101999 and 277146847 -- within the CRC network TR 183 (project A04) and SFB 1238 (project C03).
The numerical simulations were performed on the JUWELS cluster at the Forschungszentrum J\"ulich.


\bibliography{./Kitaev.bib}

\begin{thebibliography}{59}%
\makeatletter
\providecommand \@ifxundefined [1]{%
 \@ifx{#1\undefined}
}%
\providecommand \@ifnum [1]{%
 \ifnum #1\expandafter \@firstoftwo
 \else \expandafter \@secondoftwo
 \fi
}%
\providecommand \@ifx [1]{%
 \ifx #1\expandafter \@firstoftwo
 \else \expandafter \@secondoftwo
 \fi
}%
\providecommand \natexlab [1]{#1}%
\providecommand \enquote  [1]{``#1''}%
\providecommand \bibnamefont  [1]{#1}%
\providecommand \bibfnamefont [1]{#1}%
\providecommand \citenamefont [1]{#1}%
\providecommand \href@noop [0]{\@secondoftwo}%
\providecommand \href [0]{\begingroup \@sanitize@url \@href}%
\providecommand \@href[1]{\@@startlink{#1}\@@href}%
\providecommand \@@href[1]{\endgroup#1\@@endlink}%
\providecommand \@sanitize@url [0]{\catcode `\\12\catcode `\$12\catcode
  `\&12\catcode `\#12\catcode `\^12\catcode `\_12\catcode `\%12\relax}%
\providecommand \@@startlink[1]{}%
\providecommand \@@endlink[0]{}%
\providecommand \url  [0]{\begingroup\@sanitize@url \@url }%
\providecommand \@url [1]{\endgroup\@href {#1}{\urlprefix }}%
\providecommand \urlprefix  [0]{URL }%
\providecommand \Eprint [0]{\href }%
\providecommand \doibase [0]{http://dx.doi.org/}%
\providecommand \selectlanguage [0]{\@gobble}%
\providecommand \bibinfo  [0]{\@secondoftwo}%
\providecommand \bibfield  [0]{\@secondoftwo}%
\providecommand \translation [1]{[#1]}%
\providecommand \BibitemOpen [0]{}%
\providecommand \bibitemStop [0]{}%
\providecommand \bibitemNoStop [0]{.\EOS\space}%
\providecommand \EOS [0]{\spacefactor3000\relax}%
\providecommand \BibitemShut  [1]{\csname bibitem#1\endcsname}%
\let\auto@bib@innerbib\@empty
\bibitem [{\citenamefont {Balents}(2010)}]{Balents2010spin}%
  \BibitemOpen
  \bibfield  {author} {\bibinfo {author} {\bibfnamefont {L.}~\bibnamefont
  {Balents}},\ }\bibfield  {title} {{\color{Gray}\small \bibinfo {title} {{Spin
  liquids in frustrated magnets}},\ }}\href
  {http://dx.doi.org/10.1038/nature08917} {\bibfield  {journal} {\bibinfo
  {journal} {Nature}\ }\textbf {\bibinfo {volume} {464}},\ \bibinfo {pages}
  {199} (\bibinfo {year} {2010})}\BibitemShut {NoStop}%
\bibitem [{\citenamefont {{Savary}}\ and\ \citenamefont
  {{Balents}}(2017)}]{Savary2017quantum}%
  \BibitemOpen
  \bibfield  {author} {\bibinfo {author} {\bibfnamefont {L.}~\bibnamefont
  {{Savary}}}\ and\ \bibinfo {author} {\bibfnamefont {L.}~\bibnamefont
  {{Balents}}},\ }\bibfield  {title} {{\color{Gray}\small \bibinfo {title}
  {{Quantum spin liquids: a review}},\ }}\href {\doibase
  10.1088/0034-4885/80/1/016502} {\bibfield  {journal} {\bibinfo  {journal}
  {Rep. Prog. Phys.}\ }\textbf {\bibinfo {volume} {80}},\ \bibinfo {eid}
  {016502} (\bibinfo {year} {2017})}\BibitemShut {NoStop}%
\bibitem [{\citenamefont {Kalmeyer}\ and\ \citenamefont
  {Laughlin}(1987)}]{KalmeyerLaughlin1987}%
  \BibitemOpen
  \bibfield  {author} {\bibinfo {author} {\bibfnamefont {V.}~\bibnamefont
  {Kalmeyer}}\ and\ \bibinfo {author} {\bibfnamefont {R.~B.}\ \bibnamefont
  {Laughlin}},\ }\bibfield  {title} {{\color{Gray}\small \bibinfo {title}
  {{Equivalence of the resonating-valence-bond and fractional quantum Hall
  states}},\ }}\href {\doibase 10.1103/PhysRevLett.59.2095} {\bibfield
  {journal} {\bibinfo  {journal} {Phys. Rev. Lett.}\ }\textbf {\bibinfo
  {volume} {59}},\ \bibinfo {pages} {2095} (\bibinfo {year}
  {1987})}\BibitemShut {NoStop}%
\bibitem [{\citenamefont {Anderson}(1987)}]{anderson_science_235_1987}%
  \BibitemOpen
  \bibfield  {author} {\bibinfo {author} {\bibfnamefont {P.~W.}\ \bibnamefont
  {Anderson}},\ }\bibfield  {title} {{\color{Gray}\small \bibinfo {title}
  {{T}he {R}esonating {V}alence {B}ond {S}tate in {La}$_2${Cu}{O}$_4$ and
  {S}uperconductivity},\ }}\href {\doibase 10.1126/science.235.4793.1196}
  {\bibfield  {journal} {\bibinfo  {journal} {Science}\ }\textbf {\bibinfo
  {volume} {235}},\ \bibinfo {pages} {1196} (\bibinfo {year}
  {1987})}\BibitemShut {NoStop}%
\bibitem [{\citenamefont {Wen}\ \emph {et~al.}(1989)\citenamefont {Wen},
  \citenamefont {Wilczek},\ and\ \citenamefont {Zee}}]{wen_prb_39_1989}%
  \BibitemOpen
  \bibfield  {author} {\bibinfo {author} {\bibfnamefont {X.-G.}\ \bibnamefont
  {Wen}}, \bibinfo {author} {\bibfnamefont {F.}~\bibnamefont {Wilczek}}, \ and\
  \bibinfo {author} {\bibfnamefont {A.}~\bibnamefont {Zee}},\ }\bibfield
  {title} {{\color{Gray}\small \bibinfo {title} {{C}hiral spin states and
  superconductivity},\ }}\href {\doibase 10.1103/PhysRevB.39.11413} {\bibfield
  {journal} {\bibinfo  {journal} {Phys. Rev. B}\ }\textbf {\bibinfo {volume}
  {39}},\ \bibinfo {pages} {11413} (\bibinfo {year} {1989})}\BibitemShut
  {NoStop}%
\bibitem [{\citenamefont {Kitaev}(2006)}]{Kitaev2006anyons}%
  \BibitemOpen
  \bibfield  {author} {\bibinfo {author} {\bibfnamefont {A.}~\bibnamefont
  {Kitaev}},\ }\bibfield  {title} {{\color{Gray}\small \bibinfo {title}
  {{Anyons in an exactly solved model and beyond}},\ }}\href {\doibase
  10.1016/j.aop.2005.10.005} {\bibfield  {journal} {\bibinfo  {journal} {Ann.
  Phys.}\ }\textbf {\bibinfo {volume} {321}},\ \bibinfo {pages} {2} (\bibinfo
  {year} {2006})}\BibitemShut {NoStop}%
\bibitem [{\citenamefont {Yao}\ and\ \citenamefont
  {Kivelson}(2007)}]{yao-kivelson}%
  \BibitemOpen
  \bibfield  {author} {\bibinfo {author} {\bibfnamefont {H.}~\bibnamefont
  {Yao}}\ and\ \bibinfo {author} {\bibfnamefont {S.~A.}\ \bibnamefont
  {Kivelson}},\ }\bibfield  {title} {{\color{Gray}\small \bibinfo {title}
  {{Exact Chiral Spin Liquid with Non-Abelian Anyons}},\ }}\href {\doibase
  10.1103/PhysRevLett.99.247203} {\bibfield  {journal} {\bibinfo  {journal}
  {Phys. Rev. Lett.}\ }\textbf {\bibinfo {volume} {99}},\ \bibinfo {pages}
  {247203} (\bibinfo {year} {2007})}\BibitemShut {NoStop}%
\bibitem [{\citenamefont {Schroeter}\ \emph {et~al.}(2007)\citenamefont
  {Schroeter}, \citenamefont {Kapit}, \citenamefont {Thomale},\ and\
  \citenamefont {Greiter}}]{schroeter_prl_99_2007}%
  \BibitemOpen
  \bibfield  {author} {\bibinfo {author} {\bibfnamefont {D.~F.}\ \bibnamefont
  {Schroeter}}, \bibinfo {author} {\bibfnamefont {E.}~\bibnamefont {Kapit}},
  \bibinfo {author} {\bibfnamefont {R.}~\bibnamefont {Thomale}}, \ and\
  \bibinfo {author} {\bibfnamefont {M.}~\bibnamefont {Greiter}},\ }\bibfield
  {title} {{\color{Gray}\small \bibinfo {title} {{S}pin hamiltonian for which
  the {C}hiral {S}pin {L}iquid is the {E}xact {G}round {S}tate},\ }}\href
  {\doibase 10.1103/PhysRevLett.99.097202} {\bibfield  {journal} {\bibinfo
  {journal} {Phys. Rev. Lett.}\ }\textbf {\bibinfo {volume} {99}},\ \bibinfo
  {pages} {097202} (\bibinfo {year} {2007})}\BibitemShut {NoStop}%
\bibitem [{\citenamefont {Messio}\ \emph {et~al.}(2012)\citenamefont {Messio},
  \citenamefont {Bernu},\ and\ \citenamefont {Lhuillier}}]{messo_prl_108_2012}%
  \BibitemOpen
  \bibfield  {author} {\bibinfo {author} {\bibfnamefont {L.}~\bibnamefont
  {Messio}}, \bibinfo {author} {\bibfnamefont {B.}~\bibnamefont {Bernu}}, \
  and\ \bibinfo {author} {\bibfnamefont {C.}~\bibnamefont {Lhuillier}},\
  }\bibfield  {title} {{\color{Gray}\small \bibinfo {title} {{K}agome
  {A}ntiferromagnet: {A} {C}hiral {T}opological {S}pin {L}iquid?{}},\ }}\href
  {\doibase 10.1103/PhysRevLett.108.207204} {\bibfield  {journal} {\bibinfo
  {journal} {Phys. Rev. Lett.}\ }\textbf {\bibinfo {volume} {108}},\ \bibinfo
  {pages} {207204} (\bibinfo {year} {2012})}\BibitemShut {NoStop}%
\bibitem [{\citenamefont {Bauer}\ \emph {et~al.}(2014)\citenamefont {Bauer},
  \citenamefont {Cincio}, \citenamefont {Keller}, \citenamefont {Dolfi},
  \citenamefont {Vidal}, \citenamefont {Trebst},\ and\ \citenamefont
  {Ludwig}}]{bauer_ncom_5_2014}%
  \BibitemOpen
  \bibfield  {author} {\bibinfo {author} {\bibfnamefont {B.}~\bibnamefont
  {Bauer}}, \bibinfo {author} {\bibfnamefont {L.}~\bibnamefont {Cincio}},
  \bibinfo {author} {\bibfnamefont {B.}~\bibnamefont {Keller}}, \bibinfo
  {author} {\bibfnamefont {M.}~\bibnamefont {Dolfi}}, \bibinfo {author}
  {\bibfnamefont {G.}~\bibnamefont {Vidal}}, \bibinfo {author} {\bibfnamefont
  {S.}~\bibnamefont {Trebst}}, \ and\ \bibinfo {author} {\bibfnamefont
  {A.}~\bibnamefont {Ludwig}},\ }\bibfield  {title} {{\color{Gray}\small
  \bibinfo {title} {{C}hiral spin liquid and emergent anyons in a {K}agome
  lattice {M}ott insulator},\ }}\href {\doibase 10.1038/ncomms6137} {\bibfield
  {journal} {\bibinfo  {journal} {Nat. Commun.}\ }\textbf {\bibinfo {volume}
  {5}},\ \bibinfo {pages} {5137} (\bibinfo {year} {2014})}\BibitemShut
  {NoStop}%
\bibitem [{\citenamefont {Gong}\ \emph {et~al.}(2014)\citenamefont {Gong},
  \citenamefont {Zhu},\ and\ \citenamefont {Sheng}}]{gong_srep_4_2014}%
  \BibitemOpen
  \bibfield  {author} {\bibinfo {author} {\bibfnamefont {S.-S.}\ \bibnamefont
  {Gong}}, \bibinfo {author} {\bibfnamefont {W.}~\bibnamefont {Zhu}}, \ and\
  \bibinfo {author} {\bibfnamefont {D.}~\bibnamefont {Sheng}},\ }\bibfield
  {title} {{\color{Gray}\small \bibinfo {title} {{E}mergent {C}hiral {S}pin
  {L}iquid: {F}ractional {Q}uantum {H}all {E}ffect in a {K}agome {H}eisenberg
  {M}odel},\ }}\href {\doibase 10.1038/srep06317} {\bibfield  {journal}
  {\bibinfo  {journal} {Sci. Rep.}\ }\textbf {\bibinfo {volume} {4}},\ \bibinfo
  {pages} {6317} (\bibinfo {year} {2014})}\BibitemShut {NoStop}%
\bibitem [{\citenamefont {Kasahara}\ \emph
  {et~al.}(2018{\natexlab{a}})\citenamefont {Kasahara}, \citenamefont
  {Ohnishi}, \citenamefont {Mizukami}, \citenamefont {Tanaka}, \citenamefont
  {Ma}, \citenamefont {Sugii}, \citenamefont {Kurita}, \citenamefont {Tanaka},
  \citenamefont {Nasu}, \citenamefont {Motome}, \citenamefont {Shibauchi},\
  and\ \citenamefont {Matsuda}}]{kasahara_majorana_2018}%
  \BibitemOpen
  \bibfield  {author} {\bibinfo {author} {\bibfnamefont {Y.}~\bibnamefont
  {Kasahara}}, \bibinfo {author} {\bibfnamefont {T.}~\bibnamefont {Ohnishi}},
  \bibinfo {author} {\bibfnamefont {Y.}~\bibnamefont {Mizukami}}, \bibinfo
  {author} {\bibfnamefont {O.}~\bibnamefont {Tanaka}}, \bibinfo {author}
  {\bibfnamefont {S.}~\bibnamefont {Ma}}, \bibinfo {author} {\bibfnamefont
  {K.}~\bibnamefont {Sugii}}, \bibinfo {author} {\bibfnamefont
  {N.}~\bibnamefont {Kurita}}, \bibinfo {author} {\bibfnamefont
  {H.}~\bibnamefont {Tanaka}}, \bibinfo {author} {\bibfnamefont
  {J.}~\bibnamefont {Nasu}}, \bibinfo {author} {\bibfnamefont {Y.}~\bibnamefont
  {Motome}}, \bibinfo {author} {\bibfnamefont {T.}~\bibnamefont {Shibauchi}}, \
  and\ \bibinfo {author} {\bibfnamefont {Y.}~\bibnamefont {Matsuda}},\
  }\bibfield  {title} {{\color{Gray}\small \bibinfo {title} {Majorana
  quantization and half-integer thermal quantum {Hall} effect in a {Kitaev}
  spin liquid},\ }}\href {\doibase 10.1038/s41586-018-0274-0} {\bibfield
  {journal} {\bibinfo  {journal} {Nature}\ }\textbf {\bibinfo {volume} {559}},\
  \bibinfo {pages} {227} (\bibinfo {year} {2018}{\natexlab{a}})}\BibitemShut
  {NoStop}%
\bibitem [{\citenamefont {Yokoi}\ \emph {et~al.}(2020)\citenamefont {Yokoi},
  \citenamefont {Ma}, \citenamefont {Kasahara}, \citenamefont {Kasahara},
  \citenamefont {Shibauchi}, \citenamefont {Kurita}, \citenamefont {Tanaka},
  \citenamefont {Nasu}, \citenamefont {Motome}, \citenamefont {Hickey},
  \citenamefont {Trebst},\ and\ \citenamefont
  {Matsuda}}]{yokoi2020halfinteger}%
  \BibitemOpen
  \bibfield  {author} {\bibinfo {author} {\bibfnamefont {T.}~\bibnamefont
  {Yokoi}}, \bibinfo {author} {\bibfnamefont {S.}~\bibnamefont {Ma}}, \bibinfo
  {author} {\bibfnamefont {Y.}~\bibnamefont {Kasahara}}, \bibinfo {author}
  {\bibfnamefont {S.}~\bibnamefont {Kasahara}}, \bibinfo {author}
  {\bibfnamefont {T.}~\bibnamefont {Shibauchi}}, \bibinfo {author}
  {\bibfnamefont {N.}~\bibnamefont {Kurita}}, \bibinfo {author} {\bibfnamefont
  {H.}~\bibnamefont {Tanaka}}, \bibinfo {author} {\bibfnamefont
  {J.}~\bibnamefont {Nasu}}, \bibinfo {author} {\bibfnamefont {Y.}~\bibnamefont
  {Motome}}, \bibinfo {author} {\bibfnamefont {C.}~\bibnamefont {Hickey}},
  \bibinfo {author} {\bibfnamefont {S.}~\bibnamefont {Trebst}}, \ and\ \bibinfo
  {author} {\bibfnamefont {Y.}~\bibnamefont {Matsuda}},\ }\bibfield  {title}
  {{\color{Gray}\small \bibinfo {title} {{Half-integer quantized anomalous
  thermal Hall effect in the Kitaev material $\alpha$-RuCl$_3$}},\ }}\href@noop
  {} {\  (\bibinfo {year} {2020})},\ \Eprint {http://arxiv.org/abs/2001.01899}
  {arXiv:2001.01899} \BibitemShut {NoStop}%
\bibitem [{\citenamefont {Rowell}\ \emph {et~al.}(2009)\citenamefont {Rowell},
  \citenamefont {Stong},\ and\ \citenamefont {Wang}}]{Rowell2009}%
  \BibitemOpen
  \bibfield  {author} {\bibinfo {author} {\bibfnamefont {E.}~\bibnamefont
  {Rowell}}, \bibinfo {author} {\bibfnamefont {R.}~\bibnamefont {Stong}}, \
  and\ \bibinfo {author} {\bibfnamefont {Z.}~\bibnamefont {Wang}},\ }\bibfield
  {title} {{\color{Gray}\small \bibinfo {title} {{On Classification of Modular
  Tensor Categories}},\ }}\href {\doibase 10.1007/s00220-009-0908-z} {\bibfield
   {journal} {\bibinfo  {journal} {Communications in Mathematical Physics}\
  }\textbf {\bibinfo {volume} {292}},\ \bibinfo {pages} {343} (\bibinfo {year}
  {2009})}\BibitemShut {NoStop}%
\bibitem [{\citenamefont {Bruillard}\ \emph {et~al.}(2016)\citenamefont
  {Bruillard}, \citenamefont {Ng}, \citenamefont {Rowell},\ and\ \citenamefont
  {Wang}}]{Bruillard2016}%
  \BibitemOpen
  \bibfield  {author} {\bibinfo {author} {\bibfnamefont {P.}~\bibnamefont
  {Bruillard}}, \bibinfo {author} {\bibfnamefont {S.-H.}\ \bibnamefont {Ng}},
  \bibinfo {author} {\bibfnamefont {E.}~\bibnamefont {Rowell}}, \ and\ \bibinfo
  {author} {\bibfnamefont {Z.}~\bibnamefont {Wang}},\ }\bibfield  {title}
  {{\color{Gray}\small \bibinfo {title} {{On Classification of Modular Tensor
  Categories}},\ }}\href {\doibase 10.1090/jams/842} {\bibfield  {journal}
  {\bibinfo  {journal} {J. Amer. Math. Soc.}\ }\textbf {\bibinfo {volume}
  {29}},\ \bibinfo {pages} {857} (\bibinfo {year} {2016})}\BibitemShut
  {NoStop}%
\bibitem [{\citenamefont {Zhang}\ \emph {et~al.}(2012)\citenamefont {Zhang},
  \citenamefont {Grover}, \citenamefont {Turner}, \citenamefont {Oshikawa},\
  and\ \citenamefont {Vishwanath}}]{Zhang2012}%
  \BibitemOpen
  \bibfield  {author} {\bibinfo {author} {\bibfnamefont {Y.}~\bibnamefont
  {Zhang}}, \bibinfo {author} {\bibfnamefont {T.}~\bibnamefont {Grover}},
  \bibinfo {author} {\bibfnamefont {A.}~\bibnamefont {Turner}}, \bibinfo
  {author} {\bibfnamefont {M.}~\bibnamefont {Oshikawa}}, \ and\ \bibinfo
  {author} {\bibfnamefont {A.}~\bibnamefont {Vishwanath}},\ }\bibfield  {title}
  {{\color{Gray}\small \bibinfo {title} {{Quasiparticle statistics and braiding
  from ground-state entanglement}},\ }}\href {\doibase
  10.1103/PhysRevB.85.235151} {\bibfield  {journal} {\bibinfo  {journal} {Phys.
  Rev. B}\ }\textbf {\bibinfo {volume} {85}},\ \bibinfo {pages} {235151}
  (\bibinfo {year} {2012})}\BibitemShut {NoStop}%
\bibitem [{\citenamefont {Trebst}(2017)}]{Trebst2017}%
  \BibitemOpen
  \bibfield  {author} {\bibinfo {author} {\bibfnamefont {S.}~\bibnamefont
  {Trebst}},\ }\bibfield  {title} {{\color{Gray}\small \bibinfo {title}
  {{Kitaev Materials}},\ }}\href@noop {} {\bibfield  {journal} {\bibinfo
  {journal} {arXiv:1701.07056}\ } (\bibinfo {year} {2017})}\BibitemShut
  {NoStop}%
\bibitem [{\citenamefont {Wen}(1991)}]{Wen1992}%
  \BibitemOpen
  \bibfield  {author} {\bibinfo {author} {\bibfnamefont {X.}~\bibnamefont
  {Wen}},\ }\bibfield  {title} {{\color{Gray}\small \bibinfo {title} {Edge
  excitations in the fractional quantum hall states at general filling
  fractions},\ }}\href {\doibase 10.1142/S0217984991000058} {\bibfield
  {journal} {\bibinfo  {journal} {Modern Physics Letters B}\ }\textbf {\bibinfo
  {volume} {05}},\ \bibinfo {pages} {39} (\bibinfo {year} {1991})}\BibitemShut
  {NoStop}%
\bibitem [{Note1()}]{Note1}%
  \BibitemOpen
  \bibinfo {note} {An alternative parton decomposition employs complex fermions
  coupled to a $U(1)$ gauge field \cite {burnell_su2_2011}, which then leads to
  a description of the Kitaev spin liquid as a nodal superconductor. This
  picture has been particularly insightful in explaining the in-field behavior
  of the antiferromagnetic Kitaev honeycomb model which exhibits a Higgs
  transition to an intermediate gapless $U(1)$ spin liquid \cite
  {hickey_emergence_2019}.}\BibitemShut {Stop}%
\bibitem [{\citenamefont {Bernevig}\ and\ \citenamefont
  {Hughes}(2013)}]{BernevigHughes}%
  \BibitemOpen
  \bibfield  {author} {\bibinfo {author} {\bibfnamefont {B.~A.}\ \bibnamefont
  {Bernevig}}\ and\ \bibinfo {author} {\bibfnamefont {T.~L.}\ \bibnamefont
  {Hughes}},\ }\href@noop {} {\emph {\bibinfo {title} {Topological Insulators
  and Topological Superconductors}}}\ (\bibinfo  {publisher} {Princeton
  University Press},\ \bibinfo {year} {2013})\BibitemShut {NoStop}%
\bibitem [{\citenamefont {Neupert}\ and\ \citenamefont
  {Schindler}(2018)}]{neupert_tci_rev}%
  \BibitemOpen
  \bibfield  {author} {\bibinfo {author} {\bibfnamefont {T.}~\bibnamefont
  {Neupert}}\ and\ \bibinfo {author} {\bibfnamefont {F.}~\bibnamefont
  {Schindler}},\ }\bibfield  {title} {{\color{Gray}\small \bibinfo {title}
  {Topological crystalline insulators},\ }}in\ \href@noop {} {\emph {\bibinfo
  {booktitle} {Topological Matter}}}\ (\bibinfo  {publisher} {Springer},\
  \bibinfo {year} {2018})\ pp.\ \bibinfo {pages} {31--61}\BibitemShut {NoStop}%
\bibitem [{\citenamefont {Dwivedi}\ \emph {et~al.}(2018)\citenamefont
  {Dwivedi}, \citenamefont {Hickey}, \citenamefont {Eschmann},\ and\
  \citenamefont {Trebst}}]{DwivediCornerModes2018}%
  \BibitemOpen
  \bibfield  {author} {\bibinfo {author} {\bibfnamefont {V.}~\bibnamefont
  {Dwivedi}}, \bibinfo {author} {\bibfnamefont {C.}~\bibnamefont {Hickey}},
  \bibinfo {author} {\bibfnamefont {T.}~\bibnamefont {Eschmann}}, \ and\
  \bibinfo {author} {\bibfnamefont {S.}~\bibnamefont {Trebst}},\ }\bibfield
  {title} {{\color{Gray}\small \bibinfo {title} {Majorana corner modes in a
  second-order {Kitaev} spin liquid},\ }}\href {\doibase
  10.1103/PhysRevB.98.054432} {\bibfield  {journal} {\bibinfo  {journal} {Phys.
  Rev. B}\ }\textbf {\bibinfo {volume} {98}},\ \bibinfo {pages} {054432}
  (\bibinfo {year} {2018})}\BibitemShut {NoStop}%
\bibitem [{Note2()}]{Note2}%
  \BibitemOpen
  \bibinfo {note} {Note that, though all of the spin liquid ground states
  spontaneously break time-reversal symmetry, we reserve the term `chiral spin
  liquid' for those that exhibit chiral edge states (or, more technically,
  those that possess a non-zero chiral central charge).}\BibitemShut {Stop}%
\bibitem [{\citenamefont {Nasu}\ \emph {et~al.}(2014)\citenamefont {Nasu},
  \citenamefont {Udagawa},\ and\ \citenamefont
  {Motome}}]{nasu2014vaporization}%
  \BibitemOpen
  \bibfield  {author} {\bibinfo {author} {\bibfnamefont {J.}~\bibnamefont
  {Nasu}}, \bibinfo {author} {\bibfnamefont {M.}~\bibnamefont {Udagawa}}, \
  and\ \bibinfo {author} {\bibfnamefont {Y.}~\bibnamefont {Motome}},\
  }\bibfield  {title} {{\color{Gray}\small \bibinfo {title} {{Vaporization of
  Kitaev Spin Liquids}},\ }}\href {\doibase 10.1103/PhysRevLett.113.197205}
  {\bibfield  {journal} {\bibinfo  {journal} {Phys. Rev. Lett.}\ }\textbf
  {\bibinfo {volume} {113}},\ \bibinfo {pages} {197205} (\bibinfo {year}
  {2014})}\BibitemShut {NoStop}%
\bibitem [{\citenamefont {Wu}\ \emph {et~al.}(2009)\citenamefont {Wu},
  \citenamefont {Arovas},\ and\ \citenamefont {Hung}}]{WuHungGammaMatrix2009}%
  \BibitemOpen
  \bibfield  {author} {\bibinfo {author} {\bibfnamefont {C.}~\bibnamefont
  {Wu}}, \bibinfo {author} {\bibfnamefont {D.}~\bibnamefont {Arovas}}, \ and\
  \bibinfo {author} {\bibfnamefont {H.-H.}\ \bibnamefont {Hung}},\ }\bibfield
  {title} {{\color{Gray}\small \bibinfo {title} {{$\ensuremath{\Gamma}$-matrix
  generalization of the Kitaev model}},\ }}\href {\doibase
  10.1103/PhysRevB.79.134427} {\bibfield  {journal} {\bibinfo  {journal} {Phys.
  Rev. B}\ }\textbf {\bibinfo {volume} {79}},\ \bibinfo {pages} {134427}
  (\bibinfo {year} {2009})}\BibitemShut {NoStop}%
\bibitem [{\citenamefont {Shastry}\ and\ \citenamefont
  {Sutherland}(1981)}]{shastry-sutherland}%
  \BibitemOpen
  \bibfield  {author} {\bibinfo {author} {\bibfnamefont {B.~S.}\ \bibnamefont
  {Shastry}}\ and\ \bibinfo {author} {\bibfnamefont {B.}~\bibnamefont
  {Sutherland}},\ }\bibfield  {title} {{\color{Gray}\small \bibinfo {title}
  {Exact ground state of a quantum mechanical antiferromagnet},\ }}\href
  {\doibase https://doi.org/10.1016/0378-4363(81)90838-X} {\bibfield  {journal}
  {\bibinfo  {journal} {Physica B+C}\ }\textbf {\bibinfo {volume} {108}},\
  \bibinfo {pages} {1069 } (\bibinfo {year} {1981})}\BibitemShut {NoStop}%
\bibitem [{\citenamefont {Kageyama}\ \emph {et~al.}(1999)\citenamefont
  {Kageyama}, \citenamefont {Yoshimura}, \citenamefont {Stern}, \citenamefont
  {Mushnikov}, \citenamefont {Onizuka}, \citenamefont {Kato}, \citenamefont
  {Kosuge}, \citenamefont {Slichter}, \citenamefont {Goto},\ and\ \citenamefont
  {Ueda}}]{Kageyama1999}%
  \BibitemOpen
  \bibfield  {author} {\bibinfo {author} {\bibfnamefont {H.}~\bibnamefont
  {Kageyama}}, \bibinfo {author} {\bibfnamefont {K.}~\bibnamefont {Yoshimura}},
  \bibinfo {author} {\bibfnamefont {R.}~\bibnamefont {Stern}}, \bibinfo
  {author} {\bibfnamefont {N.~V.}\ \bibnamefont {Mushnikov}}, \bibinfo {author}
  {\bibfnamefont {K.}~\bibnamefont {Onizuka}}, \bibinfo {author} {\bibfnamefont
  {M.}~\bibnamefont {Kato}}, \bibinfo {author} {\bibfnamefont {K.}~\bibnamefont
  {Kosuge}}, \bibinfo {author} {\bibfnamefont {C.~P.}\ \bibnamefont
  {Slichter}}, \bibinfo {author} {\bibfnamefont {T.}~\bibnamefont {Goto}}, \
  and\ \bibinfo {author} {\bibfnamefont {Y.}~\bibnamefont {Ueda}},\ }\bibfield
  {title} {{\color{Gray}\small \bibinfo {title} {{Exact Dimer Ground State and
  Quantized Magnetization Plateaus in the Two-Dimensional Spin System
  ${\mathrm{SrCu}}_{2}({\mathrm{BO}}_{3}){}_{2}$}},\ }}\href {\doibase
  10.1103/PhysRevLett.82.3168} {\bibfield  {journal} {\bibinfo  {journal}
  {Phys. Rev. Lett.}\ }\textbf {\bibinfo {volume} {82}},\ \bibinfo {pages}
  {3168} (\bibinfo {year} {1999})}\BibitemShut {NoStop}%
\bibitem [{Note3()}]{Note3}%
  \BibitemOpen
  \bibinfo {note} {Note that this representation doubles the dimension of the
  local Hilbert space. To remedy this situation, i.e., to project down to the
  {\protect \sl physical} subspace of this extended Hilbert space, one defines
  $\Lambda _j = i c_j b_j^1 b_j^2 \protect \dots b_j^5$ and demands that the
  physical states satisfy $\Lambda _j \mathinner {|{\psi }\delimiter "526930B }
  = - \mathinner {|{\psi }\delimiter "526930B } \protect \tmspace +\thinmuskip
  {.1667em} \forall j$.}\BibitemShut {Stop}%
\bibitem [{\citenamefont {Lieb}(1994)}]{Lieb1994}%
  \BibitemOpen
  \bibfield  {author} {\bibinfo {author} {\bibfnamefont {E.~H.}\ \bibnamefont
  {Lieb}},\ }\bibfield  {title} {{\color{Gray}\small \bibinfo {title} {Flux
  phase of the half-filled band},\ }}\href {\doibase
  10.1103/PhysRevLett.73.2158} {\bibfield  {journal} {\bibinfo  {journal}
  {Phys. Rev. Lett.}\ }\textbf {\bibinfo {volume} {73}},\ \bibinfo {pages}
  {2158} (\bibinfo {year} {1994})}\BibitemShut {NoStop}%
\bibitem [{\citenamefont {Mishchenko}\ \emph {et~al.}(2017)\citenamefont
  {Mishchenko}, \citenamefont {Kato},\ and\ \citenamefont
  {Motome}}]{mishchenko_prb_96_2017}%
  \BibitemOpen
  \bibfield  {author} {\bibinfo {author} {\bibfnamefont {P.~A.}\ \bibnamefont
  {Mishchenko}}, \bibinfo {author} {\bibfnamefont {Y.}~\bibnamefont {Kato}}, \
  and\ \bibinfo {author} {\bibfnamefont {Y.}~\bibnamefont {Motome}},\
  }\bibfield  {title} {{\color{Gray}\small \bibinfo {title}
  {{F}inite-temperature phase transition to a {K}itaev spin liquid phase on a
  hyperoctagon lattice: {A} large-scale quantum {M}onte {C}arlo study},\
  }}\href {\doibase 10.1103/PhysRevB.96.125124} {\bibfield  {journal} {\bibinfo
   {journal} {Phys. Rev. B}\ }\textbf {\bibinfo {volume} {96}},\ \bibinfo
  {pages} {125124} (\bibinfo {year} {2017})}\BibitemShut {NoStop}%
\bibitem [{\citenamefont {Schnyder}\ \emph {et~al.}(2008)\citenamefont
  {Schnyder}, \citenamefont {Ryu}, \citenamefont {Furusaki},\ and\
  \citenamefont {Ludwig}}]{Schnyder2008classification}%
  \BibitemOpen
  \bibfield  {author} {\bibinfo {author} {\bibfnamefont {A.~P.}\ \bibnamefont
  {Schnyder}}, \bibinfo {author} {\bibfnamefont {S.}~\bibnamefont {Ryu}},
  \bibinfo {author} {\bibfnamefont {A.}~\bibnamefont {Furusaki}}, \ and\
  \bibinfo {author} {\bibfnamefont {A.~W.~W.}\ \bibnamefont {Ludwig}},\
  }\bibfield  {title} {{\color{Gray}\small \bibinfo {title} {{Classification of
  topological insulators and superconductors in three spatial dimensions}},\
  }}\href {\doibase 10.1103/PhysRevB.78.195125} {\bibfield  {journal} {\bibinfo
   {journal} {Phys. Rev. B}\ }\textbf {\bibinfo {volume} {78}},\ \bibinfo
  {pages} {195125} (\bibinfo {year} {2008})}\BibitemShut {NoStop}%
\bibitem [{\citenamefont {Kitaev}(2009)}]{Kitaev2009periodic}%
  \BibitemOpen
  \bibfield  {author} {\bibinfo {author} {\bibfnamefont {A.}~\bibnamefont
  {Kitaev}},\ }\bibfield  {title} {{\color{Gray}\small \bibinfo {title}
  {{Periodic table for topological insulators and superconductors}},\ }}\href
  {\doibase 10.1063/1.3149495} {\bibfield  {journal} {\bibinfo  {journal} {AIP
  Conference Proceedings}\ }\textbf {\bibinfo {volume} {1134}},\ \bibinfo
  {pages} {22} (\bibinfo {year} {2009})}\BibitemShut {NoStop}%
\bibitem [{\citenamefont {Eschmann}\ \emph {et~al.}(2020)\citenamefont
  {Eschmann}, \citenamefont {Mishchenko}, \citenamefont {O'Brien},
  \citenamefont {Bojesen}, \citenamefont {Kato}, \citenamefont {Hermanns},
  \citenamefont {Motome},\ and\ \citenamefont
  {Trebst}}]{2020EschmannThermalClassification}%
  \BibitemOpen
  \bibfield  {author} {\bibinfo {author} {\bibfnamefont {T.}~\bibnamefont
  {Eschmann}}, \bibinfo {author} {\bibfnamefont {P.~A.}\ \bibnamefont
  {Mishchenko}}, \bibinfo {author} {\bibfnamefont {K.}~\bibnamefont {O'Brien}},
  \bibinfo {author} {\bibfnamefont {T.~A.}\ \bibnamefont {Bojesen}}, \bibinfo
  {author} {\bibfnamefont {Y.}~\bibnamefont {Kato}}, \bibinfo {author}
  {\bibfnamefont {M.}~\bibnamefont {Hermanns}}, \bibinfo {author}
  {\bibfnamefont {Y.}~\bibnamefont {Motome}}, \ and\ \bibinfo {author}
  {\bibfnamefont {S.}~\bibnamefont {Trebst}},\ }\bibfield  {title}
  {{\color{Gray}\small \bibinfo {title} {{Thermodynamic classification of 3D
  Kitaev spin liquids}},\ }}\href {https://arxiv.org/abs/2006.07386} {\bibfield
   {journal} {\bibinfo  {journal} {arXiv:2006.07386}\ } (\bibinfo {year}
  {2020})}\BibitemShut {NoStop}%
\bibitem [{\citenamefont {Read}\ and\ \citenamefont
  {Sachdev}(1991)}]{Read1991}%
  \BibitemOpen
  \bibfield  {author} {\bibinfo {author} {\bibfnamefont {N.}~\bibnamefont
  {Read}}\ and\ \bibinfo {author} {\bibfnamefont {S.}~\bibnamefont {Sachdev}},\
  }\bibfield  {title} {{\color{Gray}\small \bibinfo {title} {Large-n expansion
  for frustrated quantum antiferromagnets},\ }}\href {\doibase
  10.1103/PhysRevLett.66.1773} {\bibfield  {journal} {\bibinfo  {journal}
  {Phys. Rev. Lett.}\ }\textbf {\bibinfo {volume} {66}},\ \bibinfo {pages}
  {1773} (\bibinfo {year} {1991})}\BibitemShut {NoStop}%
\bibitem [{\citenamefont {Senthil}\ and\ \citenamefont
  {Fisher}(2000)}]{Senthil2000}%
  \BibitemOpen
  \bibfield  {author} {\bibinfo {author} {\bibfnamefont {T.}~\bibnamefont
  {Senthil}}\ and\ \bibinfo {author} {\bibfnamefont {M.~P.~A.}\ \bibnamefont
  {Fisher}},\ }\bibfield  {title} {{\color{Gray}\small \bibinfo {title}
  {{${Z}_{2}$ gauge theory of electron fractionalization in strongly correlated
  systems}},\ }}\href {\doibase 10.1103/PhysRevB.62.7850} {\bibfield  {journal}
  {\bibinfo  {journal} {Phys. Rev. B}\ }\textbf {\bibinfo {volume} {62}},\
  \bibinfo {pages} {7850} (\bibinfo {year} {2000})}\BibitemShut {NoStop}%
\bibitem [{\citenamefont {Nasu}\ and\ \citenamefont {Motome}(2015)}]{Nasu2015}%
  \BibitemOpen
  \bibfield  {author} {\bibinfo {author} {\bibfnamefont {J.}~\bibnamefont
  {Nasu}}\ and\ \bibinfo {author} {\bibfnamefont {Y.}~\bibnamefont {Motome}},\
  }\bibfield  {title} {{\color{Gray}\small \bibinfo {title} {{Thermodynamics of
  Chiral Spin Liquids with Abelian and Non-Abelian Anyons}},\ }}\href {\doibase
  10.1103/PhysRevLett.115.087203} {\bibfield  {journal} {\bibinfo  {journal}
  {Phys. Rev. Lett.}\ }\textbf {\bibinfo {volume} {115}},\ \bibinfo {pages}
  {087203} (\bibinfo {year} {2015})}\BibitemShut {NoStop}%
\bibitem [{Note4()}]{Note4}%
  \BibitemOpen
  \bibinfo {note} {Such a $\pi $-flux ground state for square plaquettes is
  also generally in line with the expectation from Lieb's theorem on
  ground-state flux assignments in bipartite lattices with certain mirror
  symmetries \cite {Lieb1994}, tough it does not strictly apply to the lattice
  geometry at hand.}\BibitemShut {Stop}%
\bibitem [{\citenamefont {Altland}\ and\ \citenamefont
  {Zirnbauer}(1997)}]{Altland1997classification}%
  \BibitemOpen
  \bibfield  {author} {\bibinfo {author} {\bibfnamefont {A.}~\bibnamefont
  {Altland}}\ and\ \bibinfo {author} {\bibfnamefont {M.~R.}\ \bibnamefont
  {Zirnbauer}},\ }\bibfield  {title} {{\color{Gray}\small \bibinfo {title}
  {Nonstandard symmetry classes in mesoscopic normal-superconducting hybrid
  structures},\ }}\href {\doibase 10.1103/PhysRevB.55.1142} {\bibfield
  {journal} {\bibinfo  {journal} {Phys. Rev. B}\ }\textbf {\bibinfo {volume}
  {55}},\ \bibinfo {pages} {1142} (\bibinfo {year} {1997})}\BibitemShut
  {NoStop}%
\bibitem [{\citenamefont {Zirnbauer}(1996)}]{Zirnbauer1996}%
  \BibitemOpen
  \bibfield  {author} {\bibinfo {author} {\bibfnamefont {M.~R.}\ \bibnamefont
  {Zirnbauer}},\ }\bibfield  {title} {{\color{Gray}\small \bibinfo {title}
  {{Riemannian symmetric superspaces and their origin in random-matrix
  theory}},\ }}\href {\doibase 10.1063/1.531675} {\bibfield  {journal}
  {\bibinfo  {journal} {Journal of Mathematical Physics}\ }\textbf {\bibinfo
  {volume} {37}},\ \bibinfo {pages} {4986} (\bibinfo {year}
  {1996})}\BibitemShut {NoStop}%
\bibitem [{\citenamefont {Laumann}\ \emph {et~al.}(2012)\citenamefont
  {Laumann}, \citenamefont {Ludwig}, \citenamefont {Huse},\ and\ \citenamefont
  {Trebst}}]{ThermalMetal3}%
  \BibitemOpen
  \bibfield  {author} {\bibinfo {author} {\bibfnamefont {C.~R.}\ \bibnamefont
  {Laumann}}, \bibinfo {author} {\bibfnamefont {A.~W.~W.}\ \bibnamefont
  {Ludwig}}, \bibinfo {author} {\bibfnamefont {D.~A.}\ \bibnamefont {Huse}}, \
  and\ \bibinfo {author} {\bibfnamefont {S.}~\bibnamefont {Trebst}},\
  }\bibfield  {title} {{\color{Gray}\small \bibinfo {title} {{Disorder-induced
  Majorana metal in interacting non-Abelian anyon systems}},\ }}\href {\doibase
  10.1103/PhysRevB.85.161301} {\bibfield  {journal} {\bibinfo  {journal} {Phys.
  Rev. B}\ }\textbf {\bibinfo {volume} {85}},\ \bibinfo {pages} {161301}
  (\bibinfo {year} {2012})}\BibitemShut {NoStop}%
\bibitem [{\citenamefont {Chalker}\ and\ \citenamefont
  {Coddington}(1988)}]{ThermalMetal1}%
  \BibitemOpen
  \bibfield  {author} {\bibinfo {author} {\bibfnamefont {J.~T.}\ \bibnamefont
  {Chalker}}\ and\ \bibinfo {author} {\bibfnamefont {P.~D.}\ \bibnamefont
  {Coddington}},\ }\bibfield  {title} {{\color{Gray}\small \bibinfo {title}
  {{Percolation, quantum tunnelling and the integer Hall effect}},\ }}\href
  {\doibase 10.1088/0022-3719/21/14/008} {\bibfield  {journal} {\bibinfo
  {journal} {Journal of Physics C: Solid State Physics}\ }\textbf {\bibinfo
  {volume} {21}},\ \bibinfo {pages} {2665} (\bibinfo {year}
  {1988})}\BibitemShut {NoStop}%
\bibitem [{\citenamefont {Read}\ and\ \citenamefont
  {Ludwig}(2000)}]{ThermalMetal1b}%
  \BibitemOpen
  \bibfield  {author} {\bibinfo {author} {\bibfnamefont {N.}~\bibnamefont
  {Read}}\ and\ \bibinfo {author} {\bibfnamefont {A.~W.~W.}\ \bibnamefont
  {Ludwig}},\ }\bibfield  {title} {{\color{Gray}\small \bibinfo {title}
  {{Absence of a metallic phase in random-bond Ising models in two dimensions:
  Applications to disordered superconductors and paired quantum Hall states}},\
  }}\href {\doibase 10.1103/PhysRevB.63.024404} {\bibfield  {journal} {\bibinfo
   {journal} {Phys. Rev. B}\ }\textbf {\bibinfo {volume} {63}},\ \bibinfo
  {pages} {024404} (\bibinfo {year} {2000})}\BibitemShut {NoStop}%
\bibitem [{\citenamefont {Gruzberg}\ \emph {et~al.}(2001)\citenamefont
  {Gruzberg}, \citenamefont {Read},\ and\ \citenamefont
  {Ludwig}}]{ThermalMetal1c}%
  \BibitemOpen
  \bibfield  {author} {\bibinfo {author} {\bibfnamefont {I.~A.}\ \bibnamefont
  {Gruzberg}}, \bibinfo {author} {\bibfnamefont {N.}~\bibnamefont {Read}}, \
  and\ \bibinfo {author} {\bibfnamefont {A.~W.~W.}\ \bibnamefont {Ludwig}},\
  }\bibfield  {title} {{\color{Gray}\small \bibinfo {title} {{Random-bond Ising
  model in two dimensions: The Nishimori line and supersymmetry}},\ }}\href
  {\doibase 10.1103/PhysRevB.63.104422} {\bibfield  {journal} {\bibinfo
  {journal} {Phys. Rev. B}\ }\textbf {\bibinfo {volume} {63}},\ \bibinfo
  {pages} {104422} (\bibinfo {year} {2001})}\BibitemShut {NoStop}%
\bibitem [{\citenamefont {Chalker}\ \emph {et~al.}(2001)\citenamefont
  {Chalker}, \citenamefont {Read}, \citenamefont {Kagalovsky}, \citenamefont
  {Horovitz}, \citenamefont {Avishai},\ and\ \citenamefont
  {Ludwig}}]{ThermalMetal2}%
  \BibitemOpen
  \bibfield  {author} {\bibinfo {author} {\bibfnamefont {J.~T.}\ \bibnamefont
  {Chalker}}, \bibinfo {author} {\bibfnamefont {N.}~\bibnamefont {Read}},
  \bibinfo {author} {\bibfnamefont {V.}~\bibnamefont {Kagalovsky}}, \bibinfo
  {author} {\bibfnamefont {B.}~\bibnamefont {Horovitz}}, \bibinfo {author}
  {\bibfnamefont {Y.}~\bibnamefont {Avishai}}, \ and\ \bibinfo {author}
  {\bibfnamefont {A.~W.~W.}\ \bibnamefont {Ludwig}},\ }\bibfield  {title}
  {{\color{Gray}\small \bibinfo {title} {{Thermal metal in network models of a
  disordered two-dimensional superconductor}},\ }}\href {\doibase
  10.1103/PhysRevB.65.012506} {\bibfield  {journal} {\bibinfo  {journal} {Phys.
  Rev. B}\ }\textbf {\bibinfo {volume} {65}},\ \bibinfo {pages} {012506}
  (\bibinfo {year} {2001})}\BibitemShut {NoStop}%
\bibitem [{\citenamefont {Mildenberger}\ \emph {et~al.}(2007)\citenamefont
  {Mildenberger}, \citenamefont {Evers}, \citenamefont {Mirlin},\ and\
  \citenamefont {Chalker}}]{ThermalMetal2b}%
  \BibitemOpen
  \bibfield  {author} {\bibinfo {author} {\bibfnamefont {A.}~\bibnamefont
  {Mildenberger}}, \bibinfo {author} {\bibfnamefont {F.}~\bibnamefont {Evers}},
  \bibinfo {author} {\bibfnamefont {A.~D.}\ \bibnamefont {Mirlin}}, \ and\
  \bibinfo {author} {\bibfnamefont {J.~T.}\ \bibnamefont {Chalker}},\
  }\bibfield  {title} {{\color{Gray}\small \bibinfo {title} {{Density of
  quasiparticle states for a two-dimensional disordered system: Metallic,
  insulating, and critical behavior in the class-D thermal quantum Hall
  effect}},\ }}\href {\doibase 10.1103/PhysRevB.75.245321} {\bibfield
  {journal} {\bibinfo  {journal} {Phys. Rev. B}\ }\textbf {\bibinfo {volume}
  {75}},\ \bibinfo {pages} {245321} (\bibinfo {year} {2007})}\BibitemShut
  {NoStop}%
\bibitem [{\citenamefont {Self}\ \emph {et~al.}(2019)\citenamefont {Self},
  \citenamefont {Knolle}, \citenamefont {Iblisdir},\ and\ \citenamefont
  {Pachos}}]{Self2019}%
  \BibitemOpen
  \bibfield  {author} {\bibinfo {author} {\bibfnamefont {C.~N.}\ \bibnamefont
  {Self}}, \bibinfo {author} {\bibfnamefont {J.}~\bibnamefont {Knolle}},
  \bibinfo {author} {\bibfnamefont {S.}~\bibnamefont {Iblisdir}}, \ and\
  \bibinfo {author} {\bibfnamefont {J.~K.}\ \bibnamefont {Pachos}},\ }\bibfield
   {title} {{\color{Gray}\small \bibinfo {title} {{Thermally induced metallic
  phase in a gapped quantum spin liquid: Monte Carlo study of the Kitaev model
  with parity projection}},\ }}\href {\doibase 10.1103/PhysRevB.99.045142}
  {\bibfield  {journal} {\bibinfo  {journal} {Phys. Rev. B}\ }\textbf {\bibinfo
  {volume} {99}},\ \bibinfo {pages} {045142} (\bibinfo {year}
  {2019})}\BibitemShut {NoStop}%
\bibitem [{Note5()}]{Note5}%
  \BibitemOpen
  \bibinfo {note} {For the full Kitaev model at finite temperature, the
  disorder is far from uncorrelated, since the probability of a disorder
  realization is the Boltzmann weight corresponding to all the vison
  excitations required to arrive at it from the ground state flux
  configuration.}\BibitemShut {Stop}%
\bibitem [{Note6()}]{Note6}%
  \BibitemOpen
  \bibinfo {note} {We contrast this to a similar effect in topological
  insulators with Anderson disorder, where disorder leads to an \protect \emph
  {enhancement} of the topological phase. This difference is due to the purely
  imaginary disorder matrix in the present case, in contrast to a purely real
  one for topological insulators.}\BibitemShut {Stop}%
\bibitem [{\citenamefont {Dwivedi}\ and\ \citenamefont
  {Chua}(2016)}]{TransferMatrix}%
  \BibitemOpen
  \bibfield  {author} {\bibinfo {author} {\bibfnamefont {V.}~\bibnamefont
  {Dwivedi}}\ and\ \bibinfo {author} {\bibfnamefont {V.}~\bibnamefont {Chua}},\
  }\bibfield  {title} {{\color{Gray}\small \bibinfo {title} {Of bulk and
  boundaries: Generalized transfer matrices for tight-binding models},\ }}\href
  {\doibase 10.1103/PhysRevB.93.134304} {\bibfield  {journal} {\bibinfo
  {journal} {Phys. Rev. B}\ }\textbf {\bibinfo {volume} {93}},\ \bibinfo
  {pages} {134304} (\bibinfo {year} {2016})}\BibitemShut {NoStop}%
\bibitem [{\citenamefont {Kramer}\ and\ \citenamefont
  {MacKinnon}(1993)}]{mackinnon-kramer}%
  \BibitemOpen
  \bibfield  {author} {\bibinfo {author} {\bibfnamefont {B.}~\bibnamefont
  {Kramer}}\ and\ \bibinfo {author} {\bibfnamefont {A.}~\bibnamefont
  {MacKinnon}},\ }\bibfield  {title} {{\color{Gray}\small \bibinfo {title}
  {Localization: theory and experiment},\ }}\href
  {https://doi.org/10.1088/0034-4885/56/12/001} {\bibfield  {journal} {\bibinfo
   {journal} {Reports on Progress in Physics}\ }\textbf {\bibinfo {volume}
  {56}},\ \bibinfo {pages} {1469} (\bibinfo {year} {1993})}\BibitemShut
  {NoStop}%
\bibitem [{\citenamefont {Nasu}\ \emph {et~al.}(2015)\citenamefont {Nasu},
  \citenamefont {Udagawa},\ and\ \citenamefont {Motome}}]{Nasu2015thermal}%
  \BibitemOpen
  \bibfield  {author} {\bibinfo {author} {\bibfnamefont {J.}~\bibnamefont
  {Nasu}}, \bibinfo {author} {\bibfnamefont {M.}~\bibnamefont {Udagawa}}, \
  and\ \bibinfo {author} {\bibfnamefont {Y.}~\bibnamefont {Motome}},\
  }\bibfield  {title} {{\color{Gray}\small \bibinfo {title} {{Thermal
  fractionalization of quantum spins in a Kitaev model: Temperature-linear
  specific heat and coherent transport of Majorana fermions}},\ }}\href
  {\doibase 10.1103/PhysRevB.92.115122} {\bibfield  {journal} {\bibinfo
  {journal} {Phys. Rev. B}\ }\textbf {\bibinfo {volume} {92}},\ \bibinfo
  {pages} {115122} (\bibinfo {year} {2015})}\BibitemShut {NoStop}%
\bibitem [{\citenamefont {Yoshitake}\ \emph {et~al.}(2016)\citenamefont
  {Yoshitake}, \citenamefont {Nasu},\ and\ \citenamefont
  {Motome}}]{Yoshitake2016fractional}%
  \BibitemOpen
  \bibfield  {author} {\bibinfo {author} {\bibfnamefont {J.}~\bibnamefont
  {Yoshitake}}, \bibinfo {author} {\bibfnamefont {J.}~\bibnamefont {Nasu}}, \
  and\ \bibinfo {author} {\bibfnamefont {Y.}~\bibnamefont {Motome}},\
  }\bibfield  {title} {{\color{Gray}\small \bibinfo {title} {{Fractional Spin
  Fluctuations as a Precursor of Quantum Spin Liquids: Majorana Dynamical
  Mean-Field Study for the Kitaev Model}},\ }}\href {\doibase
  10.1103/PhysRevLett.117.157203} {\bibfield  {journal} {\bibinfo  {journal}
  {Phys. Rev. Lett.}\ }\textbf {\bibinfo {volume} {117}},\ \bibinfo {pages}
  {157203} (\bibinfo {year} {2016})}\BibitemShut {NoStop}%
\bibitem [{\citenamefont {Gao}\ \emph {et~al.}(2019)\citenamefont {Gao},
  \citenamefont {Hickey}, \citenamefont {Xiang}, \citenamefont {Trebst},\ and\
  \citenamefont {Chen}}]{Gao2019}%
  \BibitemOpen
  \bibfield  {author} {\bibinfo {author} {\bibfnamefont {Y.~H.}\ \bibnamefont
  {Gao}}, \bibinfo {author} {\bibfnamefont {C.}~\bibnamefont {Hickey}},
  \bibinfo {author} {\bibfnamefont {T.}~\bibnamefont {Xiang}}, \bibinfo
  {author} {\bibfnamefont {S.}~\bibnamefont {Trebst}}, \ and\ \bibinfo {author}
  {\bibfnamefont {G.}~\bibnamefont {Chen}},\ }\bibfield  {title}
  {{\color{Gray}\small \bibinfo {title} {{Thermal Hall signatures of non-Kitaev
  spin liquids in honeycomb Kitaev materials}},\ }}\href {\doibase
  10.1103/PhysRevResearch.1.013014} {\bibfield  {journal} {\bibinfo  {journal}
  {Phys. Rev. Research}\ }\textbf {\bibinfo {volume} {1}},\ \bibinfo {pages}
  {013014} (\bibinfo {year} {2019})}\BibitemShut {NoStop}%
\bibitem [{\citenamefont {Do}\ \emph {et~al.}(2017)\citenamefont {Do},
  \citenamefont {Park}, \citenamefont {Yoshitake}, \citenamefont {Nasu},
  \citenamefont {Motome}, \citenamefont {{Seung Kwon}}, \citenamefont {Adroja},
  \citenamefont {Voneshen}, \citenamefont {Kim}, \citenamefont {Jang},
  \citenamefont {Park}, \citenamefont {Choi},\ and\ \citenamefont
  {Ji}}]{Do2017incarnation}%
  \BibitemOpen
  \bibfield  {author} {\bibinfo {author} {\bibfnamefont {S.-H.}\ \bibnamefont
  {Do}}, \bibinfo {author} {\bibfnamefont {S.-Y.}\ \bibnamefont {Park}},
  \bibinfo {author} {\bibfnamefont {J.}~\bibnamefont {Yoshitake}}, \bibinfo
  {author} {\bibfnamefont {J.}~\bibnamefont {Nasu}}, \bibinfo {author}
  {\bibfnamefont {Y.}~\bibnamefont {Motome}}, \bibinfo {author} {\bibfnamefont
  {Y.}~\bibnamefont {{Seung Kwon}}}, \bibinfo {author} {\bibfnamefont {D.~T.}\
  \bibnamefont {Adroja}}, \bibinfo {author} {\bibfnamefont {D.~J.}\
  \bibnamefont {Voneshen}}, \bibinfo {author} {\bibfnamefont {K.}~\bibnamefont
  {Kim}}, \bibinfo {author} {\bibfnamefont {T.-H.}\ \bibnamefont {Jang}},
  \bibinfo {author} {\bibfnamefont {J.-H.}\ \bibnamefont {Park}}, \bibinfo
  {author} {\bibfnamefont {K.-Y.}\ \bibnamefont {Choi}}, \ and\ \bibinfo
  {author} {\bibfnamefont {S.}~\bibnamefont {Ji}},\ }\bibfield  {title}
  {{\color{Gray}\small \bibinfo {title} {{Incarnation of Majorana Fermions in
  Kitaev Quantum Spin Lattice}},\ }}\href {https://arxiv.org/abs/1703.01081}
  {\bibfield  {journal} {\bibinfo  {journal} {arXiv:1703.01081}\ } (\bibinfo
  {year} {2017})}\BibitemShut {NoStop}%
\bibitem [{\citenamefont {Kasahara}\ \emph
  {et~al.}(2018{\natexlab{b}})\citenamefont {Kasahara}, \citenamefont {Sugii},
  \citenamefont {Ohnishi}, \citenamefont {Shimozawa}, \citenamefont
  {Yamashita}, \citenamefont {Kurita}, \citenamefont {Tanaka}, \citenamefont
  {Nasu}, \citenamefont {Motome}, \citenamefont {Shibauchi},\ and\
  \citenamefont {Matsuda}}]{Kasahara2018}%
  \BibitemOpen
  \bibfield  {author} {\bibinfo {author} {\bibfnamefont {Y.}~\bibnamefont
  {Kasahara}}, \bibinfo {author} {\bibfnamefont {K.}~\bibnamefont {Sugii}},
  \bibinfo {author} {\bibfnamefont {T.}~\bibnamefont {Ohnishi}}, \bibinfo
  {author} {\bibfnamefont {M.}~\bibnamefont {Shimozawa}}, \bibinfo {author}
  {\bibfnamefont {M.}~\bibnamefont {Yamashita}}, \bibinfo {author}
  {\bibfnamefont {N.}~\bibnamefont {Kurita}}, \bibinfo {author} {\bibfnamefont
  {H.}~\bibnamefont {Tanaka}}, \bibinfo {author} {\bibfnamefont
  {J.}~\bibnamefont {Nasu}}, \bibinfo {author} {\bibfnamefont {Y.}~\bibnamefont
  {Motome}}, \bibinfo {author} {\bibfnamefont {T.}~\bibnamefont {Shibauchi}}, \
  and\ \bibinfo {author} {\bibfnamefont {Y.}~\bibnamefont {Matsuda}},\
  }\bibfield  {title} {{\color{Gray}\small \bibinfo {title} {Unusual thermal
  hall effect in a kitaev spin liquid candidate
  $\ensuremath{\alpha}\text{\ensuremath{-}}{\mathrm{rucl}}_{3}$},\ }}\href
  {\doibase 10.1103/PhysRevLett.120.217205} {\bibfield  {journal} {\bibinfo
  {journal} {Phys. Rev. Lett.}\ }\textbf {\bibinfo {volume} {120}},\ \bibinfo
  {pages} {217205} (\bibinfo {year} {2018}{\natexlab{b}})}\BibitemShut
  {NoStop}%
\bibitem [{\citenamefont {Burnell}\ and\ \citenamefont
  {Nayak}(2011)}]{burnell_su2_2011}%
  \BibitemOpen
  \bibfield  {author} {\bibinfo {author} {\bibfnamefont {F.~J.}\ \bibnamefont
  {Burnell}}\ and\ \bibinfo {author} {\bibfnamefont {C.}~\bibnamefont
  {Nayak}},\ }\bibfield  {title} {{\color{Gray}\small \bibinfo {title} {{SU}(2)
  slave fermion solution of the {Kitaev} honeycomb lattice model},\ }}\href
  {\doibase 10.1103/PhysRevB.84.125125} {\bibfield  {journal} {\bibinfo
  {journal} {Phys. Rev. B}\ }\textbf {\bibinfo {volume} {84}},\ \bibinfo
  {pages} {125125} (\bibinfo {year} {2011})}\BibitemShut {NoStop}%
\bibitem [{\citenamefont {Hickey}\ and\ \citenamefont
  {Trebst}(2019)}]{hickey_emergence_2019}%
  \BibitemOpen
  \bibfield  {author} {\bibinfo {author} {\bibfnamefont {C.}~\bibnamefont
  {Hickey}}\ and\ \bibinfo {author} {\bibfnamefont {S.}~\bibnamefont
  {Trebst}},\ }\bibfield  {title} {{\color{Gray}\small \bibinfo {title}
  {Emergence of a field-driven {U}(1) spin liquid in the {Kitaev} honeycomb
  model},\ }}\href {\doibase 10.1038/s41467-019-08459-9} {\bibfield  {journal}
  {\bibinfo  {journal} {Nature Communications}\ }\textbf {\bibinfo {volume}
  {10}},\ \bibinfo {pages} {1} (\bibinfo {year} {2019})}\BibitemShut {NoStop}%
\bibitem [{\citenamefont {Fukui}\ \emph {et~al.}(2005)\citenamefont {Fukui},
  \citenamefont {Hatsugai},\ and\ \citenamefont {Suzuki}}]{Fukui_2005}%
  \BibitemOpen
  \bibfield  {author} {\bibinfo {author} {\bibfnamefont {T.}~\bibnamefont
  {Fukui}}, \bibinfo {author} {\bibfnamefont {Y.}~\bibnamefont {Hatsugai}}, \
  and\ \bibinfo {author} {\bibfnamefont {H.}~\bibnamefont {Suzuki}},\
  }\bibfield  {title} {{\color{Gray}\small \bibinfo {title} {{Chern Numbers in
  Discretized Brillouin Zone: Efficient Method of Computing (Spin) Hall
  Conductances}},\ }}\href {\doibase 10.1143/jpsj.74.1674} {\bibfield
  {journal} {\bibinfo  {journal} {Journal of the Physical Society of Japan}\
  }\textbf {\bibinfo {volume} {74}},\ \bibinfo {pages} {1674} (\bibinfo {year}
  {2005})}\BibitemShut {NoStop}%
\bibitem [{\citenamefont {Mahan}(1980)}]{Mahan}%
  \BibitemOpen
  \bibfield  {author} {\bibinfo {author} {\bibfnamefont {G.~D.}\ \bibnamefont
  {Mahan}},\ }\href@noop {} {\emph {\bibinfo {title} {Many Particle Physics}}}\
  (\bibinfo  {publisher} {Plenum},\ \bibinfo {address} {New York},\ \bibinfo
  {year} {1980})\BibitemShut {NoStop}%
\end{thebibliography}%


\widetext
\appendix

\section{Average Chern number}
\label{app:chern}

\begin{figure}[h!]
\includegraphics[width=0.245\columnwidth]{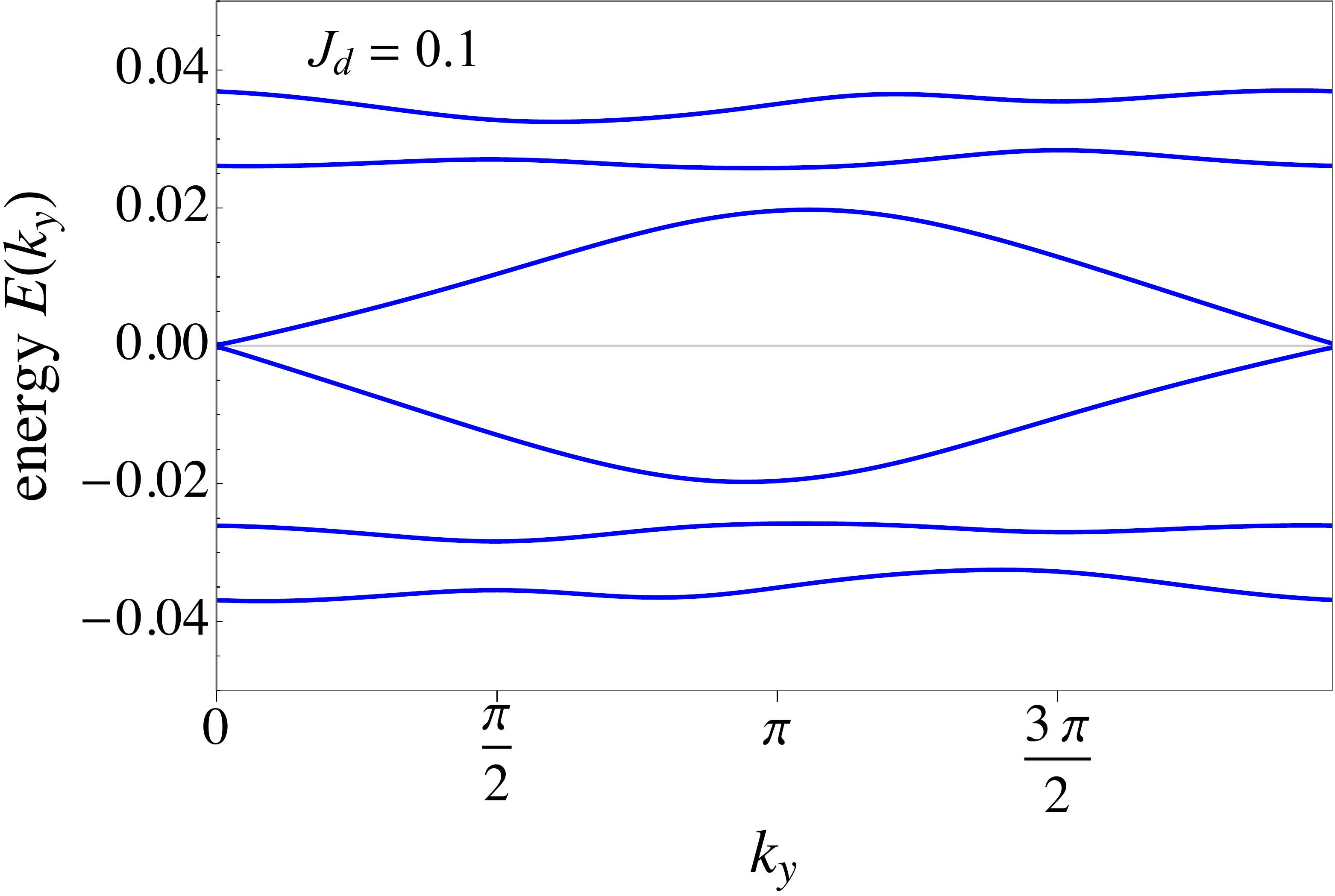}
\includegraphics[width=0.245\columnwidth]{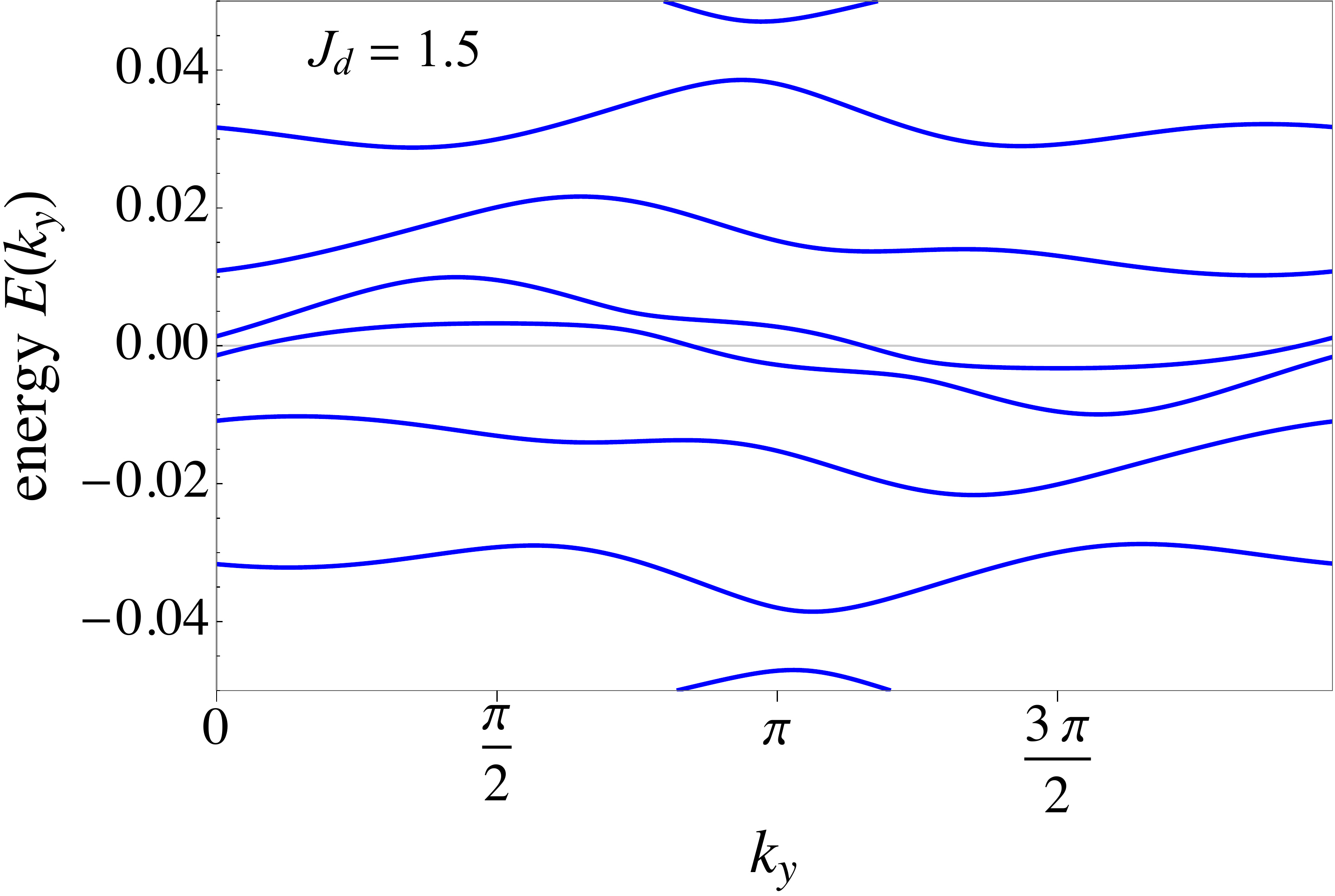}
\includegraphics[width=0.245\columnwidth]{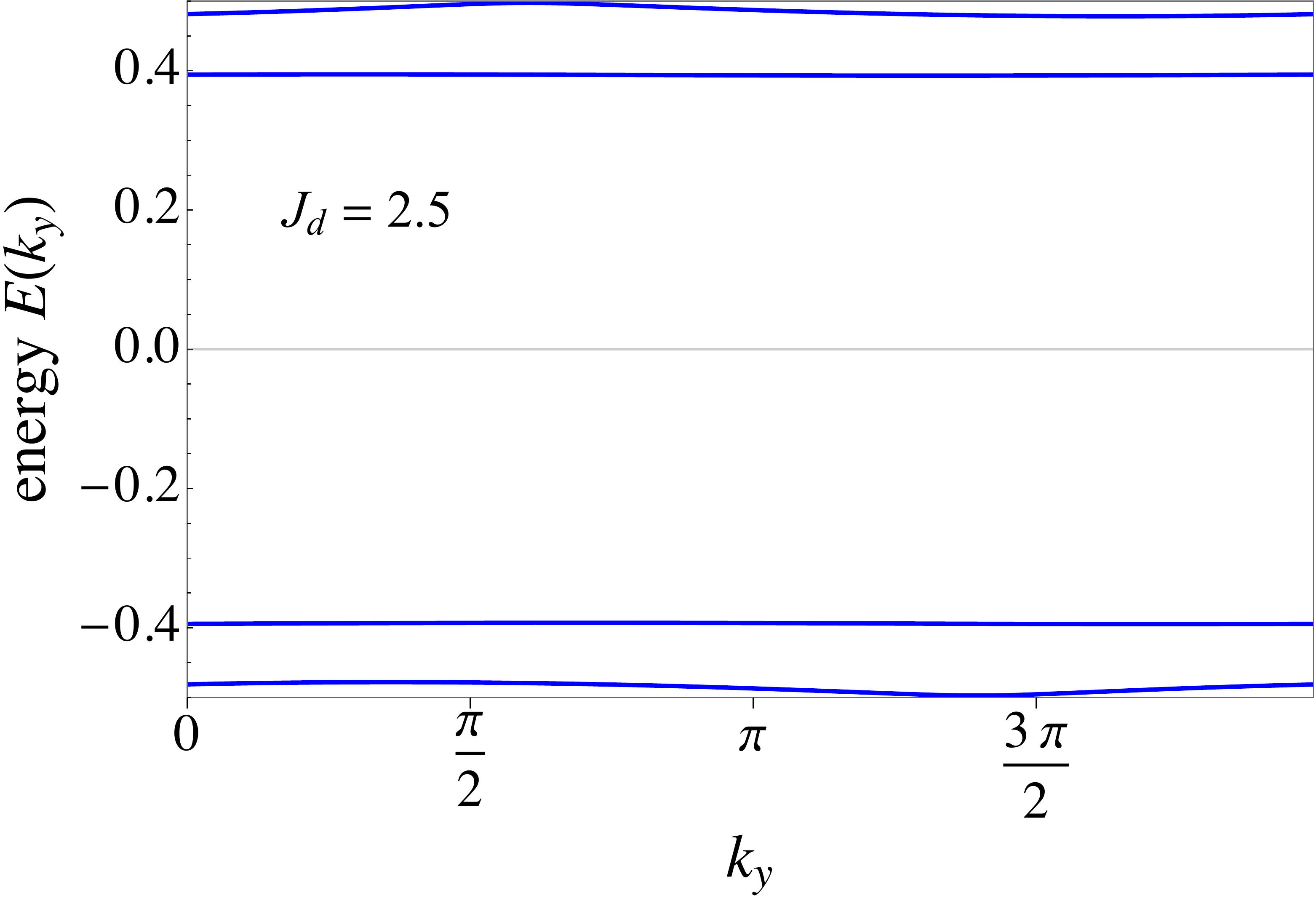}
\includegraphics[width=0.245\columnwidth]{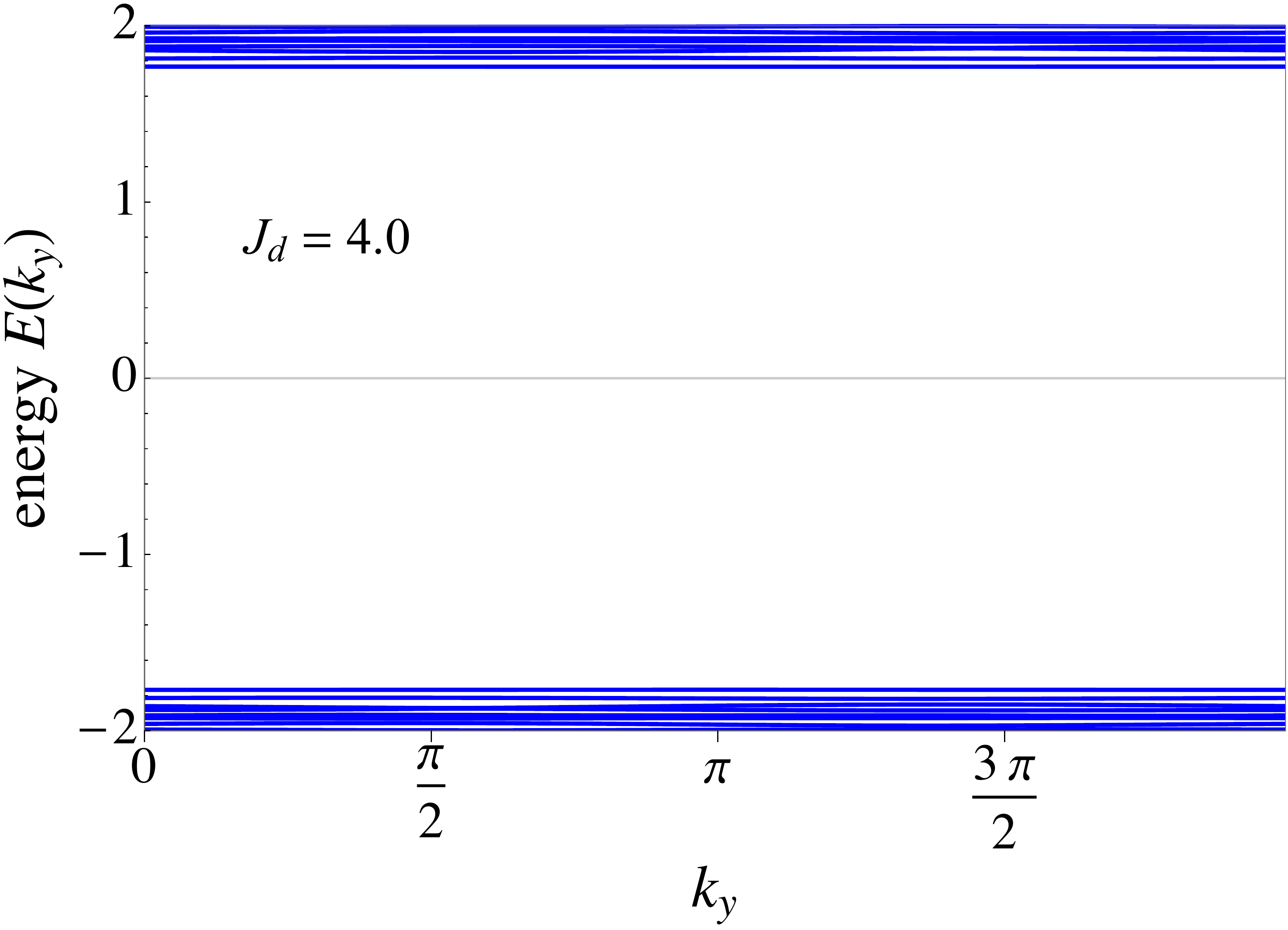}

\caption{{\bf Band structures near $E = 0$} for different parameters values $J_d$ at $T = \infty$ (in phase diagram \ref{fig:PhaseDiagramShastryChern}), showing the transition from the gapless to the gapped high-temperature phase.}
\label{fig:ChernNumbers4}

\end{figure}

To characterize the Majorana band structures in the intermediate temperature regime, we have numerically calculated an {\sl average} Chern number $\braket{|\nu|}$ in our Monte Carlo simulations. A single measurement contributing to this average is done for a given $\mathbb{Z}_2$ gauge field configuration (which effectively acts as a disorder potential for the Majorana fermions in this intermediate temperature regime). The calculation is performed according to the method for non-Abelian Berry connections, which was introduced in Ref. \onlinecite{Fukui_2005}: Here, the real-space $\mathbb{Z}_2$ gauge field configuration $\{u_{ij}\}$ of the size $L = 10$ ($N = 400$ sites) is used for the construction of a supercell in momentum space, described by the Hamiltonian $\mathcal{H}_{\bf k}(\{u_{ij}\})$. The Brillouin zone of the system is discretized with a mesh of $10^2$ ${\bf k}$-points. For each pair of nearest neighbor ${\bf k}$-points $({\bf k}_l, {\bf k}_l + {\bf q}_i)$, the $U(N)$ gauge variable
\begin{equation}
U_{{\bf q}_i} = \frac{\det(\psi^\dagger({\bf k}_l) \psi({\bf k}_l + {\bf q}_i))}{|\det(\psi^\dagger({\bf k}_l) \psi({\bf k}_l + {\bf q}_i))|},
\end{equation}
is calculated, where $\psi$ is the $N/2 \times N/2$ matrix that is composed of the multiplet $(\ket{u_1({\bf k})} \dots \ket{u_{N/2}({\bf k})})$ of eigenstates, belonging to the lower half of eigenvalues $\epsilon_{1}, \dots, \epsilon_{N/2}$ of $\mathcal{H}_{\bf k}$. Here, the relaxed gap opening condition $\epsilon_{n{\bf k}} \neq \epsilon_{m{\bf k}}$ for $n \leq N/2$ and $m > N/2$ has to be fulfilled.
From the $U(N)$ gauge variables $U_{{\bf q}_i}$, we calculate a (gauge-invariant) plaquette field strength
\begin{equation}
F_{12} ({\bf k}_l) = \log \left( \frac{U_{q_1}({\bf k}_l) U_{q_2}({\bf k}_l + {\bf q}_1)}{U_{q_1}({\bf k}_l + {\bf q}_2) U_{q_2}({\bf k}_l)}  \right),
\label{eq:PlaquetteFieldStrengthChernNumberCalculation}
\end{equation}
which is defined on the square plaquettes of the discretized Brillouin zone, and for which
\begin{equation}
-\pi < \frac{1}{i} F_{12}({\bf k}_l) \leq \pi.
\end{equation}
The (numerical) Chern number of the band structure can now be expressed by the sum over all plaquette field strengths $F_{12}({\bf k}_l)$ of the lattice,
\begin{equation}
\tilde{\nu} = \frac{1}{2\pi i} \sum_l F_{12}({\bf k}_l),
\end{equation}
such that $\tilde{\nu} \in \mathbb{Z}$ and $\tilde{\nu} \rightarrow \nu$ for a sufficiently fine discretization of the Brillouin zone. 
\begin{figure}[h!]
\includegraphics[width=0.3\columnwidth]{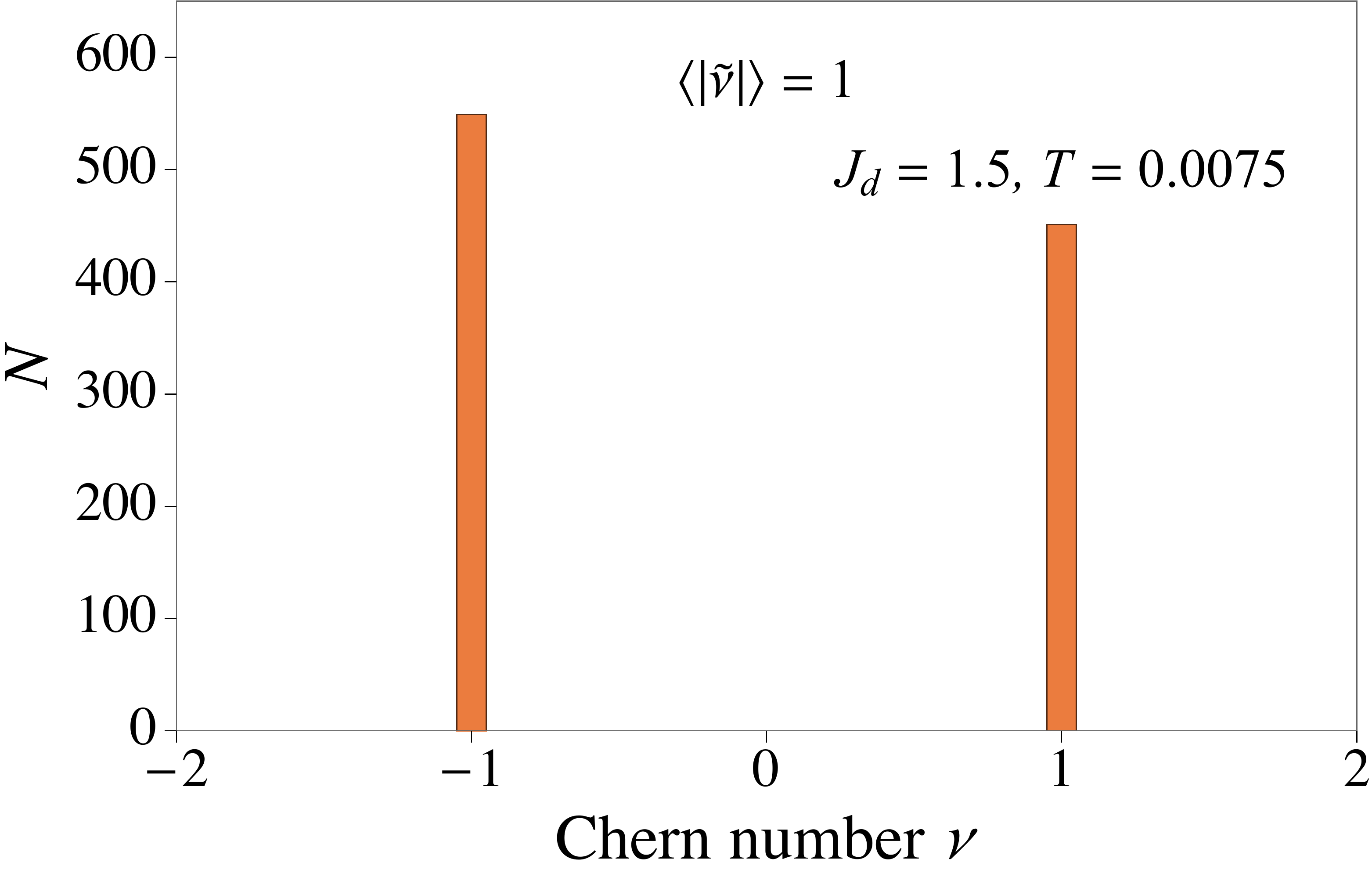}
\hskip 0.1 \columnwidth
\includegraphics[width=0.3\columnwidth]{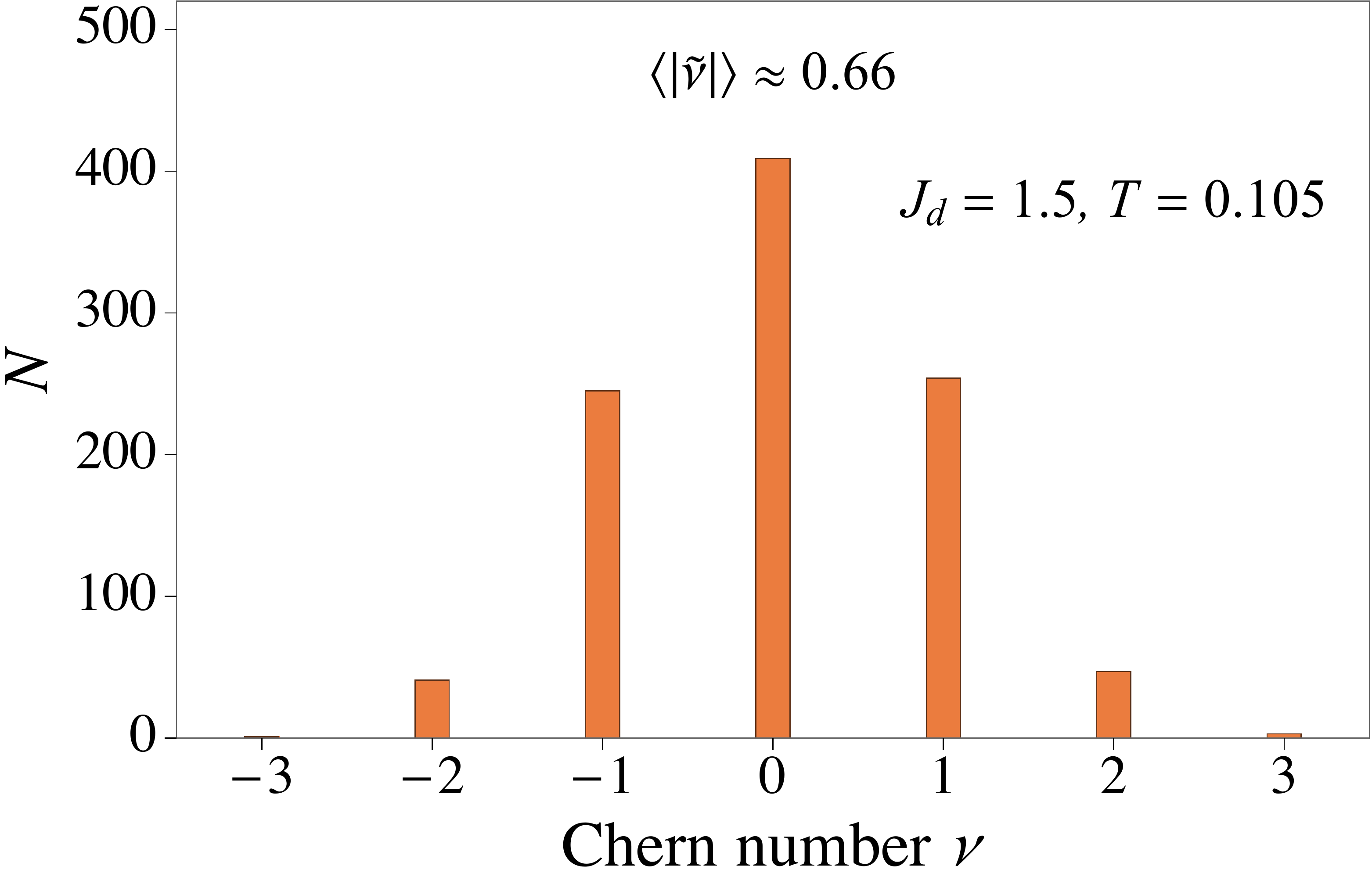}

\caption{{\bf Chern number histograms} for two different temperatures at the parameter point $J_d = 1.5$  (in phase diagram \ref{fig:PhaseDiagramShastryChern}). In the intermediate-temperature regime, the occurrence of integer Chern number results is an artifact, since here, the system is in a gapless phase.}
\label{fig:ChernNumbers3}
\end{figure}
Note that the condition $\tilde{\nu} \in \mathbb{Z}$ follows from the way the plaquette field strength \eqref{eq:PlaquetteFieldStrengthChernNumberCalculation} is constructed, but the integer value only has a physical meaning if the band structure is {\it gapped}. While this is the case for the ground state phases of the Shastry-Sutherland Kitaev system and the high-temperature phase at $J_d > 2$ (Fig. \ref{fig:PhaseDiagramShastryChern}), the different integer results for the {\it gapless} high-temperature phase at $J_d < 2$ are non-physical, and a mere ``breakdown'' indicator for the gapless phase -- see the Chern number histograms in Fig. \ref{fig:ChernNumbers3} and the band structures in Fig. \ref{fig:ChernNumbers4}. The same is true for the high-temperature results shown in Fig. \ref{fig:PhaseDiagramShastrySOSLChern}.

\begin{figure}[h!]
\includegraphics[width=0.245\columnwidth]{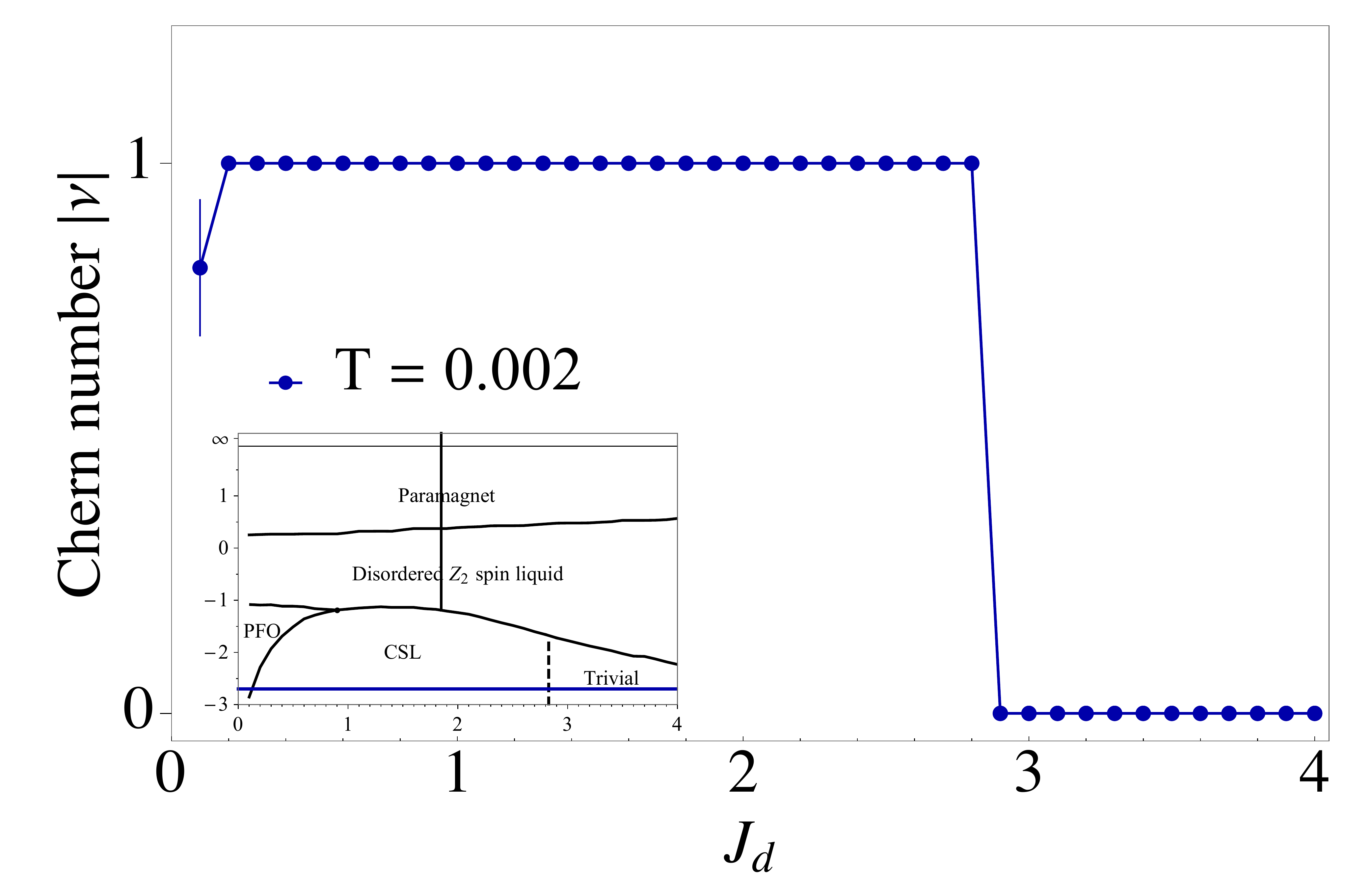}
\includegraphics[width=0.245\columnwidth]{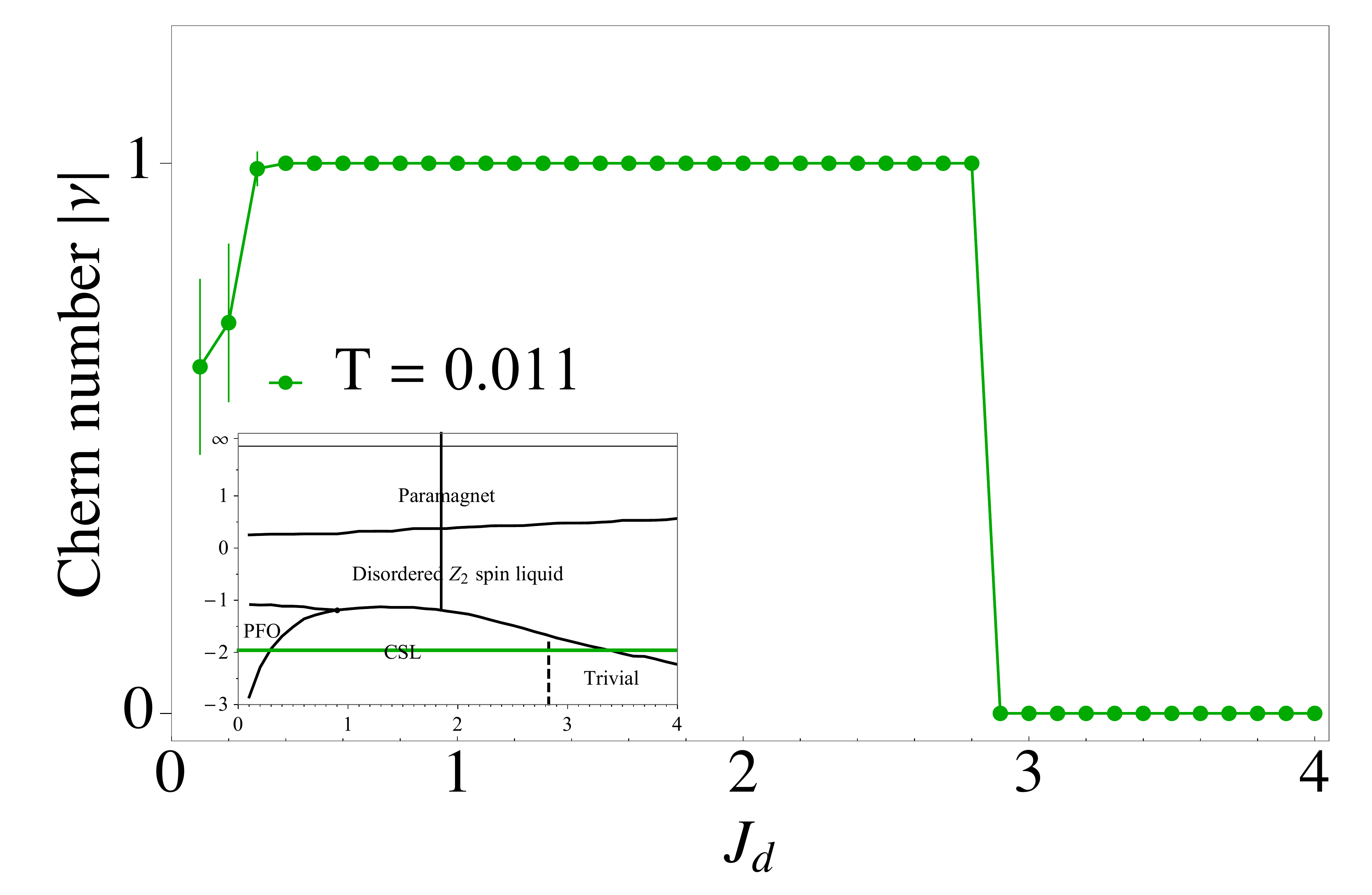}
\includegraphics[width=0.245\columnwidth]{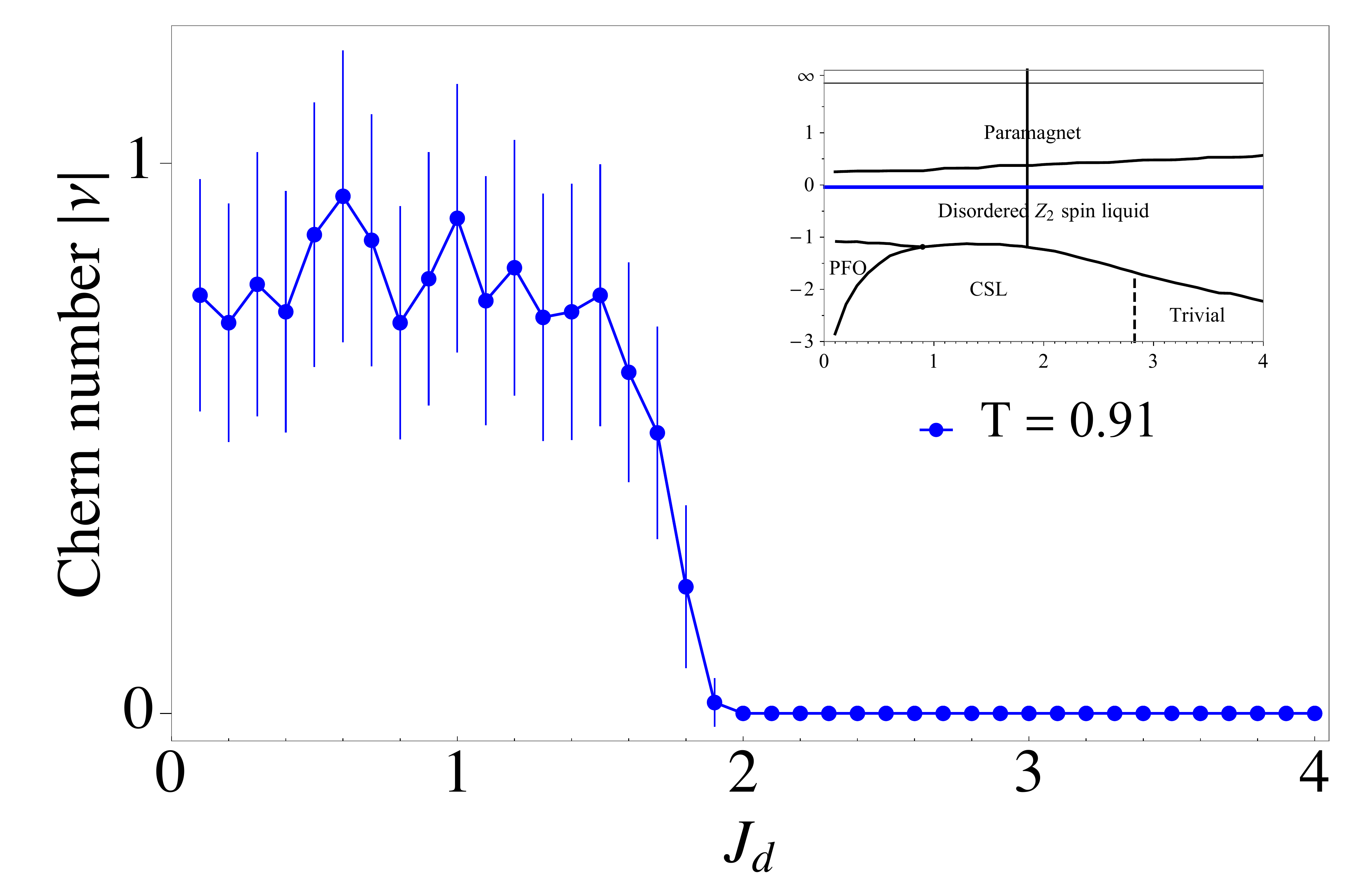}
\includegraphics[width=0.245\columnwidth]{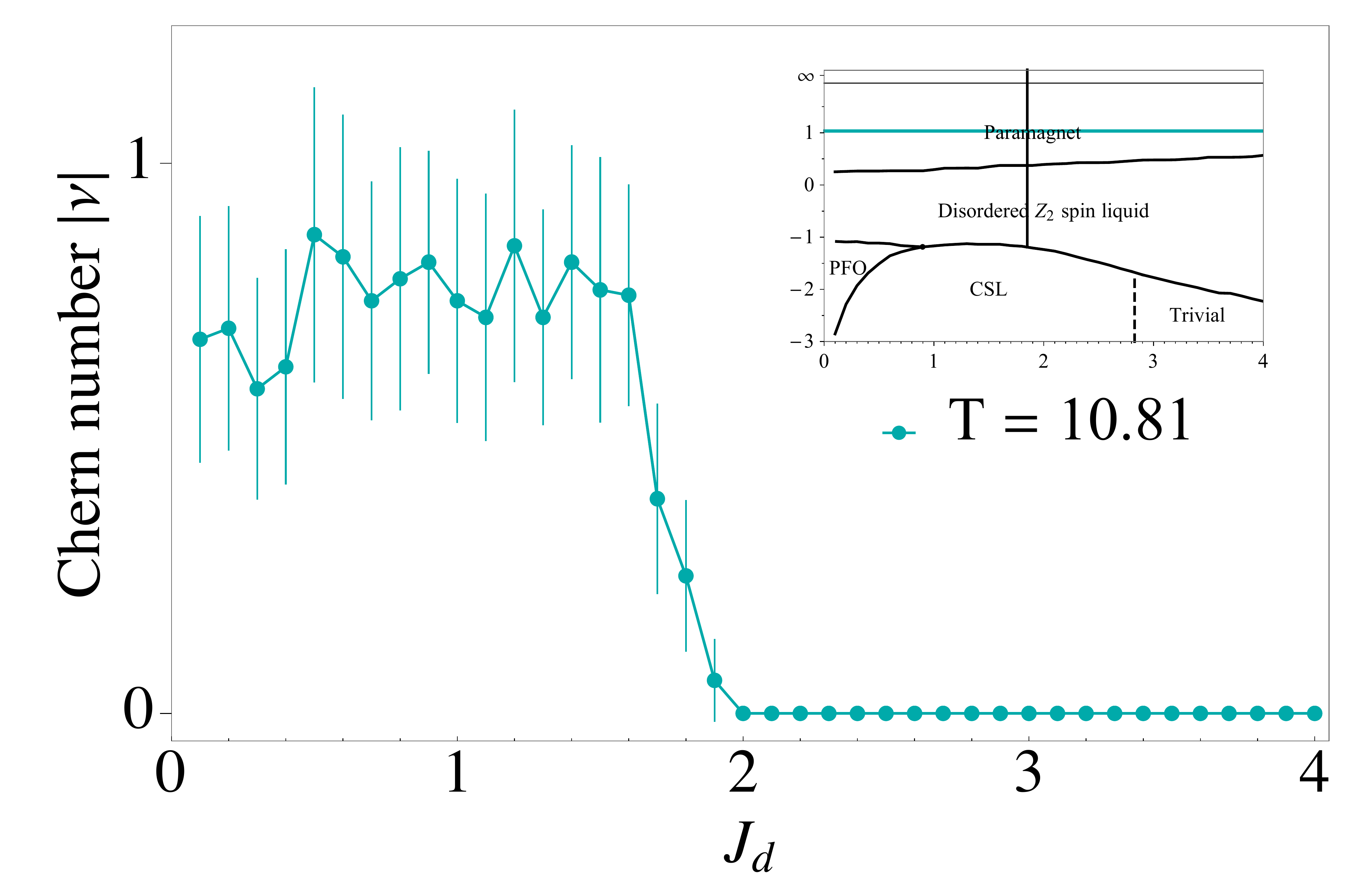}

\caption{{\bf Average chern number}  $\braket{|\nu|}$ for different horizontal (constant temperature) cuts in phase diagram \ref{fig:PhaseDiagramShastryChern}. }
\label{fig:ChernNumbers1}
\end{figure}

\begin{figure}[h!]
\includegraphics[width=0.245\columnwidth]{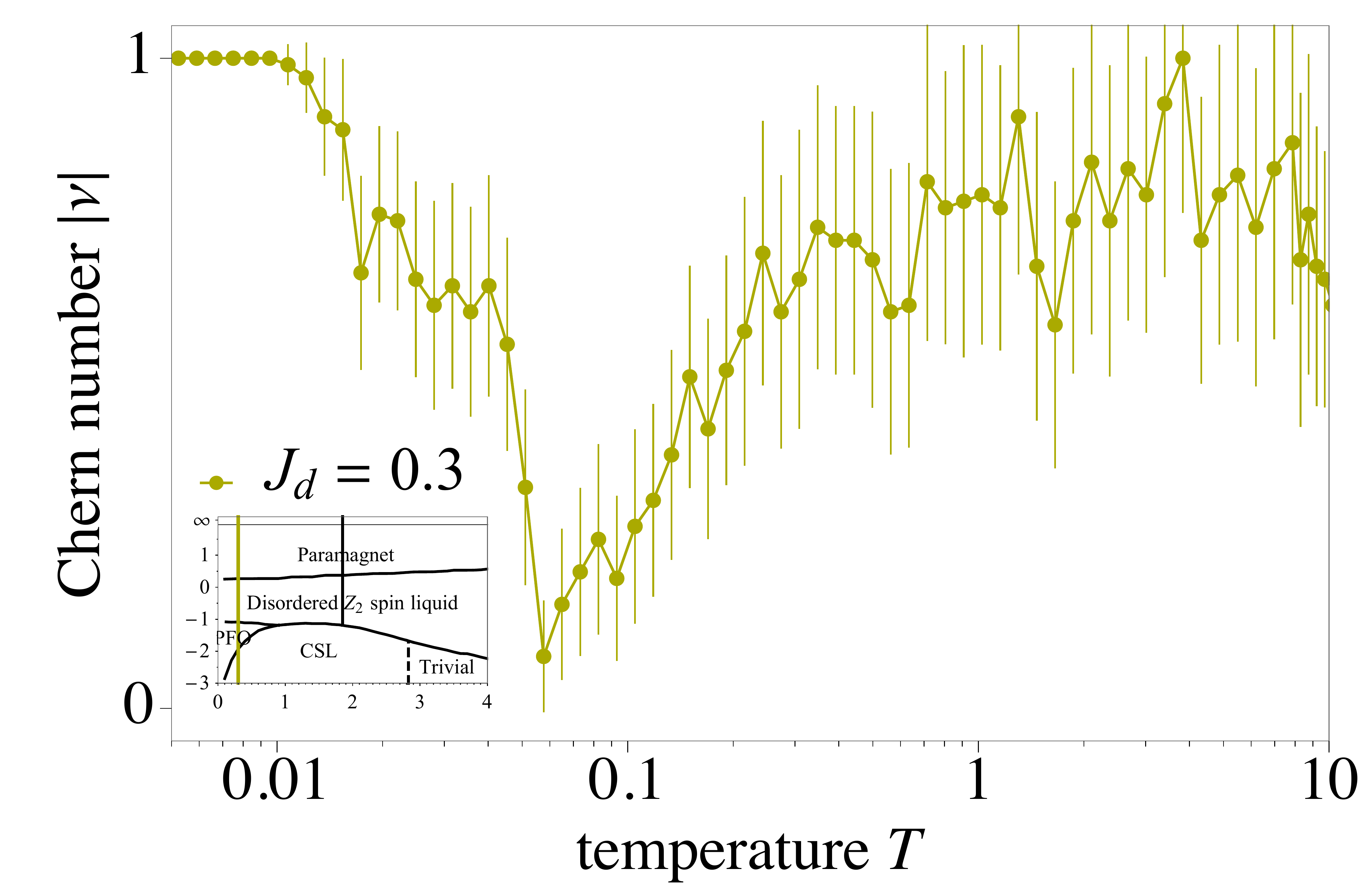}
\includegraphics[width=0.245\columnwidth]{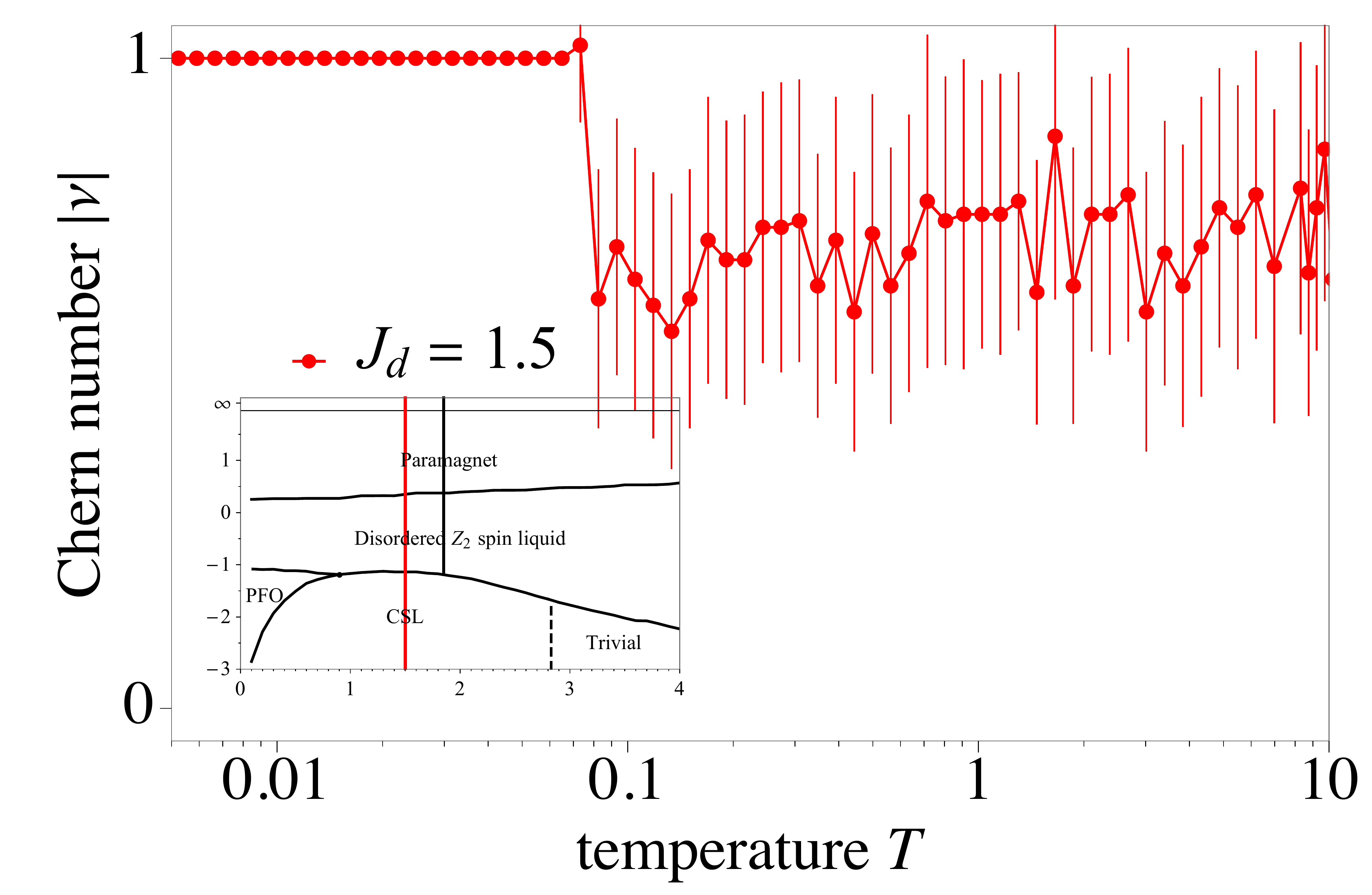}
\includegraphics[width=0.245\columnwidth]{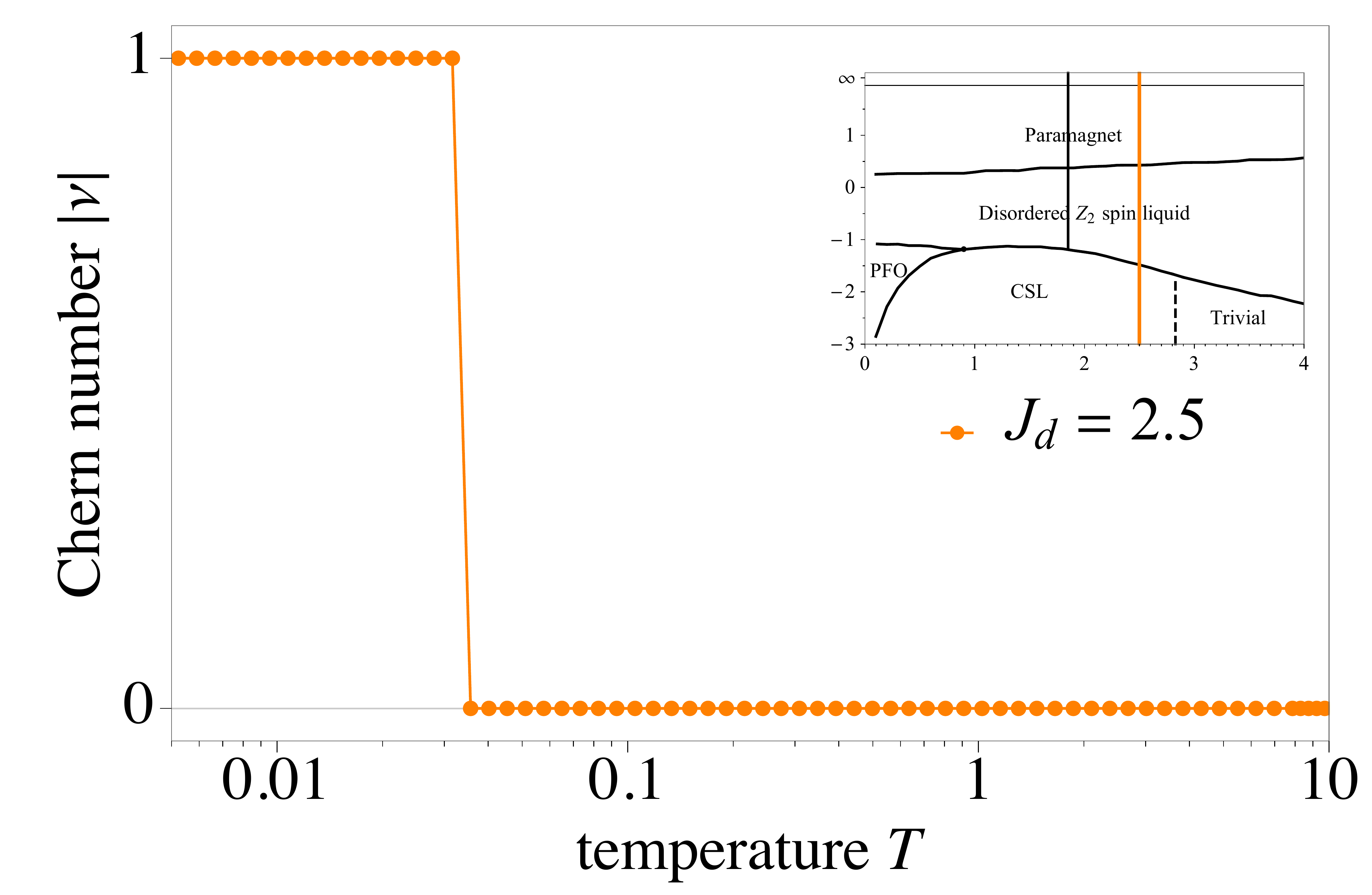}
\includegraphics[width=0.245\columnwidth]{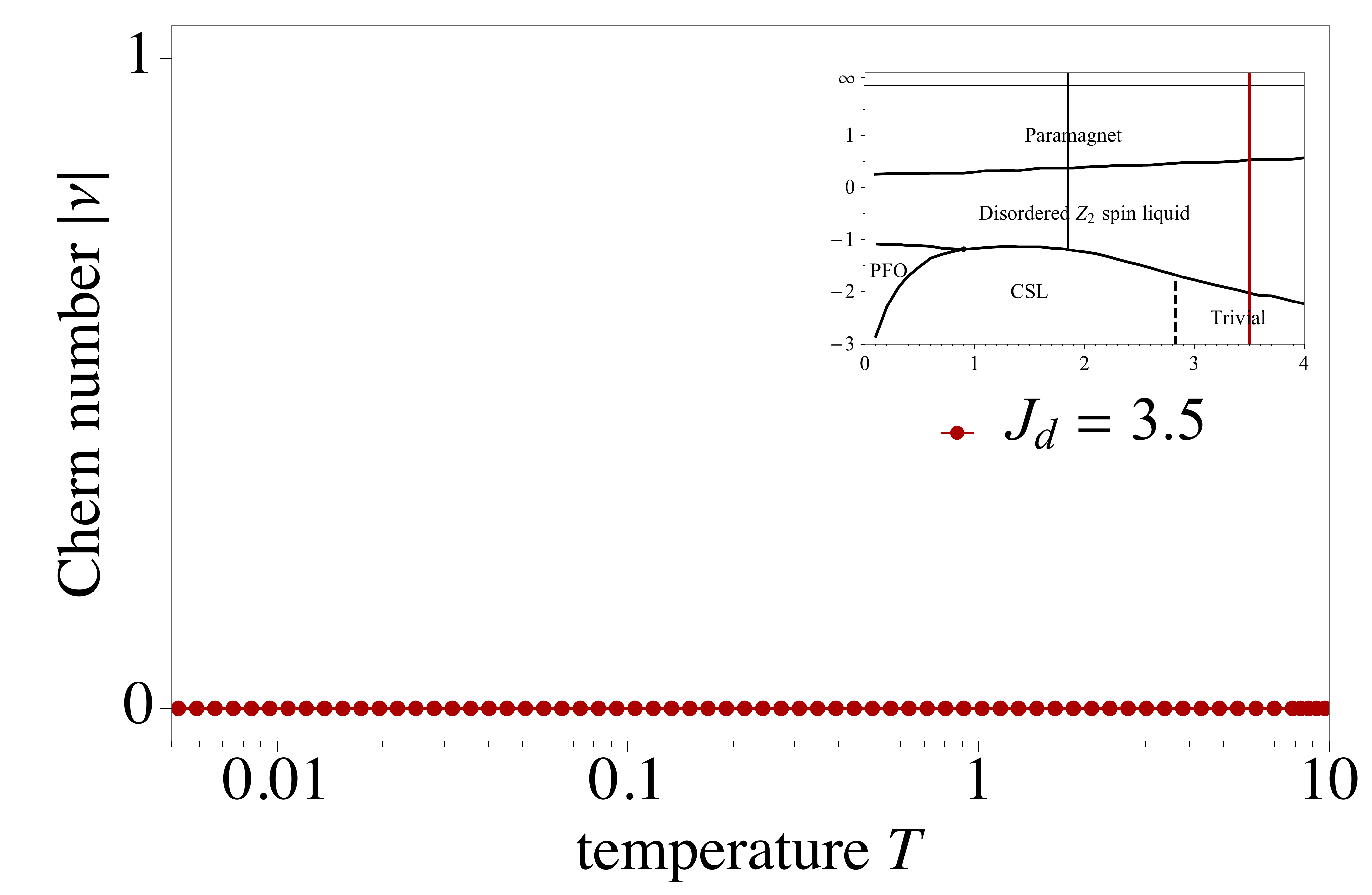}

\caption{{\bf Average chern number} $\braket{|\nu|}$ for different vertical (constant $\Jz$) cuts in phase diagram \ref{fig:PhaseDiagramShastryChern}.}
\label{fig:ChernNumbers2}
\end{figure}

Detailed scans of the average Chern number calculations are shown in Figs. \ref{fig:ChernNumbers1}, \ref{fig:ChernNumbers2}.

\clearpage


\section{Disorder-induced renormalization}
\label{app:Tmatrix}
The Shastry-Sutherland lattice is described by a 4-site unit cell, periodic with translation vectors $\vec{a}_1 = \hat{x}$ and $\vec{a}_2 = \hat{y}$. Thus, the Shastry-Sutherland-Kitaev model, after a Majorana decomposition, can be written in the momentum space as $\hlt = \sum_\vec{k} \hlt_B(\vec{k}) c(\vec{k}) c(-\vec{k})$, with 
\begin{equation}
  \hlt_B(\vec{k}) = i 
  \begin{pmatrix}
      0				                & J_x' - J_x \e^{-i k_x}	 & -J_z \e^{-i(k_x-k_y)}	  &  J_y' - J_y \e^{ik_y}		\\
      -J_x' + J_x \e^{ik_x}	  & 0				                 & J_y' - J_y \e^{ik_y}		  &  -J_z				            \\
      J_z \e^{i(k_x-k_y)}	    &  -J_y' + J_y \e^{-i k_y} & 0				                &  J_x' - J_x \e^{i k_x}	\\
      -J_y' + J_y \e^{-i k_y}	& J_z				               & -J_x' + J_x \e^{-i k_x}	&  0		
  \end{pmatrix},
  \label{eq:bloch_hlt}
\end{equation}
where we set $J_x = J_y = J + \delta J$, $J_x' = J_y' = J - \delta J$ and $J_z = \Jz$. In the following, we restrict to the case of $\delta J = 0$.

\subsection{Vison excitations as additive disorder}

The Bloch Hamiltonian of Eq.~\eqref{eq:bloch_hlt} corresponds to the ground state flux sector with $\pi$-flux through each 4-plaquette. Flux-disordered configurations, corresponding to the model at finite temperatures, can be obtained starting from this Hamiltonian and flipping the gauge field on a set of bonds, i.e., flipping the sign of one of the $J$ bonds. This leads to the creation of two 4-vison excitations on the two plaquettes of which the bond is a part. Note that this can be done in four different ways, owing to the four inequivalent $J$ bonds, which can be taken as the four intracell hoppings. We think of these vison excitations as an additive disorder and write it schematically as $\hlt = \hlt_0+\sum_{i,\alpha} \mathcal{U}^{(\alpha)}(\vec{r}_i)$, where $\mathcal{U}^{(\alpha)}(\vec{r}_i)$ is an impurity potential matrix at position $\vec{r}_i$ and $\alpha$ indicates the matrix structure corresponding to one of the four inequivalent vison excitations. For instance, such a matrix for flipping the 1--2 bond can be written as 
\begin{equation}
X^{12} = 2 i J 
  \begin{pmatrix}
      0		 & -1	 & 0	 & 0		\\
      1 	 & 0	 & 0   & 0		\\
      0	   & 0	 & 0   & 0		\\
      0	   & 0	 & 0   & 0		
  \end{pmatrix}, 
\end{equation}
and $X^{23}$, $X^{34}$ and $X^{41}$ take a similar form, which can be compactly written as $[X^{ab}]_{ij} = 2iJ (\delta_{ja}\delta_{ib}-\delta_{ia}\delta_{jb})$. A typical disorder configuration is obtained by flipping each $J$ bond with probability $p$. For small $p$, the disorder corresponds to a set of localized scatterers, which can be analyzed via perturbation theory. 

The average effect of disorder is encoded by the self-energy matrix $\vec\Sigma_p(i\omega,\vec{k})$, that appears in the definition of the disorder-averaged Green's function:
\begin{equation}
\mathcal{G}(i\omega,\vec{k}) 
\equiv \left\langle \left[i\omega - \hlt(\vec{k}) \right]^{-1} \right\rangle_\text{dis} 
= \left[i\omega - \hlt_0(\vec{k}) - \vec\Sigma_p(i \omega,\vec{k})\right]^{-1} , 
\end{equation}
where $ \left\langle \, . \, \right\rangle_\text{dis}$ represents averaging over the position and bond direction of the disorder. In the following, we are only interested in the self-energy at $\omega = 0$. Owing to the particle-hole symmetry, the entries in the self-energy  --- as in the real-space Hamiltonian --- must be purely imaginary. We can further identify its matrix structure by noting that the perturbation series above consists of terms of the form $X^{ab} \hlt_0^{-1}(\vec{k}) X^{ab}$, where $X^{ab}$ is one of the disorder matrices defined above. Given the form of $\hlt$, such a product has only two nonzero components at the same position as $X^{ab}$, \emph{viz}, at indices $ab$ and $ba$. Since each type of disorder is equally likely, the disorder-averaged self-energy takes the form 
\begin{equation}
\vec\Sigma_p(0,\vec{k}) = i\Sigma_p(0,\vec{k})
  \begin{pmatrix}
      0		 & -1	 & 0	 & -1		\\
      1 	 & 0	 & -1  & 0		\\
      0	   & 1	 & 0   & -1		\\
      1	   & 0	 & 1   & 0			
  \end{pmatrix}, 
  \label{eq:sigma_mat}
\end{equation}
where $\Sigma_p(0,\vec{k}) \in\mathbb{R}$. Since the coefficient (matrix) of $J$ in the Bloch Hamiltonian takes precisely this form for any $k_x = k_y$, the effect of disorder is essentially a renormalization of $J$. In particular, at the quantum critical point $\Jz = 2\sqrt2 J_0$, the spectrum consists of a single Dirac point at $\vec{k}_0 = (\pi,\pi)$, so that we write $J \rightarrow J_\text{eff}=J-\Sigma_p(0, \vec{k}_0)$. This can also be interpreted as renormalizing the mass of this Dirac fermion, and the phase boundary corresponds to the new $\Jz$ for which this mass vanishes.

To compute the disorder averaged self-energy, we expand it in powers of $p$ as 
\begin{align}
\vec\Sigma_{p}(\omega, \vec k)&=p\left(\:
\vcenter{\hbox{\includegraphics[scale=0.35]{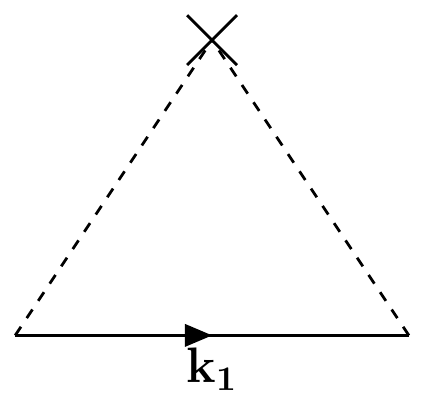}}}
+\vcenter{\hbox{\includegraphics[scale=0.35]{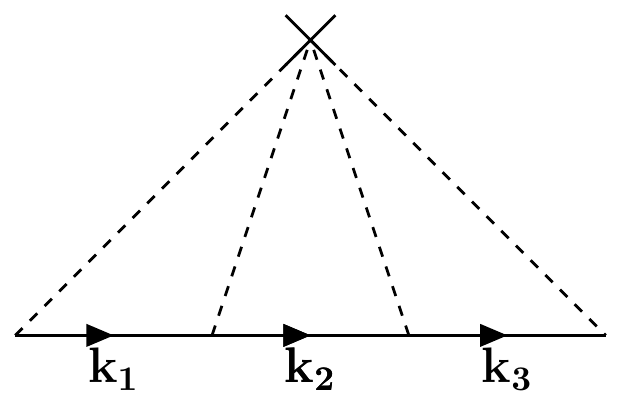}}}
+\dots\right) 
+p^2\left(\vcenter{\hbox{\includegraphics[scale=0.35]{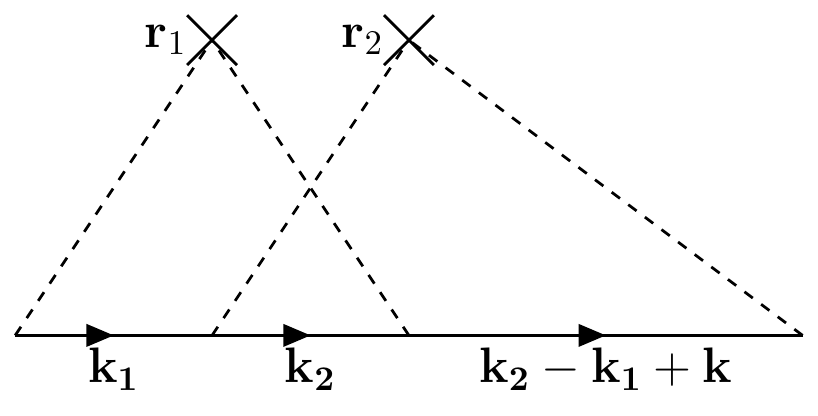}}}+\dots\right)+\mathrm{O}(p^3),\label{ImpExp}\\ 
&=p \vec{T}^{(1)}+p^2 \vec{T}^{(2)}+\mathrm{O}(p^3),  \nonumber
\end{align}
where $\vec{T}^{(n)}$ has the same matrix structure as $\vec\Sigma_p$ (Eq.~\eqref{eq:sigma_mat}) with coefficient $i T^{(n)}$. 
Physically, this corresponds to an expansion in the number of scattering events \cite{Mahan}. Note that we only have terms with an even number of scattering events, since those with an odd number cancel after disorder averaging. The self consistent version of Eq.~\eqref{ImpExp} is obtained by simply replacing the bare Green's functions with the full disorder-averaged Green's functions\cite{Mahan}.

\subsection{The Born and T-matrix approximations} 
At low disorder densities, we can truncate the perturbation series of Eq.~\eqref{ImpExp} using the T-matrix approximation to compute the disorder-averaged self-energy at linear order in $p$  i.e., only including terms corresponding to a single impurity scattering, so that $\Sigma_p(0,\vec{k}) \approx p T ^{(1)}$. This can be further simplified by taking the Born approximation to the full T-matrix $T ^{(1)}\approx T ^{(1)}_{\rm Born}$, wherein we retain only the first Feynman diagram of Eq.~\eqref{ImpExp}. The self-consistent versions of these approximations can be calculated by replacing the bare Green's function in Eq.~\eqref{ImpExp} with their disorder-averaged counterparts (which contains $\vec\Sigma_p(i\omega, \vec{k})$). Setting $J=1$ throughout this subsection, the renormalization of the coupling constant $J$ for all these approximations take the form 
\begin{equation}
  J_\text{eff}(p) = 1 - p T ^{(1)}, \qquad  T^{(1)}>0.
\end{equation}
The system is now gapless when $\Jz = 2\sqrt2 J_\text{eff}(p)$, which yields the phase boundary on the $\Jz$--$p$ plane. The net effect of the disorder is a shift of the phase boundary towards smaller $\Jz$.  

The validity of the approximations discussed above is, at first sight, unclear. This is because our system is gapped and so the usual small parameter for a Fermi sea, $1/(k_F \ell)$ -- with $\ell$ being the mean free path -- is not well defined. Nonetheless, the results for the phase boundary obtained using these approximations agree qualitatively with the numerical results obtained from a transfer matrix calculation, as shown in Fig.~\ref{fig:Tmat}. Furthermore, they also suffice to explain the origin of a ``thermal metal'' phase (Fig.~\ref{fig:PhaseDiagramShastryChern}) and predict its location in the phase diagram.

The basic mechanism for the appearance of the thermal metal is the breakdown of the T-matrix approximation as $p$ is increased and the interference effects from by multiple-impurity scattering become relevant. More precisely, since the disorder strength (i.e., the prefactor in $X^{ab}$) is $2$ and each impurity must have an even number of scattering events, the leading order term for an $n$-impurity scattering scales as $(4 p)^n$. These terms can be neglected only when $4p \ll1$. However, as the disorder density increases to $p \sim 1/4$, the diagrams at all orders in Eq.~\eqref{ImpExp} become relevant, so that the coefficients of $p^n$ in $\Sigma_p(0,\vec{k}_0)$ must be taken into accout for all $n$. This means that for each value of $\Jz$, the equation $\Jz = 2\sqrt2 J_\text{eff}(p)$ can have an infinite number of solutions for $p$. Thus, once this density is reached, it is possible for the phase boundary to become a region on the $\Jz$--$p$ of the disorder phase diagram and a disorder induced metal can be formed. 

Using the various approximations to the T-matrix, we can estimate the coupling constant for which the thermal metal appears by computing the value of $J_\text{eff}(p)$ for which the phase boundary hits the line $p = 1/4 $, i.e., as $  \Jz = 2\sqrt2\left(1 - T^{(1)}/4 \right)$. It turns out that $T^{(1)}_{\rm Born}$ can be computed analytically for $\Jz = 2$, since the corresponding integral can be evaluated using methods for rational trigonometric functions. In fact, for this special value of the coupling constant, we find that
\begin{equation}
  T^{(1)}_{\rm Born}= 8 \int \frac{d^2 \vec k}{(2\pi)^2} \vec G_0(0,\vec k)=4-2\sqrt{2},  
\end{equation}
which exactly satisfies the condition for the phase boundary obtained above. From the Born approximation, we therefore obtain an entirely analytic estimate of the critical value of $\Jz$ where the thermal metal emerges, which is already fairly close to the numerically observed value (see Fig.~\ref{fig:Tmat}). Better approximations can be obtained by including more diagrams in the perturbative expansion of Eq.~\eqref{ImpExp} and/or by using the self-consistent version of this expansion. For example, the self-consistent T-matrix estimates the position of the transition as $\Jz \approx 1.8 $, which is comparable to that found by the transfer matrix technique (Fig~\ref{fig:Tmat}) and quantum Monte Carlo (Fig.~\ref{fig:PhaseDiagramShastryChern}).


\section{Vison gaps}
\label{app:vison}

\begin{figure}[t]
\includegraphics[width=0.47\columnwidth]{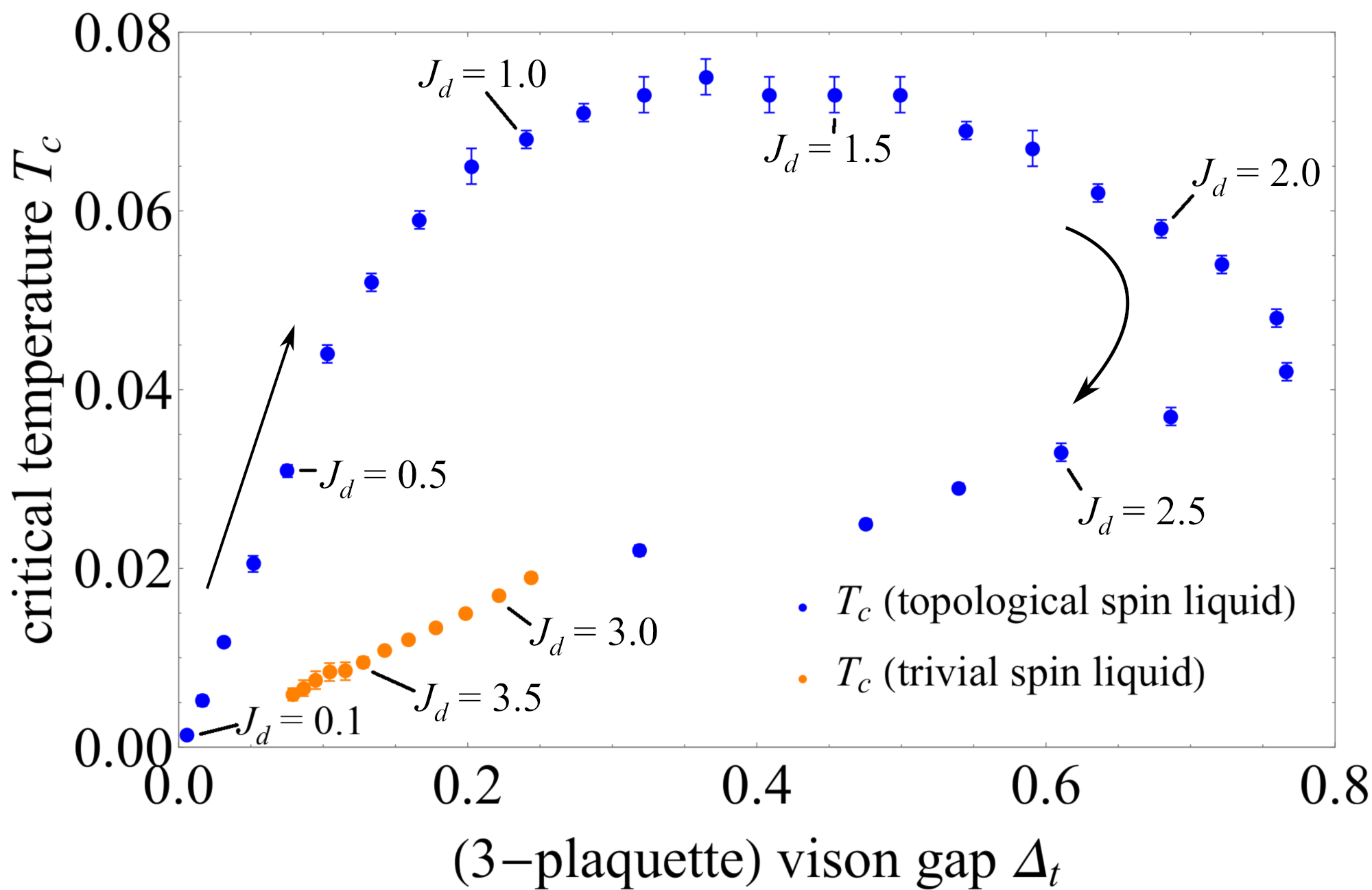}
\hskip 0.05\columnwidth
\includegraphics[width=0.47\columnwidth]{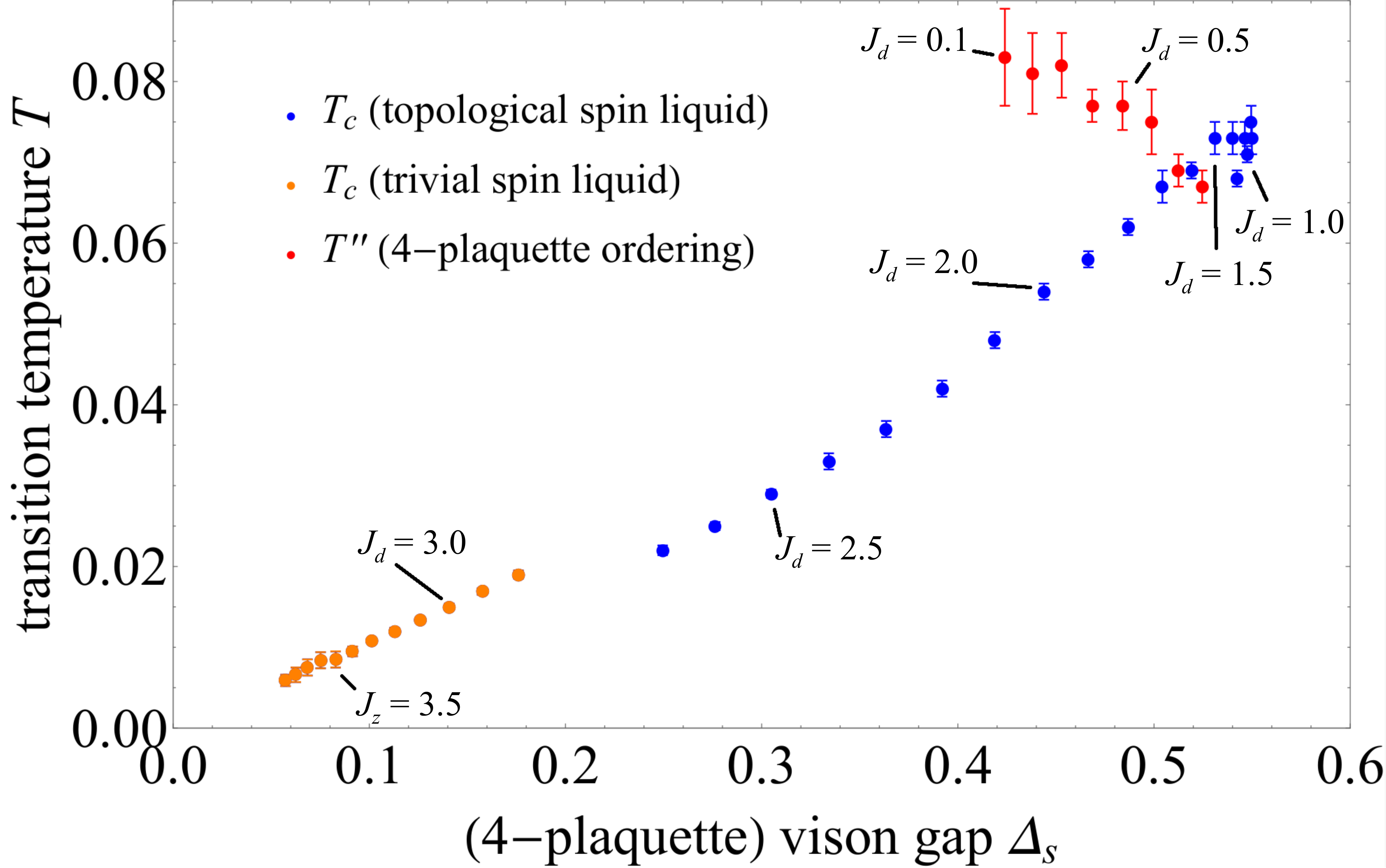}
\caption{{\bf Transition temperatures} of the triangle- / square plaquettes as a function of the vison gap $\Delta_t$ / $\Delta_s$. Both curves suggest correlations between the quantities, which supports results from 3D Kitaev systems. The arrows indicate increasing values of $\Jz / J$.}
\label{fig:VisonGapsShastry}
\end{figure}

To complement the discussion of the phase diagram inFig.~\ref{fig:SchematicPhaseDiagram} of the main text we here report on the relation between the transition temperatures and the vison gaps of the system (summarized in Fig.~\ref{fig:VisonGapsShastry}). Corresponding to the two sets of elementary plaquettes in the Shastry-Sutherland lattice, we can distinguish two kinds of vison excitations. A pair of visons on the {\it triangle plaquettes} is created by flipping a diagonal $\Jz$ bond, whereas flipping a horizontal or vertical bond generates a pair of visons on the {\it square plaquettes}. Fig. \ref{fig:VisonGapsShastry} a shows the values of the critical temperature $T_c$ as a function of the triangle-plaquette vison gap $\Delta_t$. Fig. \ref{fig:VisonGapsShastry} b shows the values of the square-ordering transition temperature, which is $T''$ for $\Jz / J \leq 0.9$ and $T_c$ for $\Jz / J \geq 0.9$, as a function of the square-plaquette vison gap $\Delta_s$.

We can determine a pronounced linear correlation between $T_c$ and $\Delta_s$ for a wide range of values $\Delta_s \leq 0.5$, which corresponds to coupling parameter values $\Jz / J > 1.5$ (orange and blue data points). For the partial-flux order limit $\Jz / J < 1$, there is apparently also a linear correlation between the square-ordering temperature $T''$ and $\Delta_s$ with a different (negative) slope (red data points). This implies that a larger vison gap corresponds to a lower transition temperature $T''$ in this limit. For $1 \leq \Jz/J \leq 1.5$, where $\Delta_s$ is the largest, the data points are too close to each other to determine a functional relation. It is here that the $T$-$\Delta_s$ curve ``U-turns'' after the gap $\Delta_s$ reaches its largest value.

For the 3-plaquettes, we see a linear correlation between $T_c$ and $\Delta_t$ for the trivial phase $\Jz / J > 2.8$ and parts of the chiral phase $\Jz / J \geq 2.3$, where
$\Delta_t$ has its maximum value. The pronounced ``U-turn'' of the $T_c$-$\Delta_t$-curve thereafter corresponds to moving $\Jz$ to lower values. For $\Jz / J \leq 1$, there is again a linear correlation between both quantities.

We can state that the sections with a linear correlation between the transition temperature and the vison gap are consistent with results from 3D Kitaev systems \cite{2020EschmannThermalClassification}. However, the ``U-turn''-behavior that is witnessed for both the $T$-$\Delta_s$- and the $T_c$-$\Delta_t$-curve suggests that the relation between transition temperature and gap is not a simple, global linear function $T(\Delta) = m \Delta$. Instead, the slope $m$ is changed in different parameter regions, whereas in the region of extremal $\Delta$-values, there is no linear correlation at all. Nonetheless, it can be stated that a general correlation between both quantities is verified by these results.

\end{document}